\documentclass[11pt]{article}
\usepackage{comment,soul}
\usepackage[hidelinks]{hyperref}
\usepackage{enumitem}
\usepackage{amsmath,amssymb,epsf,cite,graphicx,subfigure}
\usepackage{amsmath,braket,tikzsymbols}
\usepackage[bbgreekl]{mathbbol}
\usepackage{comment}
\numberwithin{equation}{section}

\definecolor{airforceblue}{rgb}{0.36, 0.54, 0.66}
\newcommand{\beq}{\begin{equation}}
\newcommand{\eeq}{\end{equation}}

\begin{document}
\baselineskip=15.5pt
\pagestyle{plain}
\setcounter{page}{1}

\begin{center}
{\LARGE \bf Wormholes and black hole microstates in AdS/CFT}
\vskip 1cm

\textbf{Jordan Cotler$^{1,a}$ and Kristan Jensen$^{2,b}$}

\vspace{0.5cm}

{\it ${}^1$ Society of Fellows, Harvard University, Cambridge, MA 02138, USA \\}

\vspace{0.3cm}

{\it ${}^2$ Department of Physics and Astronomy, University of Victoria, Victoria, BC V8W 3P6, Canada\\}

\vspace{0.1cm}

{\tt  ${}^a$jcotler@fas.harvard.edu, ${}^b$kristanj@uvic.ca\\}

\medskip

\end{center}

\vskip1cm

\begin{center}
{\bf Abstract}
\end{center}
\hspace{.3cm} 
It has long been known that the coarse-grained approximation to the black hole density of states can be computed using classical Euclidean gravity. In this work we argue for another entry in the dictionary between Euclidean gravity and black hole physics, namely that Euclidean wormholes describe a coarse-grained approximation to the energy level statistics of black hole microstates. To do so we use the method of constrained instantons to obtain an integral representation of wormhole amplitudes in Einstein gravity and in full-fledged AdS/CFT. These amplitudes are non-perturbative corrections to the two-boundary problem in AdS quantum gravity. The full amplitude is likely UV sensitive, dominated by small wormholes, but we show it admits an integral transformation with a macroscopic, weakly curved saddle-point approximation. The saddle is the ``double cone'' geometry of Saad, Shenker, and Stanford, with fixed moduli. In the boundary description this saddle appears to dominate a smeared version of the connected two-point function of the black hole density of states, and suggests level repulsion in the microstate spectrum. Using these methods we further study Euclidean wormholes in pure Einstein gravity and in IIB supergravity on Euclidean AdS$_5\times\mathbb{S}^5$. We address the perturbative stability of these backgrounds and study brane nucleation instabilities in 10d supergravity. In particular, brane nucleation instabilities of the Euclidean wormholes are lifted by the analytic continuation required to obtain the Lorentzian spectral form factor from gravity. Our results indicate a factorization paradox in AdS/CFT.

\newpage

\tableofcontents

\section{Introduction}

String theory and the AdS/CFT correspondence provide powerful frameworks for studying quantum mechanical aspects of black holes.  In some scenarios string theory gives a microscopic counting of black hole microstates~\cite{strominger1996microscopic}, and holographic duality has allowed us to bring CFT knowledge to bear on black hole physics.  Of course it is useful to study aspects of black holes directly in quantum gravity without using a dual description.  In this paper we address a particular question~\cite{Cotler:2016fpe} directly in the bulk: what are the energy level statistics of black hole microstates?

As a concrete example, consider the original version of AdS/CFT for type IIB string theory on AdS$_5 \times \mathbb{S}^5$~\cite{maldacena1999large}, where the AdS factor has an $\mathbb{S}^3$ spatial boundary.  String theory on this background is dual to $\mathcal{N} = 4$ super Yang-Mills on $\mathbb{S}^3$; this CFT has a large number of heavy states dual to black hole microstates, labeled by quantum numbers whose details are beyond our reach. Indeed, determining the values of these quantum numbers, chiefly the energies, amounts to a deeply non-perturbative question from either the bulk or boundary points of view.

However, while the precise details of the black hole spectrum are inaccessible in semiclassical gravity, we can access the coarse-grained approximation to the black hole density of states by a controlled computation in Euclidean gravity. Famously, one obtains the black hole equation of state~\cite{hawking1974black, hawking1975particle, gibbons1978black} by computing the action of the Euclidean continuation of a Lorentzian black hole, imposing that the Euclidean section is smooth. This particular result belongs to the tradition of using the Euclidean gravitational path integral to unearth features of Lorentzian black holes~\cite{gibbons1978black}.  

While the density of states encodes the coarse-grained profile of the microstate spectrum, what if we wanted to calculate the coarse-grained two-point level statistics?  In this paper, we find a Euclidean gravity computation that addresses this question for a variety of AdS black holes.  In particular, we argue that such two-point energy level statistics are encoded by Euclidean wormholes with the same boundary topology as the black hole being studied.  This adds a new entry to the dictionary between Euclidean gravity and black hole physics. 

This result has some precedent in the study of simple models of low-dimensional gravity. Wormhole amplitudes have been computed in nearly AdS$_2$ Jackiw-Teitelboim (JT) gravity~\cite{Saad:2019lba, stanford2019jt} and for pure AdS$_3$ gravity~\cite{cotler2020ads, cotler2020ads2}. These Euclidean wormholes have two asymptotic regions, and as such give a connected contribution to the two-boundary problem in AdS gravity. This two-boundary problem encodes the two-point function of the black hole density of states in these models.  In each case, the statistics encoded in the wormhole amplitudes reveal that the energies exhibit long-range level repulsion in a manner quantitatively matching random matrix theory.  Interestingly, the statistics have a distinctive signature indicating that the underlying theories are each ensemble-averaged; this is understood in detail for JT gravity which is dual to a double-scaled matrix model.  The AdS$_3$ setting is more mysterious, and the details of an ensemble-averaged dual, or even the consistency of the theory itself, are not presently known. There has been much recent work on formulating and reasoning about ensemble-averaged holography, in particular in two and three dimensions. See e.g.~\cite{pollack2020eigenstate, belin2020random, 1800406, 1800422, maxfield2020path, bousso2020gravity}.

Level repulsion is a generic feature of many-body chaotic quantum systems, often emulating  random matrix theory~\cite{mehta2004random, haake1991quantum}. This genericity is a good reason to expect that the spectrum of black hole microstates exhibits level repulsion in general~\cite{Cotler:2016fpe}, and not only in the simple models of quantum gravity mentioned above. For this reason we would like to compute wormhole amplitudes in higher-dimensional gravity and especially full-fledged AdS/CFT. 

A core problem is that it is a priori impossible to adapt the JT and 3d computations to higher-dimensional gravity.  JT gravity and pure 3d gravity are power-counting renormalizable as well as topological. However, in four and higher dimensions, gravity is nonrenormalizable and there are propagating gravitons, and so it is not clear how many of the lessons from low-dimensional gravity carry over.

At a technical level, in JT and AdS$_3$ gravity, the wormhole amplitudes are non-perturbative objects with no saddle point approximation.\footnote{In Euclidean AdS$_3$ gravity there are wormhole solutions to the field equations when the boundaries are surfaces with genus $g\geq 2$~\cite{maldacena2004wormholes}. These uplift to Euclidean wormholes in the $D1/D5$ system, although it is not entirely clear if the dual CFT exists when placed on higher genus surfaces.} Nevertheless one can compute the amplitudes, ultimately because JT and pure 3d gravity do not have bulk excitations, but only boundary excitations and moduli. In four and higher dimensions, the generic situation is that wormhole amplitudes are akin to those of lower-dimensional gravity, in that they are non-perturbative and inherently off-shell, but without any simplifications.\footnote{
There are known wormhole solutions in pure Einstein gravity or supergravity which are stable in certain sectors, although many such solutions are non-generic. For instance there are axion wormholes in various supergravities~\cite{giddings1988axion, lavrelashvili1987disruption, arkani2007euclidean}.
There are also wormhole solutions of pure Einstein gravity with negative cosmological constant, but only when the boundary is negatively curved~\cite{maldacena2004wormholes}. Finally, there are certain solutions in 10- and 11-dimensional supergravity where the boundary is positively curved but there is a non-trivial $R$-symmetry background in the dual CFT~\cite{maldacena2004wormholes, marolf2021ads}.} Consider the simplest possible setting where the boundary is flat or positively curved and we do not turn on sources for any other operators. Then the Witten-Yau theorem states that the Einstein field equations do not admit Euclidean wormhole solutions~\cite{witten1999connectedness}. In this ``vanilla'' setting the wormhole amplitude, assuming it exists, is then intrinsically non-perturbative and unlike an ordinary instanton has no saddle point approximation.

Despite these difficulties, progress was made in~\cite{Saad:2018bqo} on spectral statistics in higher-dimensional gravity through a new family of solutions in Einstein gravity coming from timelike orbifolds of two-sided AdS black holes.  These saddles were coined ``double cones'' to reflect their shape and orbifold singularity. A key observation is that these saddles come with a zero mode whose volume is proportional to the length of the $\mathbb{S}^1$ factor of the boundary (the timelike orbifolded direction), and that the wormhole amplitude is proportional to this length. This result is consistent with the long-range energy level repulsion of black hole microstates. However, the approach of~\cite{Saad:2018bqo} came with some puzzles, including the orbifold singularity, subtleties with infinite-temperature physics, and flat directions at tree level (the mass and angular momenta of the black hole before orbifolding).  In~\cite{Saad:2018bqo} there is a proposal for the dual description of the double cone and how to stabilize the mass via an excursion into complex metrics, but one might worry about the rules of the game when it comes to complex saddles in quantum gravity.

In light of historical and recent work in Euclidean quantum gravity, we are emboldened to take Euclidean gravity seriously as an effective field theory; our ideology is to imitate the rules of effective field theory as much as possible, with the exceptions that in weakly coupled gravity we expect to sum over topologies as long as the metrics at hand are macroscopic and smooth. Our approach is to identify a computational method in both the nearly-AdS$_2$ JT and AdS$_3$ examples which can be adapted to higher-dimensional gravity, namely the method of constrained instantons which has a long history in ordinary field theory~\cite{affleck1981constrained,affleck1984dynamical}, allowing the computation of instanton corrections in scenarios where there are no instanton solutions to the field equations. (See Section~\ref{S:constrainedInstantons} for a review.)

This technology allows us to obtain Euclidean wormhole amplitudes in four and higher dimensions that indeed generalize those of JT and pure AdS$_3$ gravity, and furthermore appear to encode the two-point energy level statistics of black hole microstates in a way consistent with level repulsion. These wormholes have a pseudomodulus, which can be understood as the energy perceived on the boundary, and on general grounds we find that the wormhole amplitude at zero fixed spin can be written as an integral over this energy, of the schematic form
\beq
\label{E:theAmplitude}
	Z_{\rm wormhole}(\beta_1,\beta_2) = \int_{E_0}^{\infty} dE\, f(E;\beta_1,\beta_2) \,e^{-(\beta_1+\beta_2)E} (1 + O(G))\,.
\eeq
Here $\beta_1$ and $\beta_2$ are the sizes of the Euclidean time circles on the two boundaries and $E$ is the energy carried by the wormhole. The domain of integration $E\geq E_0$ precisely corresponds to the allowed energies of the microstates of non-rotating black holes in AdS. The exponential piece $(\beta_1+\beta_2)E$ comes from a classical gravity computation, $f(E;\beta_1,\beta_2)$ from one-loop effects at fixed energy, and the corrections from two and higher loops. The integrand admits an effective field theory approximation far from the spectral edge, but the Boltzmann suppression indicates that the full amplitude is dominated by the low-energy, small bottleneck limit, where curvatures blow up and presumably one requires the details of the ultraviolet completion. Nevertheless we can extract level statistics away from the spectral edge with a controlled EFT approximation, by taking the integral transform of $Z_{\rm wormhole}$ employed in~\cite{Saad:2018bqo}, morally a microcanonical version of the amplitude. This observable, closely related to the two-point function of the density of states, admits a saddle-point approximation in gravity, and the saddle is in fact the double cone of~\cite{Saad:2018bqo} with its moduli stabilized.\footnote{In terms of the Euclidean data $\beta_1$ and $\beta_2$, the double cone is an analytic continuation of the Euclidean wormhole with $\beta_1 = - \beta_2 = iT$ and $T$ the size of the orbifolded circle. The Boltzmann-like factor $e^{-(\beta_1+\beta_2)E}$ then vanishes. However in order to stabilize the moduli it is crucial that one integrates over fluctuations of $\beta_1$ and $\beta_2$ away from these values, so that the Boltzmann factor is non-trivial. The ensuing wormholes are off-shell configurations in gravity, exactly the sort accessible with the method of constrained instantons.}  Our results provide new evidence for level repulsion in the black hole microstate spectrum, with the advantage that we work with smooth Euclidean geometries throughout. (We arrive at the double cone only after finding the microcanonical version of the Euclidean amplitude.)  We discuss another bulk observable, whose boundary interpretation is not yet clear, which admits a saddle-point approximation where the saddle is a genuine macroscopic Euclidean wormhole.

In order for the integral representation~\eqref{E:theAmplitude} to be sensible, it is important that the wormholes under consideration are stable against quadratic perturbations. We continue the perturbative analysis of our previous work~\cite{cotler2020gravitational} with the result that the simple wormholes we study with $\beta_1=\beta_2$ are perturbatively stable.

We perform a similar, albeit more involved analysis for the paradigmatic example of AdS/CFT, type IIB string theory on $\text{AdS}_5 \times \mathbb{S}^5$; we find gravitational constrained instantons in type IIB supergravity with $N$ units of 5-form flux. These are likewise wormholes which appear to encode level repulsion in black hole microstate statistics. To confirm that these gravitational constrained instantons are sensible, we check that they are stable with respect to the most dangerous fluctuations of the supergravity fields. We also study contributions to the wormhole amplitude arising from the nucleation of $D3-\overline{D3}$ brane pairs in the wormhole. Wormholes with nucleated 3-brane pairs are stable and in fact dominate over the wormholes for a wide range of moduli space. However, in order to study whether or not there is level repulsion it is convenient to perform a certain analytic continuation of the wormhole amplitude. This continuation is the one relevant to obtain the Lorentzian spectral form factor as well as to find the double cone of~\cite{Saad:2018bqo}.  We find that this continuation lifts the 3-brane nucleation instability, so that the wormhole is the most dominant contribution among those presently known.  This suggests that we can robustly study black hole microstate level statistics in string theory.

In all of these cases the level statistics we find have the structure indicative of a disorder-averaged theory.  This is in contrast with the conventional expectation that in stringy examples of holographic duality the boundary theory is a single CFT, not an ensemble. So these wormholes imply a sharp factorization paradox in AdS/CFT. However, the ``paradoxical'' contributions from wormholes encode useful physics we have every right to expect from black holes, namely level repulsion. If factorization is fixed by other, perhaps non-geometric contributions to the two-boundary amplitude, then those contributions must exactly cancel out these physically reasonable contributions from the wormholes.\footnote{For an example of similar effects in a simple model of AdS/CFT, the tensionless string on $\mathcal{M}_3 \times \mathbb{S}^3 \times \mathbb{T}^4$, see~\cite{eberhardt2021summing}.} This paradox is made all the stranger by the stabilized version of the double cone, which gives a non-perturbatively suppressed but still rather large violation of factorization, proportional to the length of the orbifolded boundary circle. There are also other non-vanilla yet still stable wormhole saddles~\cite{marolf2021ads} which are inconsistent with factorization. All told, we have to wonder if these wormholes are a feature rather than a bug. Toward this end, in the Discussion we suggest a mechanism for an ensemble interpretation even in AdS$_5\times\mathbb{S}^5$.

The rest of the paper is organized as follows. In Section~\ref{S:reviewRMTgrav} we review energy level statistics, focusing on random matrix theory and low-dimensional gravity. In Section~\ref{S:reviewWormholes} we review and expand upon the calculus of constrained instantons, and develop a new method for finding important off-shell configurations in gauge theory and gravity. These are our main technical tools for studying wormhole amplitudes. We provide an intermediate summary of these results in Section~\ref{S:fromWormholesToStatistics}, focusing on the integral representation of the wormhole amplitude we mentioned above. Using this general form we show how to extract observables, closely related to the two-point function of the density of states, which admit a controlled semiclassical approximation around a saddle point. One of these saddles is in fact the double cone geometry of~\cite{Saad:2018bqo}. 

In Section~\ref{S:stability} we perform a perturbative stability analysis around our Euclidean wormholes.  We find that these wormholes are perturbatively stable when $\beta_1=\beta_2$.  In Section~\ref{S:stringwormholes} we adapt our methods to embed wormholes into full-fledged AdS/CFT, focusing on type IIB supergravity on $\text{AdS}_5 \times \mathbb{S}^5$, and provide evidence that these wormholes are perturbatively stable by studying the fluctuations of the most dangerous instability channels. In Section~\ref{S:branes}, we study instanton contributions to the wormhole amplitude which correspond to brane nucleation and brane dynamics. Embedded into supergravity, these wormholes are generically unstable to the nucleation of brane-antibrane pairs, which screen the Ramond-Ramond flux supporting the wormhole. This instability is a non-perturbative one: wormholes with nucleated brane pairs have lower action with the wormholes without. However, we find that this nucleation instability disappears when we extract the late-time spectral form factor of the dual theory on $\mathbb{S}^3$ from the two-boundary amplitude. We conclude with a discussion in Section~\ref{S:discussion}, focusing on factorization in AdS/CFT and the holographic dictionary more broadly, and future directions.

\emph{Note}: As this manuscript was nearing completion we were made aware of the work of~\cite{Mahajan2021wormholes} which also studies aspects of double cones and brane nucleation.

\section{Spectral form factor in RMT and low-dimensional gravity}
\label{S:reviewRMTgrav}
\subsection{Overview of long-range level repulsion in RMT}
\label{S:RMT}

Let us be more precise about the kind of energy level statistics we seek to investigate.  For simplicity, consider a $d \times d$ Hamiltonian $H$ with eigenvalues $E_1,...,E_d$.  We can write the density of states as the delta comb $\rho(E) = \frac{1}{d}\sum_{i = 1}^d \delta(E- E_i)$ and the 2-point correlations can be similarly written as $\rho_2(E,E') = \frac{1}{d^2} \sum_{i,j=1}^d \delta(E - E_i) \delta(E' - E_j)$.  We will be primarily interested in $\rho_2(E,E')$ for energy eigenstates corresponding to black hole microstates.  In this setting, what do we expect to find?

A useful proxy for black holes are quantum chaotic systems, which have characteristic patterns of energy level statistics that emulate random matrix theory~\cite{Cotler:2016fpe}.  As such, let us consider $\rho_2(E,E')$ for a random matrix theory.  Suppose we have some measure $e^{-d \, \text{tr}(V(H))}\,dH$ over $d \times d$ Hamiltonians, and define the ensemble-averaged quantities $\rho(E) = \left\langle \frac{1}{d}\sum_{i = 1}^d \delta(E- E_i) \right\rangle$ and $\rho_2(E,E') = \left\langle\frac{1}{d^2} \sum_{i,j=1}^d \delta(E - E_i) \delta(E' - E_j)\right\rangle$.  Then for a wide range of random matrix ensembles, we find the universal result~\cite{mehta1960statistical, gaudin1961loi, dyson1962statistical}
\begin{equation}
\label{E:levelrepulse1}
\rho_2(E, E') - \rho(E) \rho(E') \sim - \frac{1}{d^2(E- E')^2}
\end{equation}
which indicates long-range level repulsion between eigenvalues.

\begin{figure}[t!]
\begin{center}
\includegraphics[width=4in]{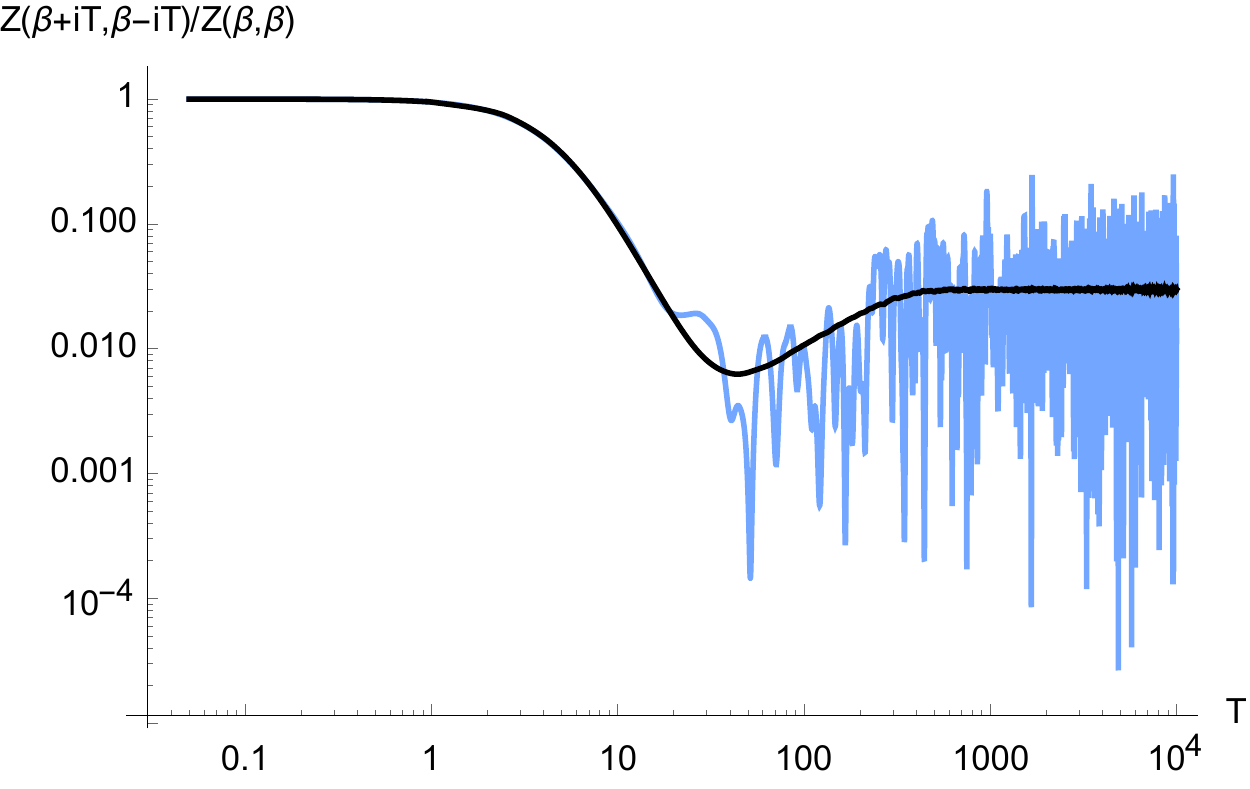}
\caption{\label{F:GUE1} A log-log plot of the spectral form factor $Z(\beta + i T, \beta - i T)$ (normalized by $Z(\beta, \beta)$) as a function of $T$ for the Gaussian Unitary Ensemble (GUE).  We have considered the GUE for $500 \times 500$ matrices with the real and imaginary parts of each matrix element sampled from a Gaussian distribution with variance $1/500$.  The black curve is averaged over $10000$ samples and the light blue curve is a single sample.}
\end{center}
\end{figure}

For our purposes, there is a convenient repackaging of the same information.  Namely, we define the spectral form factor (see e.g.~\cite{mehta2004random})
\begin{equation}
Z(\beta_1, \beta_2) = \left\langle \sum_{i,j=1}^d e^{- \beta_1 E_i} e^{-\beta_2 E_j} \right\rangle = \left\langle \text{tr}(e^{- \beta_1 H}) \,\text{tr}(e^{- \beta_2 H})\right\rangle
\end{equation}
which is related to $\rho_2(E, E')$ by a double Laplace transform and rescaling by $d^2$.  The long-range level repulsion exhibited in~\eqref{E:levelrepulse1} has a clear signature in the spectral form factor for certain complex arguments, namely $Z(\beta + i T, \beta - i T)$.  A log-log plot of the (normalized) spectral form factor for a canonical random matrix ensemble, namely the Gaussian Unitary Ensemble (GUE), is shown in Fig.~\ref{F:GUE1}.  Let us use the terminology from~\cite{Cotler:2016fpe}.  The black curve is disorder-averaged over the GUE, and has an initial downward slope ending in a dip (the minimum of the curve), followed by a linear ramp which terminates in a plateau.  The initial sloping behavior is due to the disconnected contributions to the spectral form factor, i.e.~$\langle \text{tr}(e^{- \beta_1 H})\rangle \langle \text{tr}(e^{- \beta_2 H})\rangle$.  The feature of interest is the linear ramp, which persists for a time which scales as the dimension $d = 500$ of the Hilbert space on which the Hamiltonians act.  This ramp is a direct consequence of the long-range level repulsion in~\eqref{E:levelrepulse1}, and is the feature we will look for in AdS quantum gravity.  The ultimate plateau is a more subtle effect, and is due to the finiteness of the level spacings.

Another notable feature of the spectral form factor is exhibited by the light blue curve in Fig.~\ref{F:GUE1}.  This curve is the spectral form factor for a single instance of the GUE ensemble.  While the light blue curve coincides with the disorder-averaged spectral form factor at early times, this is not the case at later times relevant for the ramp and plateau.  In this sense, the spectral form factor is not self-averaging at late times~\cite{prange1997spectral}.  However, by smearing the single instance via a running time average, one can obtain a curve which is close to the ensemble-averaged one.  Accordingly, at late times we can thing of a single instance as looking like the ensemble-averaged curve with large fluctuations.

For the ramp region which is our main interest, there is evidently a stark difference between the spectral form factor for a single theory versus an ensemble average over many theories.  Since known theories of gravity in four and higher dimensions are expected to be individual theories rather than ensemble averages, we might expect such theories of quantum gravity to provide us with a spectral form factor like the light blue curve in Fig.~\ref{F:GUE1}. Curiously, we do not find evidence that this is the case which has potentially profound consequences to be discussed later.

\begin{figure}[t!]
\begin{center}
\includegraphics[width=3in]{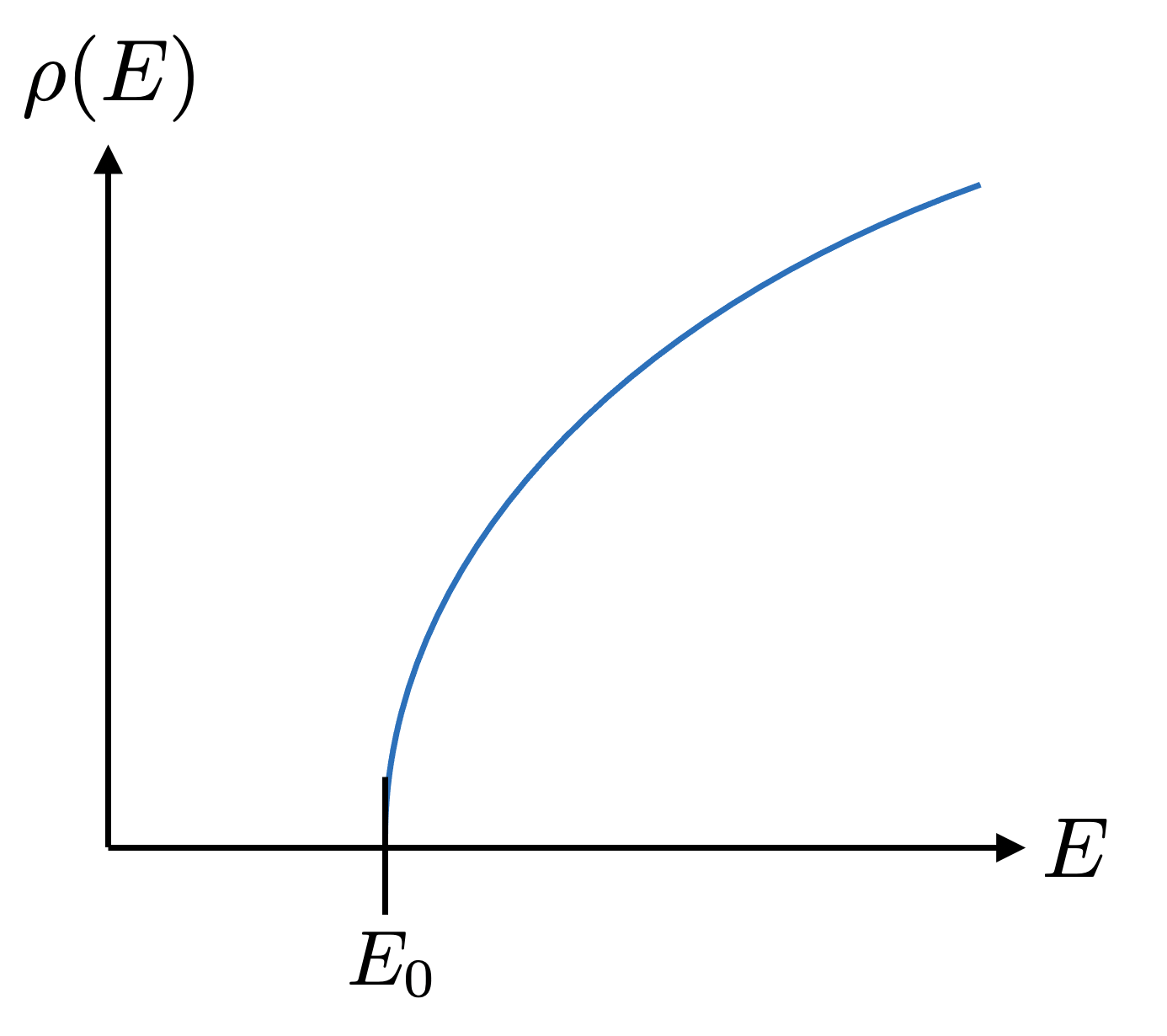}
\caption{\label{F:DOS1} Depiction of a double-scaled density of states.}
\end{center}
\end{figure}

There is a setting where we can directly compute the ramp in the spectral form factor which will be 
important to keep in mind when we turn to wormhole amplitudes in gravity. This setting is double-scaled random matrix theory, in which the ramp contribution has a universal expression~\cite{brezin1993universality, eynard2007invariants, eynard2015random}. Suppose we have a model with a single Hermitian matrix $H$ where expectation values are computed as
\begin{equation}
\langle O(H) \rangle = \int dH \, e^{-d \,\text{tr}(V(H,d))}
\end{equation}
where $V(H,d)$ is a power series in $H$ with coefficients depending on $d$.  A double scaled limit is one for which taking $d \to \infty$ results in a density of states $\rho(E)$ which has a cut at some minimum energy $E_0$ and trails off to infinity for $E > E_0$.  Such a density of states is depicted in Fig.~\ref{F:DOS1}. This is like taking an ordinary matrix model and zooming in on the left edge of the spectrum (alternatively, one can zoom in on the right edge).  If the density of states has a square root edge, i.e.~$\rho(E) \sim C (E-E_0)^{1/2}$ for $E \approx E_0$, then the connected contribution to the spectral form factor has the universal form~\cite{brezin1993universality, eynard2007invariants}
\begin{equation}
\label{E:SFF1}
Z_{\text{conn}}(\beta_1, \beta_2) = \langle \text{tr}(e^{- \beta_1 H}) \, \text{tr}(e^{- \beta_2 H})\rangle_{\text{conn}} = \frac{1}{2\pi} \frac{\sqrt{\beta_1 \beta_2}}{\beta_1 + \beta_2} \,e^{-(\beta_1 + \beta_2) E_0}\,\,+\, [\text{non-perturbative}]\,.
\end{equation}
It follows that
\begin{equation}
\label{E:SFF2}
Z_{\text{conn}}(\beta + iT, \beta - iT) = \frac{1}{2\pi} \frac{\sqrt{\beta^2 + T^2}}{2\beta} \,e^{-2\beta E_0}\sim \frac{T}{4\pi \beta} \,e^{-2\beta E_0}
\end{equation}
for large $T$, which provides the linear ramp.  Our findings suggest an expression similar to the latter in gravity.

\subsection{Black hole microstate energy statistics in JT and AdS$_3$ gravity}

There are several settings in low-dimensional gravity in which one can analytically compute long-range energy level repulsion for black hole microstates.  The original example is nearly-AdS$_2$ Jackiw-Teitelboim (JT) gravity, and the closely related Sachdev-Ye-Kitaev (SYK) model. The SYK model is a theory of quantum mechanical Majorana fermions with disordered all-to-all interactions, which at large $N$ and low-temperatures can be written in a way that suggests a 2d stringy interpretation~\cite{kitaev, Maldacena:2016hyu}.  A novel two-replica saddle which encodes the ramp of the spectral form factor was discovered in~\cite{Saad:2018bqo}.  Nearly-AdS$_2$ JT gravity~\cite{Jensen:2016pah, Maldacena:2016upp, Engelsoy:2016xyb} was shown to be dual to a double scaled matrix model~\cite{Saad:2019lba}; since the density of states has a square root edge, the connected contribution to the spectral form factor is precisely given by Eq.~\eqref{E:SFF1} (for $E_0 = 0$) and the ramp is given by the continuation in Eq.~\eqref{E:SFF2}.  This amplitude manifests itself geometrically in the gravitational description as a wormhole with topology $\mathbb{S}^1 \times I$ with two boundaries having renormalized lengths $\beta_1$ and $\beta_2$, respectively.  There are a number of variants of JT gravity which are likewise dual to double scaled random matrix models, and so have connected spectral form factors either identical to Eq.~\eqref{E:SFF1} or of a similar form if the matrix model falls into a different universality class~\cite{stanford2019jt, maxfield2020path, witten2020matrix}.

In the above examples, smooth ramps arise since the theories in question are disordered or in some specific cases are dual to matrix models.  The connection between disordered physics and 2d gravity goes back many years (see e.g.~\cite{brezin1993large}); however, it is conventionally thought that quantum gravity in three and especially four and higher dimensions do not have ensemble descriptions.

In previous work~\cite{cotler2020ads, cotler2020ads2}, we have computed the analog of the ramp in pure 3d Einstein gravity with a negative cosmological constant, strongly suggesting that it bears an ensemble description which generalizes random matrix theory. Since the details (and in fact the existence) of such an ensemble dual are presently unknown, our approach has been to directly work with 3d gravity. By analogy with the nearly-AdS$_2$ JT setting in which a $\mathbb{S}^1 \times I$ wormhole amplitude provides the dominant contribution to the connected spectral form factor for black hole microstate level statistics~\cite{Saad:2019lba}, in pure AdS$_3$ quantum gravity we computed the $\mathbb{T}^2 \times I$ amplitude with two asymptotically Euclidean AdS$_3$ regions with torus boundaries of complex structures $\tau_1$ and $\tau_2$, respectively. The result to at least (and likely beyond) one-loop order is~\cite{cotler2020ads}
\begin{equation}
\label{E:AdS3amp1}
Z_{\mathbb{T}^2 \times I}(\tau_1, \tau_2) = \frac{1}{2\pi^2} \frac{1}{\sqrt{\text{Im}(\tau_1)} |\eta(\tau_1)|^2} \frac{1}{\sqrt{\text{Im}(\tau_2)} |\eta(\tau_2)|^2} \sum_{\gamma \in \text{PSL}(2;\mathbb{Z})} \frac{\text{Im}(\tau_1) \text{Im}(\gamma \tau_2)}{|\tau_1 + \gamma \tau_2|^2}
\end{equation}
where $\eta(\tau)$ is the Dedekind eta function.  Notice that the amplitude is invariant under independent modular transformations of each boundary torus (as ought to be the case in quantum gravity), namely $Z_{\mathbb{T}^2 \times I}(\gamma \tau_1, \gamma' \tau_2) = Z_{\mathbb{T}^2 \times I}(\tau_1, \tau_2)$.  To interpret~\eqref{E:AdS3amp1}, suppose for the moment that pure AdS$_3$ gravity is an ensemble average over CFT$_2$'s.  In such a putative ensemble we can examine $\langle Z_{\mathbb{T}^2}(\tau_1) \,Z_{\mathbb{T}^2}(\tau_2)\rangle_{\text{conn}}$, i.e.~the connected, ensemble-averaged expectation value of the product of two torus partition functions.  By analogy with the JT calculation, we might expect that
\begin{equation}
\langle Z_{\mathbb{T}^2}(\tau_1) \,Z_{\mathbb{T}^2}(\tau_2)\rangle_{\text{conn}} = Z_{\mathbb{T}^2 \times I}(\tau_1, \tau_2) \,\,+\,[\text{non-perturbative}]\,.
\end{equation}

To simplify the answer, we elect to extract the contribution on the left-hand side from primary operators; this corresponds to stripping off the infinite products in the Dedekind eta functions from~\eqref{E:AdS3amp1}.  Further, we Fourier transform the result in $\text{Re}(\tau_1), \text{Re}(\tau_2)$ to work at fixed spin, and take $\text{Im}(\tau_1) = \beta_1$, $\text{Im}(\tau_2) = \beta_2$ in the low-temperature regime to obtain~\cite{cotler2020ads}
\begin{equation}
\langle Z_{s_1}^P(\tau_1) \, Z_{s_2}^P(\tau_2) \rangle_{\text{conn}} = \frac{1}{2\pi} \frac{\sqrt{\beta_1 \beta_2}}{\beta_1 + \beta_2} \, e^{- (\beta_1 + \beta_2) E_{s_1}}\left(\delta_{s_1, s_2} + O(\beta^{-1})\right)\,,
\end{equation}
with $E_s = 2\pi \left(|s| - \frac{1}{12}\right)$ the threshold energy for BTZ black holes with spin $s$.  Note that the left-hand side is interpretational (and as such, can be regarded as suggestive notation) whereas the right-hand side is a due to natural manipulations of~\eqref{E:AdS3amp1} which are suggested by the interpretation.  The left-hand side quantifies the connected correlations between black hole microstates corresponding to primary states, with spin $s_1$ and inverse temperature $\beta_1$, and spin $s_2$ and inverse temperature $\beta_2$, respectively.  The right-hand side, due to our gravity computation, remarkably results in level statistics which exactly match the form of double-scaled random matrix theory, up to corrections suppressed in the inverses of $\beta_1$ and $\beta_2$.  Since we can find a smooth ramp in the Lorentzian spectral form factor by
taking $\beta_1 = \beta + i T$ and $\beta_2 = \beta - i T$, we may conclude that the BTZ microstates corresponding to Virasoro primaries at fixed spin repel.  More generally, one can show that the contribution of $Z_{\mathbb{T}^2 \times I}(\tau_1, \tau_2)$ to the level statistics of BTZ microstates at low temperatures, even those that are non-primaries, are precisely described by a double-scaled random matrix theory with Virasoro symmetry.  These findings provide strong evidence that pure AdS$_3$ gravity is dual to an ensemble, which is corroborated by other recent findings~\cite{maxfield2020path}. 

While JT and pure AdS$_3$ gravity may be dual to ensembles, this could be potentially written off as a peculiarity of low-dimensional gravity.  What hope do we have for studying the spectral statistics of black hole microstates in higher-dimensional gravity, and especially in stringy examples of AdS/CFT?  One clue is that both the nearly-AdS$_2$ JT and pure AdS$_3$ computations of black hole microstate statistics can be arrived at using the method of constrained instantons~\cite{affleck1981constrained, Saad:2019lba, cotler2020ads, cotler2020gravitational}.  We aim to study higher-dimensional generalizations of these objects in pure Einstein gravity and ultimately in string theory.  In the next Section, we review and clarify the framework of constrained instantons~\cite{affleck1981constrained}, and specialize to the setting of gravitational constrained instantons~\cite{cotler2020gravitational} that are relevant for the computation of wormhole amplitudes in higher-dimensional gravity.

\section{Wormholes and constrained instantons}
\label{S:reviewWormholes}

In this Section we review the wormholes obtained in~\cite{affleck1981constrained} and expand on the constrained instanton calculus that leads to them. We go into much greater depth here than in our prior work, presenting a number of pedagogical comments and new insights along the way. The output of this analysis is a family of near-saddles of Einstein gravity, along with a semiclassical approximation to the integral over wormhole metrics. In Subsection~\ref{S:complement} we present another perspective on these wormholes, new to this work, whereby the wormholes are solutions to the field equations of the gauge-fixed theory where the gauge-fixing auxiliary fields pick up nonzero profiles. 

\subsection{Constrained instantons}
\label{S:constrainedInstantons}

We begin with a review of the constrained instanton machinery, originally developed for field theory in~\cite{affleck1981constrained} and adapted for Einstein gravity in~\cite{cotler2020gravitational} (see also~\cite{Stanford:2020wkf}).\footnote{There are many other places where the constrained instanton calculus has been used without calling it as such. Examples include the study of non-BPS multi-instanton configurations; Jackiw-Teitelboim gravity~\cite{Saad:2019lba}; and more recently still, pure gravity in three dimensions~\cite{cotler2020ads}.} The basic idea is to take a field theory with quantum fields $\varphi$, and then for a real constraint functional $\mathcal{C}[\varphi]$ to insert the resolution of the identity
\beq
	1 = \int \frac{d\lambda d\zeta}{2\pi} \,e^{i \lambda\left( \mathcal{C}[\varphi] - \zeta\right)}
\eeq
into the path integral. Here $\lambda$ is integrated over a contour parallel to the real axis, and $\zeta$ is integrated over the possible values of the constraint. Inserting the above into the partition function $Z = \int [d\varphi] e^{-S[\varphi]}$ of a Euclidean field theory, we have the formal identity
\beq
	Z = \int [d\varphi] \frac{d\lambda d\zeta}{2\pi} \,e^{-S_E[\varphi] + i \lambda \left( \mathcal{C}[\varphi] - \zeta\right)}\,.
\eeq
Treating the argument of the exponential as an action $S_{\rm tot}$\,, the saddle points obey
\beq
\label{E:deltaStot}
	\frac{\delta S_E}{\delta \varphi} - i \lambda \frac{\delta \mathcal{C}}{\delta \varphi} = 0\,, \qquad \mathcal{C}[\varphi] - \zeta = 0\,, \qquad \lambda = 0\,.
\eeq
However, suppose we save until last the integration over the possible values $\zeta$. That is, one strategy for performing the path integral is to study physics at fixed value of the constraint $\mathcal{C}$, and then to integrate over the possible values of the constraint. If done correctly we have merely sliced the field integration in a way that depends on the constraint $\mathcal{C}$, but the final result for the partition function is independent of that choice. The trick is to make a useful choice of constraint.

We use the term ``constrained instanton'' to refer to a solution to this constrained problem, i.e. a solution to the first two equations in~\eqref{E:deltaStot},
\beq
\label{E:constraineddeltaS}
	\frac{\delta S_E}{\delta \varphi} - i \lambda \frac{\delta \mathcal{C}}{\delta \varphi} = 0\,, \qquad \mathcal{C}[\varphi] - \zeta = 0\,,
\eeq
where the Lagrange multiplier $\lambda$ enforcing the constraint acquires a nonzero, imaginary value. In fact, when such a constrained instanton exists with $\varphi = \varphi_0$, $\lambda = \lambda_0$, and $\zeta = \zeta_0$ it is in general continuously connected to a line of constrained instantons with $\varphi = \varphi_0 + \delta \varphi$, $\lambda = \lambda_0+ d\lambda$ , and $\zeta = \zeta_0+d\zeta$ with $d\lambda$ imaginary. The argument for this is to fix $d\lambda$ and then linearize the first equation in~\eqref{E:constraineddeltaS} around the original constrained instanton. One may solve for the perturbation $\delta \varphi$ sourced by the perturbation $d\lambda$. Feeding the perturbation $\delta \varphi$ into the second equation determines the perturbation in $\zeta$.

When the field theory is weakly coupled there is now a new candidate semiclassical expansion. Let $g$ denote the weak coupling. Then we may perform the standard weak coupling expansion at fixed $\zeta$, and integrate over $\zeta$ at the end. Schematically, we have
\beq
\label{E:semiZ}
	Z = \int d\zeta \,e^{-S_E[\varphi_{\zeta}]} Z_{1}[\varphi_{\zeta}]\left( 1 + g Z_2[\varphi_{\zeta}] + O(g^2)\right)\,,
\eeq
where $\varphi_{\zeta}$ denotes a constrained instanton at fixed $\zeta$, $S_{E}[\varphi_{\zeta}]$ is the action of the constrained instanton, $Z_{1}[\varphi_{\zeta}]$ is the one-loop determinant from integrating out the quantum fields and $\lambda$ around the constrained instanton, and $Z_2$, $Z_3$, are the two-loop, three-loop, etc. corrections, all at fixed $\zeta$.

\subsubsection{Some finite-dimensional examples}

It is instructive to consider finite-dimensional examples. Take
\beq
	I(N;a) = \int_0^{\infty} d\zeta \int_{-\infty}^{\infty} dy \, \rho(\zeta)\,e^{-N\left( \zeta +(a+\zeta^2)\cosh(y)\right)}\,,
\eeq
which we can analyze at large $N$. The full integral does not admit a saddle point approximation, but the integral at fixed $\zeta$ does around $y=0$. We can think of each point on the ray $y = 0$, $\zeta \geq 0$ as being a constrained instanton, with action $N(\zeta+(a+\zeta^2) \cosh(y))$ and $\rho(\zeta)$ a determinant that arises from integrating out other degrees of freedom at fixed $\zeta$. Using
\beq
	\int_{-\infty}^{\infty} dy\, e^{-N (a+\zeta^2) \cosh(y)} \approx e^{-N(a+\zeta^2)}\sqrt{\frac{2\pi}{N(a+\zeta^2)}}\left( 1 -\frac{1}{8N (a+\zeta^2)} + O\left(\frac{1}{N^2}\right)\right)\,,
\eeq
we have a candidate large $N$ expansion\footnote{Assuming $\rho(\zeta)$ is such that the integral is dominated near $\zeta = 0$, one may systematically expand the integral in large $N$ by rescaling $\zeta \to \zeta/N$, expanding the integrand in powers of $1/N$, and integrating term by term. The leading result is $e^{-N a} \sqrt{\frac{2\pi}{N^3a}}$.}
\beq
\label{E:intexample1}
	I_{\rm semi}(N;a)= \sqrt{\frac{2\pi}{N}} \int_0^{\infty} d\zeta \, e^{-N(\zeta+(a+\zeta^2))}\frac{\rho(\zeta)}{\sqrt{a+\zeta^2}}\left( 1 - \frac{1}{8N(a+\zeta^2)} + O\left( \frac{1}{N^2}\right)\right)\,.
\eeq
Comparing with~\eqref{E:semiZ} we see that $\frac{ \rho(\zeta)}{\sqrt{a+\zeta^2}}$ is playing the role of the one-loop determinant $Z_{\rm 1-loop}[\varphi_{\zeta}]$ around the constrained instanton, the $O(1/N)$ correction maps to the two-loop correction $Z_2[\varphi_{\zeta}]$, and one expects the integral to be dominated by the behavior at $\zeta= 0$.

To test the effectiveness of this approximation we can fix $\rho, a,$ and $N$, numerically integrate the full integral $I$ to arbitrarily high accuracy, and compare with the ``two-loop'' approximation (numerically integrated over $\zeta$). We find for $\rho=a=1$ and $N=100$
\beq
	\sqrt{\frac{N^3a}{2\pi}} \,e^{Na}I = 0.979786\,, \qquad \sqrt{\frac{N^3a}{2\pi}}\,e^{Na} I_{\rm semi} =  0.979779\,,
\eeq
a relative error of only $7\times 10^{-6}$. (We have normalized the integrals so that they go to unity as we take $N\to\infty$ for fixed $a$ and $\rho=1$.) This approximation works even down to $N=2$ where we find $N^{\frac{3}{2}}\sqrt{\frac{a}{2\pi}} \,e^{Na}I =0.59575$ and $N^{\frac{3}{2}}\sqrt{\frac{a}{2\pi}} \,e^{Na}I_{\rm semi} = 0.58914$, again for $\rho=a=1$.  This gives an error of $1.11\%$. 

This example is also useful as it helps us to understand how the approximation can break down. One way is if the saddle-point approximation at fixed $\zeta$ itself breaks down. In our example the saddle-point expansion parameter is not $N$, but $N(a+\zeta^2)$. For $\zeta \leq O(1/\sqrt{N})$ and $a \leq O(1/N)$, the expansion parameter is $O(1)$. In this scenario, the saddle point approximation breaks down at small $\zeta$, the very region that dominates the integral. If we take the strict $a=0$ limit and, say, take $\rho(\zeta) = c\,\zeta$, we would have
\beq
	I_{\rm semi} = \sqrt{\frac{2\pi}{N}} \,c\int_0^{\infty} d\zeta \,e^{-N(\zeta+\zeta^2)} \left( 1 - \frac{1}{8N\zeta^2} + O\left( \frac{1}{N^2}\right)\right)\,.
\eeq
In this case the integral defining the ``one-loop approximation'' (i.e.~where we drop the $1/N$ term and higher corrections) is convergent, but the leading two-loop correction is not.  Indeed, taking $\rho=1$, $a=10^{-3}$, and $N=100$ we have (again comparing to ``two-loop order'', where the two-loop correction now converges on account of the small nonzero $a$)
\beq
	N^{\frac{3}{2}}\sqrt{\frac{a}{2\pi}}\,e^{Na} I =  0.6372\,, \qquad N^{\frac{3}{2}}\sqrt{\frac{a}{2\pi}}\,e^{Na} I_{\rm semi} =  -0.1200\,,
\eeq
so that the semiclassical approximation does not even have the correct sign.

Another possible way the expansion can break down is if the ``one-loop determinant'' $\frac{\rho(\zeta)}{\sqrt{a+\zeta^2}}$ is sufficiently singular as $\zeta \to 0$, or diverges sufficiently fast as $\zeta \to\infty$. However, in this finite-dimensional example, either one of these behaviors also results in a failure of convergence of the integral we started with.

A different finite dimensional example is
\beq
	I = \int_0^{\zeta} d\zeta \int_{-\infty}^{\infty} dy \,\rho(\zeta) e^{-N (\zeta + V)}\,, \qquad V = \frac{m^2(\zeta)}{2}y^2 + \frac{y^4}{4}\,,
\eeq
where $m^2(\zeta)$ varies smoothly between a negative value at $\zeta=0$, and a positive one for $\zeta >\zeta_0$ for some $\zeta_0$. Treating $\zeta$ as the constraint, for $\zeta < \zeta_0$ there are three lines of constrained instantons, at $y=0,y = \pm \sqrt{-m^2(\zeta)}$. $y=0$ is a line of unstable constrained instantons, while the ones at $y= \pm \sqrt{-m^2}$ are stable. These three lines meet at $\zeta_0$, and for $\zeta > \zeta_0$ there is only one line of constrained saddles with $y=0$. This last line is stable. It is easy to check numerically for a simple profile for $m^2$ that the full integral is well-approximated by the ``one-loop'' approximation where one integrates over the stable constrained saddles,
\beq
	I \approx 2\int_0^{\zeta_0} d\zeta \,\rho(\zeta)e^{-N b} e^{\frac{N m^4(\zeta)}{4}}\sqrt{\frac{\pi}{N m^2(\zeta)}} + \int_{\zeta_0}^{\infty} d\zeta \,\rho(\zeta)e^{-N b} \sqrt{\frac{2\pi}{N m^2(\zeta)}}\,.
\eeq
The first integral is taken over the two lines of saddles with $y = \pm \sqrt{-m^2}$ for $\zeta < \zeta_0$, and the second corresponds to the line $y=0$ for $\zeta > \zeta_0$. The part of the integrand that is not $\rho(\zeta) e^{-N b}$ arises from the Gaussian approximation to the integral over $y$ at fixed $\zeta$. This approximation works rather well even for modest $N$. The lesson from this example is that, just like ordinary saddle point integration, the most important configurations are the perturbatively stable saddles (now at fixed constraint) with smallest action. 

The example~\eqref{E:intexample1}, and the way the semiclassical approximation may break down, will be useful to bear in mind when we study constrained instantons in Einstein gravity. 

\subsubsection{Formal expression for the one-loop approximation}
\label{S:formal}

Here we endeavor to obtain a general expression for the one-loop approximation to the integral over constrained instantons. This analysis leads to an important lesson for our study of perturbative stability of wormholes in Einstein gravity and string theory.

Starting from
\beq
	Z = \int [d\varphi] \frac{d\lambda d\zeta}{2\pi} \, e^{-S[\varphi] + i \lambda(\mathcal{C}[\varphi]-\zeta)}\,,
\eeq
suppose we have a continuous set of constrained instantons satisfying $\frac{\delta S_{\rm tot}}{\delta \varphi} = \frac{\partial S_{\rm tot}}{\partial\lambda} = 0$, i.e.
\beq
	\frac{\delta S}{\delta \varphi} - i \lambda \frac{\delta \mathcal{C}}{\delta \varphi} = 0\,, \qquad \mathcal{C}[\varphi] - \zeta = 0\,.
\eeq
Let us label these configurations by $\varphi_{\zeta}$ and $\lambda_{\zeta}$\,. We have in mind a perturbative field theory with a weak coupling expansion parameter $G$, where the on-shell action $S$ is of $O(1/G)$, as is $\lambda\,\mathcal{C}$. Expanding around the constrained instanton at fixed $\zeta$,
\beq
	\varphi = \varphi_{\zeta} + \delta \varphi\,, \qquad \lambda = \lambda_{\zeta} + \delta \lambda\,,
\eeq
we have
\beq
	S_{\rm tot} = S[\varphi_{\zeta}] + \frac{1}{2} \int d^{d+1}x \,d^{d+1}y \,\delta \varphi(x) \frac{\delta^2S_{\rm tot}}{\delta \varphi(x)\delta\varphi(y)}\delta \varphi(y) - i \delta \lambda  \int d^{d+1}x \,\frac{\delta\mathcal{C}}{\delta \varphi} \delta\varphi  + O(\text{fluctuations}^3)\,,
\eeq
where
$\frac{\delta^2S_{\rm tot}}{\delta \varphi^2}$ and $\frac{\delta\mathcal{C}}{\delta\varphi}$ are evaluated on the instanton at fixed $\zeta$. We then integrate out the fluctuations $\delta \varphi$ and $\delta \lambda$. Clearly we must diagonalize the kernel $\frac{\delta^2S_{\rm tot}}{ \delta \varphi^2}$. The fields $\varphi$ in general include bosons and fermions, although the variation of the constraint $\int d^{d+1}x \frac{\delta \mathcal{C}}{\delta\varphi}\delta\varphi$ is purely bosonic. Rather than integrating out $\delta\lambda$ first, which enforces the constraint $\mathcal{C}[\varphi]=\zeta$ at first order in fluctuations, it is convenient to first integrate out the fluctuations of the quantum fields. Zero modes must be treated separately in the usual way, while nonzero modes lead to Gaussian integrals with a linear term owing to the constraint. To write simple expressions we consider a kernel with discrete spectrum, with the obvious generalization to a continuous spectrum.  Integrating out the quantum fields then produces a Gaussian distribution for $\delta \lambda$. Denote the determinant of the fermionic part of the kernel as $\mathcal{D}_{\rm F}'(\zeta)$ and the determinant of the bosonic part as $\mathcal{D}'_{\rm B}(\zeta)$, in both cases omitting zero modes. Let $v_{i}$ be the bosonic eigenfunctions of $\frac{\delta^2S_{\rm tot}}{\delta \varphi^2}$ with eigenvalue $\chi_i$ normalized as $\int d^{d+1}x\,d^{d+1}y \,v_i(x) \frac{\delta^2 S_{\rm tot}}{\delta \varphi(x)\delta\varphi(y)} v_j(y) =\chi_i  \delta_{ij}$, and we further denote
\beq
	\int d^{d+1}x \frac{\delta \mathcal{C}}{\delta \varphi} v_i = \kappa_i\,.
\eeq
Then to one-loop order we have
\beq
	\int [d\varphi]\, e^{-S[\varphi]+i \lambda(\mathcal{C}[\varphi]-\zeta)}  = V_{\rm zm}(\zeta)\frac{\mathcal{D}_{\rm F}'(\zeta)}{\sqrt{\mathcal{D}'_{\rm B}(\zeta)}} \, e^{- \frac{\delta\lambda^2}{2 \sigma(\zeta)}}\left( 1 + \hdots\right)\,, \qquad \sigma(\zeta)= \sum_i \frac{\chi_i}{\kappa_i^2}\,,
\eeq
with $V_{\rm zm}(\zeta)$ the zero mode volume at fixed $\zeta$. The dots indicate higher loop corrections as well as cubic and higher order terms in $\delta \lambda$. We denote the effective width of the distribution for $\delta\lambda$ by $\sigma(\zeta)$, since the data $(\chi_i,\kappa_i)$ of the kernel depends on the instanton under consideration, i.e.~on $\zeta$. Note that the Gaussian action for $\delta\lambda$ is right-sign if the bosonic spectrum of $\frac{\delta^2S_{\rm tot}}{\delta \varphi^2}$ is non-negative, which we implicitly required to integrate out the bosonic fluctuations in the first place. Then to one-loop order at fixed $\zeta$ we have
\beq
	Z = \int d\zeta \,e^{-S[\varphi_{\zeta}]} V_{\rm zm}(\zeta)\frac{\mathcal{D}_{\rm F}'(\zeta)}{\sqrt{\mathcal{D}'_{\rm B}(\zeta)}} \sqrt{\frac{\sigma(\zeta)}{2\pi}}\left( 1+ \hdots\right)
\eeq

This general derivation teaches us that there is a notion of stability for constrained instantons at fixed $\zeta$. Stability requires that the bosonic part of the kernel $\frac{\delta^2 S_{\rm tot}}{\delta \varphi^2}$ has a non-negative spectrum, as one might expect. We have seen that this also implies perturbative stability for the fluctuations $\delta\lambda$ of the Lagrange multiplier. Later we will use this result to study the perturbative stability of the wormholes above with torus and $\mathbb{S}^1\times\mathbb{S}^3$ cross-section, finding that they are indeed stable. 

\subsection{Gravitational constrained instantons}
\label{S:review}

Now we move on from this rather general discussion to consider constrained instantons in Einstein gravity following~\cite{cotler2020gravitational}. Since in this manuscript we want to study Euclidean wormholes in the context of AdS/CFT, we focus on gravity in Euclidean signature with negative cosmological constant. As we outlined in the Introduction, we make the working assumption that consistent theories of quantum gravity include, at weak Newton's constant, an integral over non-singular metrics, and that the difficulties one encounters along the way are resolved in a \emph{bona fide} ultraviolet completion like string theory. That is, we take the effective gravitational field theory seriously as a path integral,
\beq
	Z = \int \frac{[dg]}{\text{Diff}} \,e^{-S_{\rm grav}} \,, \qquad S_{\rm grav} = -\frac{1}{16\pi G}\int d^{d+1}x\sqrt{g} \left( R - 2\Lambda\right) + S_{\rm bdy}\,, \qquad \Lambda = -\frac{d(d-1)}{2}\,.
\eeq
We work in units where the AdS radius is equal to one.  Above, $S_{\rm bdy}$ includes the Gibbons-Hawking term and those boundary terms required by holographic renormalization. The division by diffeomorphisms is accomplished at one-loop order through the standard Faddeev-Popov procedure. Usually for Einstein gravity with a cosmological constant, one does this by fixing a background, expanding in fluctuations around the background, and choosing a gauge-fixing condition for the fluctuations along with a concomitant ghost term. Letting $g_{\mu\nu}$ denote the metric of the background, $h_{\mu\nu}$ the metric perturbation, and $\mathcal{G}_{\mu}[h]$ the gauge-fixing condition, the gauge-fixing and ghost actions are
\beq
\label{E:gaugeghost}
	S_{\rm gauge-fixing} + S_{\rm ghost} =\int d^{d+1}x \sqrt{g} \left( i  f^{\mu}\mathcal{G}_{\mu} + \frac{\xi}{2}f^{\mu}f_{\mu} + b^{\mu} \frac{\delta \mathcal{G}_{\mu}}{\delta \xi^{\nu}}c^{\nu}\right)\,,
\eeq
with $\xi\geq 0$ a parameter, $f^{\mu}$ the gauge-fixing auxiliary field, and $(b^{\mu},c^{\nu})$ the Faddeev-Popov ghosts. For $\xi=0$ we have a delta function gauge, and for $\xi>0$ we have an $R_{\xi}$ gauge, in which the gauge-fixing condition is broadened into a Gaussian average. The gauge-fixed path integral is, at least to one-loop order around the background metric,
\beq
	Z = \int [dg][df][db][dc] e^{-S}\,, \qquad S = S_{\rm grav} + S_{\rm gauge-fixing} + S_{\rm ghost}\,.
\eeq
Gauge-invariance survives as a vestigial memory in the gauge-fixed path integral through a BRST symmetry, which among other things implies that the final result for $Z$ is independent of the gauge-fixing condition.

It is into the gauge-fixed path integral that we now insert a resolution of the identity with a constraint functional $\mathcal{C}[g]$ so that formally
\beq
	Z = \int [dg][df][db][dc] \frac{d\lambda d\zeta}{2\pi} \,e^{-S + i \lambda \left( \mathcal{C}[g]-\zeta\right)}\,.
\eeq
Note that the constraint need not be gauge-invariant, and indeed we will benefit from a non-gauge-invariant constraint. What we do require is that the insertion of the constraint does not spoil BRST invariance. Of course, since all we have done is inserted $1$ into the path integral, it is clear that we have not spoiled BRST, but for completeness we give a simple proof that this is the case. The action of the BRST transformation on the metric is
\beq
	\delta_Q g_{\mu\nu} = D_{\mu}(\varepsilon c_{\nu}) + D_{\nu}(\varepsilon c_{\mu})\,,
\eeq
where $\varepsilon$ is a Grassmann-odd constant and $D_{\mu}$ is the covariant derivative. The new term $-i \lambda (\mathcal{C}[g]-\zeta)$ in the action is trivially invariant under BRST if we define the action on $(\lambda,\zeta)$ as
\beq
	\delta_Q \lambda = 0\,, \qquad \delta_Q \zeta = \delta_Q \mathcal{C}[g]=\int d^{d+1}x\,\frac{\delta \mathcal{C}}{\delta g_{\mu\nu}}\delta_Q g_{\mu\nu}\,.
\eeq

We are interested in Euclidean wormholes with metrics of the form
\beq
	ds^2 = d\rho^2 + g_{ij}(x,\rho) dx^i dx^j\,,
\eeq
where $\rho$ is a radial coordinate and the transverse directions are labeled by $x^i$ with $i=1,...,d$.  We have in mind ``cross-sections'' at constant $\rho$ which are rather symmetric spaces, such as a torus, sphere, or $\mathbb{S}^1\times \mathbb{S}^{d-1}$. The spaces of interest will be asymptotically hyperbolic, with a conformal boundary attained as $|\rho|\to \infty$ with
\beq
	ds^2 \approx d\rho^2 + e^{2|\rho|}g^{(0)}_{ij}(x)dx^idx^j\,.
\eeq
We fix two cutoff slices near the boundaries by demanding that the induced metric on the $\rho \sim \ln \left( \frac{1}{\varepsilon}\right) \gg 1$ slice is $\frac{\gamma_{1ij}(x)}{\varepsilon^2}$ and the induced metric on the $\rho \sim -\ln\left(\frac{1}{\varepsilon}\right) \ll -1$ slice is $\frac{\gamma_{2ij}(x)}{\varepsilon^2}$, with $\gamma_{1ij}$ and $\gamma_{2ij}$ fixed.

Wormholes of this sort contribute to the two-boundary problem in Einstein gravity, in particular to the sum over connected geometries which interpolate between the two boundaries. There are also disconnected spaces, and the genuine saddles of the two-boundary problem are of this sort; namely two copies of one-boundary solutions to the field equations. Effectively, the distance between the two boundaries is infinite for the disconnected configurations. With this in mind (and following~\cite{Saad:2019lba,cotler2020gravitational}) we introduce a constraint related to the length between the two boundaries,
\beq
\label{E:constraint}
	\mathcal{C}[g] = \int d^{d+1}x \,\sqrt{g_{\rho \rho}} \,F(x)\,,
\eeq
weighted by a function $F(x)$ to be explained momentarily. This constraint is invariant under radial reparameterizations, $\rho = \rho(r)$, but not under more general diffeomorphisms. Following our discussion around Eqn.~\eqref{E:constraineddeltaS}, we look for solutions to
\beq
\label{E:constrainedEOMs}
	R_{\mu\nu}-\frac{R}{2}g_{\mu\nu}+\Lambda g_{\mu\nu} -  8\pi i \,G\lambda F(x) \frac{\sqrt{g_{\rho \rho}}}{\sqrt{g}}g_{\rho\rho} \delta_{\mu}^{\rho}\delta_{\nu}^{\rho} = 0\,, \qquad \int d^{d+1}x \,\sqrt{g_{\rho \rho}}\,F(x) - \zeta = 0\,.
\eeq
We presently review two such classes of solutions: (i) solutions with $\mathbb{S}^1\times\mathbb{T}^{d-1}$ cross-section in $d \geq 2$ dimensions, and (ii) solutions with $\mathbb{S}^1\times\mathbb{S}^3$ cross-section in five dimensions. Other examples may be found in~\cite{cotler2020gravitational}. We focus on examples with $\mathbb{S}^1\times X$ cross-section, so that the disconnected solutions include Euclidean black holes with horizon $X$. 

\subsubsection{$\mathbb{S}^1\times\mathbb{T}^{d-1}$ cross-section}

Now we let the $x^i$ parameterize a $d$-dimensional torus. We have in mind a scenario where the two ``boundary CFT's'' are on the same spatial torus $\mathbb{T}^{d-1}$ but at potentially different inverse temperatures $\beta_1$ and $\beta_2$. As such we separate one of the $x^i$ from the rest and regard it as Euclidean time, with $x^i=(\tau,\vec{x}_{\perp})$ and $\tau \sim \tau+1$. 

We fix the boundary metrics
\beq
\label{E:torusBC}
	 \gamma_{1ij}dx^i dx^j = \beta_1^2 d\tau^2 + d\vec{x}_{\perp}\cdot d\vec{x}_{\perp}\,, \qquad \gamma_{2ij}dx^idx^j = \beta_2^2 d\tau^2 + d\vec{x}_{\perp}\cdot d\vec{x}_{\perp}\,.
\eeq
The disconnected geometries with these boundary conditions are ones where a circle of the boundary torus contracts to zero size in the bulk. For example, the Euclidean version of the non-rotating toroidal black hole at inverse temperature $\beta$ has a line element
\beq
\label{E:torusBH}
	ds^2 = d\rho^2 +\left(\frac{4\pi}{d}\right)^2\cosh^{\frac{4}{d}}\left( \frac{d\,\rho}{2}\right)\left( \tanh^2\left( \frac{d\,\rho}{2}\right) d\tau^2 +\beta^{-2}d\vec{x}_{\perp}\cdot d\vec{x}_{\perp}\right)\,,
\eeq
and a renormalized action
\beq
	S_{\rm grav} = -\frac{V}{4 dG}\left( \frac{4\pi}{d\beta}\right)^{d-1}\,.
\eeq
There are other geometries where a spatial circle contracts, or linear combinations of spatial and temporal circles. For instance, if the spatial torus has $x^2 \sim x^2+ L$, the geometry where the spatial circle $x^2$ contracts is
\beq
	ds^2 = d\rho^2 + \left(\frac{4\pi}{dL}\right)^2\cosh^{\frac{4}{d}}\left( \frac{d\,\rho}{2}\right)\left( \tanh^2\left( \frac{d\,\rho}{2}\right)(dx^2)^2 +\beta^2 d\tau^2 + \sum_{i=3}^d (dx^i)^2\right)\,,
\eeq
with a renormalized action
\beq
	S_{\rm grav}  = -\frac{V\beta}{4d GL}\left( \frac{4\pi}{dL}\right)^{d-1}\,,
\eeq
and this is sometimes called an AdS soliton. The AdS soliton dominates at low temperature, and there is a gap to the black hole threshold. A computation of the gap, either through the boundary stress tensor of the Lorentzian black hole, or an inverse Laplace transform of the black hole partition function with respect to $\beta$, reveals that there is a continuum of black hole states with energies $E\geq 0$. 

The disconnected solutions to the two-boundary problem are just two copies of these basic geometries, with an action parametrically of the form $S_{\rm grav} \sim - \frac{1}{G}$. 

Now we look for connected constrained instantons. We consider metrics which preserve translation invariance along the cycles of the torus, and also preserve time-reversal. While the identifications of the spatial torus break spatial rotational invariance, we further ansatz that the metric is invariant under spatial rotations. (Said another way, we look for translationally and rotationally invariant wormholes with $\mathbb{S}^1\times\mathbb{R}^{d-1}$ cross-section, and then quotient by a $\mathbb{Z}^{d-1}$ subgroup of spatial translations.) These are of the form
\beq
	ds^2 = d\rho^2 + e^{2f_1(\rho)} d\tau^2 + e^{2f_2(\rho)}d\vec{x}_{\perp}\cdot d\vec{x}_{\perp}\,.
\eeq
In this case it is easy to solve the modified field equations~\eqref{E:constrainedEOMs} subject to the boundary conditions~\eqref{E:torusBC}. There is a one-parameter family of solutions labeled by a parameter $b$,
\begin{align}
\begin{split}
\label{E:torusWormhole}
	ds^2 &= d\rho^2 + \frac{b^2}{4}\left( 2\cosh\left( \frac{d\,\rho}{2}\right)\right)^{\frac{4}{d}} \left( \left( \frac{\beta_1e^{\frac{d\,\rho}{2}} + \beta_2e^{-\frac{d\,\rho}{2}}}{2\cosh\left( \frac{d\,\rho}{2}\right)}\right)^2 d\tau^2 + d\vec{x}_{\perp}^2\right)\,, 
	\\
	 \lambda &=i \frac{d(d-1)\left(\frac{b}{2}\right)^d}{32\pi G}(\beta_1+\beta_2)\,,
\end{split}
\end{align}
where we have taken $F=1$. This is a bottleneck geometry, where the torus reaches a minimal size $\sim b^d(\sqrt{\beta_1}+\sqrt{\beta_2})^2V$, with $V$ the volume of the spatial torus. So in order to have a non-singular geometry we require $b>0$, and the wormhole pinches off as $b\to 0$.

The two boundaries are located at 
\beq
	\rho_1 = \ln \left( \frac{2}{b\varepsilon}\right)\,, \qquad \rho_2 = - \ln \left( \frac{2}{b\varepsilon}\right)\,,
\eeq
and the constraint, proportional to the length of the wormhole, reads
\beq
	\mathcal{C} = \int d^{d+1}x \,\sqrt{g_{\rho \rho}} = V \ln \left( \frac{1}{\varepsilon}\right) + 2V \ln \left( \frac{2}{b}\right)  = \zeta\,.
\eeq
This length diverges as $\varepsilon \to 0$, but there is a meaningful finite part proportional to $\ln b$. As a result, fixing the constraint to be $V \ln (1/\varepsilon) + [\text{finite}]$ amounts to fixing the parameter $b$. 

The parameter $b$ governs not only the length of the wormhole and the size of the bottleneck, but also the boundary stress tensor and the renormalized action. The latter is
\beq
\label{E:torusWHaction}
	S_{\rm grav} = (\beta_1+\beta_2)E\,, \qquad E = V \mathcal{E}\,, \qquad \mathcal{E} = \frac{(d-1)\left( \frac{b}{2}\right)^d}{4\pi G}\,,
\eeq
where $\mathcal{E}=T^{\tau}{}_{\tau}$ is the energy density contained in the boundary stress tensor (that energy density is the same on both boundaries), and $E = \int d^{d-1}x_{\perp} \,T^{\tau}{}_{\tau}$ is the boundary energy. So we see that there is a family of wormholes, with actions going as $S_{\rm grav}\sim \frac{b^d}{G}$. Since the action depends on $b$ we see that this parameter is best thought of as a pseudomodulus labeling the wormhole. These geometries are clearly subdominant compared to the disconnected saddles, whose action is large and negative. Note that since $b\geq 0$, the energy is necessarily positive, $E\geq 0$, the same regime where one finds non-rotating black holes. Unfortunately, the wormhole action is smallest at small $b$; this is the regime where the bottleneck pinches off, curvatures blow up, and effective field theory breaks down. We therefore expect that our semiclassical analysis is sensible only for fixed $b$ as $G\to 0$.

The parameter $b$ leads to another interesting effect: the boundary stress tensors acquire a nonzero trace $T^i{}_i \propto \frac{b^d(\beta_1+\beta_2)}{G}$. This trace vanishes for a saddle, corresponding to boundary Weyl invariance. These wormholes spontaneously break the diagonal part of the doubled scale symmetry of the two boundary problem, while preserving its axial subgroup.\footnote{The boundary stress tensors obey $\mathcal{T}=\int d^dx \sqrt{\gamma_1}\,(T_1)^i{}_i = \int d^dx \sqrt{\gamma_2}\,(T_2)^i{}_i$, so that under constant infinitesimal Weyl rescalings $\delta \gamma_{1ij} = 2\sigma_1 \gamma_{1ij}$ and $\delta \gamma_{2ij} = 2\sigma_2 \gamma_{2ij}$ we have $\delta S_{\rm grav} = (\sigma_1+\sigma_2) \mathcal{T}$.} The spontaneously broken scale symmetry generates the pseudomodulus $b$.

There are two extreme limits in which these  wormholes become genuine saddles. The first is $b\to 0$, in which case the geometry becomes singular. The second is $\beta_1+\beta_2= 0$, for any $b$. For real $\beta_1, \beta_2$, we then have $\beta_1 = - \beta_2 = \beta$, so that the metric becomes
\beq
	ds^2 = d\rho^2 + \frac{b^2\beta^2}{4}\left( 2\cosh\left( \frac{d\,\rho}{2}\right)\right)^{\frac{4}{d}} \left( \tanh^2\left( \frac{d\,\rho}{2}\right)d\tau^2 + \beta^{-2}d\vec{x}_{\perp}^2\right)\,,
\eeq
which for $\frac{b^2\beta^2}{4} = \left( \frac{4\pi}{d}\right)^2$ is precisely the Euclidean black hole~\eqref{E:torusBH}. There is another way to achieve $\beta_1 + \beta_2 = 0$ and have a sensible geometry, namely to analytically continue $\beta_1 = iT$ and $\beta_2 = -i T$. This results in a Lorentzian geometry with a metric 
\beq
\label{E:doubleCone}
	ds^2 = d\rho^2 + \frac{b^2}{4}\left(2\cosh\left( \frac{d\,\rho}{2}\right)\right)^{\frac{4}{d}} \left( -\tanh^2\left( \frac{d\,\rho}{2}\right)dt^2 + d\vec{x}_{\perp}^2\right)\,,
\eeq
where we have defined $t =  T \tau$ so that $t \sim t + T$. If we were to ignore the periodicity of $t$, this line element describes the exterior of the two-sided toroidal black hole with energy density $\mathcal{E} \sim \frac{b^d}{G}$ given by~\eqref{E:torusWHaction}, where $\rho >0$ corresponds to the right exterior and $\rho <0$ to the left exterior. The periodically identified geometry~\eqref{E:doubleCone} is the orbifold of the exterior of the two-sided black hole by $t \sim t + T$. This orbifold has a fixed surface at $\rho = 0$, and so the geometry is singular there, of the local form
\beq
	ds^2 \approx d\rho^2 - B^2 \rho^2 dt^2 + d\vec{y}_{\perp}^2\,,
\eeq
where $B = 2^{\frac{2}{d}-2}bd$ and $y_{\perp} \propto x_{\perp}$. If it were not for the identification $t\sim t + T$ this would simply be flat space in Rindler coordinates, but the orbifold leads to a singularity at $\rho = 0$, the Lorentzian analogue of a conical singularity. This orbifold has appeared in the literature before: it is precisely the double cone geometry of Saad, Shenker, and Stanford~\cite{Saad:2018bqo}. So the wormhole~\eqref{E:torusWormhole} generalizes the double cone geometry~\cite{cotler2020gravitational} (see also~\cite{Stanford:2020wkf}), and indeed we can think of the wormhole as an analytic continuation of the double cone to real $\beta_1, \beta_2$. 

There are presumably also ``spinning'' wormholes where the boundary stress tensor has nonzero spin. One expects that the spins on the two boundaries are equal like the energies, by a gravitational Gauss' law. If this is the case it would be interesting to study the allowed energies of wormholes at fixed spin, to see if they coincide with the spectrum of black hole energies at the same spin. Precisely this effect happens in AdS$_3$ gravity, and for static $\mathbb{S}^1\times\mathbb{S}^3$ wormholes in five dimensions as we recall below.

\subsubsection{$\mathbb{S}^1\times\mathbb{S}^3$ cross-section}

Here we consider geometries with $\mathbb{S}^1\times\mathbb{S}^{d-1}$ cross-section. We focus our attention on the cases $d=2$, where we have a torus cross-section, and $d=4$ as we are only able to find analytic expressions for wormhole geometries in those dimensions. Since the $d=2$ case was discussed above, we proceed to $d=4$, i.e.~a $5d$ bulk.

We impose that the boundary condition on the two boundary metrics are
\beq
	\gamma_{1ij}dx^idx^j = \beta_1^2d\tau^2 + d\Omega_3^2 \,, \qquad \gamma_{2ij}dx^idx^j = \beta_2^2 d\tau^2 + d\Omega_3^2\,,
\eeq
where again $\tau\sim \tau+1$. The boundary metrics are for $\mathbb{S}^1\times\mathbb{S}^3$, where one is effectively at inverse temperature $\beta_1$ at large positive $\rho$ and the other is effectively at inverse temperature $\beta_2$ at large negative $\rho$.

We begin with the disconnected saddles, which are two copies of a saddle with a single boundary. Imposing translation/rotational invariance along with time-reversal, there are two such geometries: Euclidean global AdS and the non-rotating Euclidean black hole. These have the line elements
\begin{align}
\begin{split}
	ds^2_{\rm global} & = d\rho^2 + \beta^2 \cosh^2(\rho)d\tau^2 + \sinh^2(\rho)d\Omega_3^2\,,
	\\
	ds^2_{\rm BH} & = d\rho^2 + \frac{b^2 \cosh(2\rho)-1}{2}\left(   \left( \frac{\beta\sinh(2\rho)}{\cosh(2\rho)-\frac{1}{b^2}}\right)^2d\tau^2 + d\Omega_3^2\right)\,,
\end{split}
\end{align}
and renormalized actions\footnote{We are working in a convention where we only add boundary counterterms proportional to the volume and integrated scalar curvature of the cutoff slice. There is of course the freedom to redefine the action by finite boundary counterterms involving the integral of a quadratic polynomial in the curvatures, which would have the effect of shifting $S_{\rm global}$ and $S_{\rm BH}$ by an additive constant proportional to $\beta$. The difference $S_{\rm global}-S_{\rm BH}$ is scheme independent.}
\begin{align}
\begin{split}
	S_{\rm global} & = \frac{3\pi \beta}{32G}\,,
	\\
	S_{\rm BH} & = -\frac{\pi b^2(b^2-4)\beta}{32G}\,.
\end{split}
\end{align}
Regularity of the Euclidean black hole at $\rho = 0$ requires
\beq
	\frac{2b^4\beta^2}{b^2-1} = (2\pi)^2\,, 
\eeq
which implies
\beq
	b^2 = \frac{\pi^2}{\beta^2} \pm \frac{\pi^2\sqrt{1-\frac{2\beta^2}{\pi^2}}}{\beta^2}\,.
\eeq
For all these metrics $b^2\geq 1$ and so the 5d geometry is regular, although we see that the black holes only exist for $\beta <\pi/\sqrt{2}$. The negative branch is thermodynamically unstable, but the positive branch is not. In the canonical ensemble, the low-temperature thermodynamics is dominated by the Euclidean global AdS saddle, and at high temperatures by the positive branch of the black hole saddle; the two are separated by the first-order Hawking-Page transition. However, when it comes to the spectrum of black holes, the black hole spectrum is continuous and is not separated by a gap from the vacuum, with $E\geq E_0=\frac{3\pi}{32 G}$. The minimum mass black hole is attained as $b\to 1$.

Now we look for wormholes which are translationally invariant in Euclidean time, rotationally invariant, and time-reversal invariant. We pick a constraint consistent with the rotational symmetry, so that $F(x) = \sqrt{g_{\mathbb{S}^3}}$\,. Then there is a one-parameter family with
\begin{align}
\begin{split}
\label{E:S1xS3wormhole}
	ds^2 & = d\rho^2 + \frac{b^2\cosh(2\rho)-1}{2}\left( \left(\frac{\beta_1e^{2\rho}+\beta_2e^{-2\rho}}{2\left( \cosh(2\rho)-\frac{1}{b^2}\right)}\right)^2d\tau^2 + d\Omega_3^2\right)\,,
	\\
	\lambda & = i\frac{3b^4}{32\pi G}(\beta_1+\beta_2)\,,
\end{split}
\end{align}
where non-singularity of the geometry requires $b\geq 1$. Note that this geometry becomes precisely the Euclidean black hole for real $\beta_1 = -\beta_2 =\beta$. The action and energy of the wormhole are
\beq
	S_{\rm grav} = (\beta_1+\beta_2)E\,, \qquad E = b^4 E_0 = \frac{3\pi b^4}{32 G}\,,
\eeq
and so we see that there is a spectrum of non-spinning wormholes with energies $E \geq  E_0$, coinciding with the spectrum of energies of non-rotating black holes.

Like the wormhole with torus cross-section, this geometry becomes an honest saddle if $\beta_1 +\beta_2=0$, in the same way as our discussion of the toroidal black hole. If $\beta_1 = - \beta_2 = \beta$, then for a particular value of $b$ we recover the Euclidean black hole, and ~if $\beta_1 =i T$ and $\beta_2=-i T$, the wormhole becomes a timelike orbifold of two-sided black hole with $\mathbb{S}^3$ horizon. That orbifold is an example of the double cone studied in~\cite{Saad:2018bqo}. The wormhole therefore again generalizes the double cone~\cite{cotler2020gravitational}.

\subsection{Another perspective: $\lambda$-solutions}
\label{S:complement}

From our presentation above it is clear that the constrained instanton calculus is, at its heart, a strategy to isolate new non-perturbative features in the path integral; furthermore, it may be employed in any field theory. However, it is an interesting fact that the wormholes we presented above may be discovered without introducing the constrained instanton machinery in the first place. This fact is new to this work and has an analogue in any gauge theory.

Suppose that we pick a delta function gauge for Einstein gravity, rather than an $R_{\xi}$ gauge. In particular let us fix a radial gauge, so that the gauge-fixing condition reads\footnote{This gauge-fixing condition is actually achievable only locally, but not globally. The situation here is analogous to axial gauge for Yang-Mills theory on $\mathbb{R}^3\times\mathbb{S}^1$. The existence of a gauge-invariant Wilson loop around the $\mathbb{S}^1$ means that the component of the gauge field in that direction, $A_{\tau}$, can only be set to zero locally, but not globally. To enact the analogue of axial gauge in that case, one expands around a value for the Wilson loop, imposing as a gauge condition that the fluctuations respect $\delta A_{\tau} = 0$, and then sums over all possible choices of the Wilson loop. Here, the analogue of the Wilson loop is a relative twist between the two boundaries, enacted by a large diffeomorphism $x^i \to x^i + f^i(\rho)$ where $\lim_{\rho\to \infty}f^i - \lim_{\rho\to-\infty} f^i$ parameterizes the twist. As in the Yang-Mills example, the analogue of radial gauge is to work around a geometry with fixed twist, impose that the fluctuations obey $\delta g_{\mu \rho} = 0$, and then sum over all values of the twist.}
\beq
	g_{\mu \rho} = \delta_{\mu}^{\rho}\,,
\eeq
where here $\mu$ is a spacetime index and $\rho$ refers to the radial direction. The basic idea now is that the gauge-fixing term in the gravity action,
\beq
	S_{\rm gauge-fixing} = i \int d^{d+1} x \sqrt{g} f^{\mu}(g_{\mu\rho}-\delta_{\mu}^{\rho})\,,
\eeq
acts like a source term in the Einstein's equations. Including the effect of the gauge-fixing term, the Einstein's equations are modified as
\beq
\label{E:lambdaEinsteinEqs}
	R_{\mu\nu} - \frac{R}{2}g_{\mu\nu}+ \Lambda g_{\mu\nu} = 8 \pi i G (f_{\mu}\delta_{\nu}^{\rho} + f_{\nu}\delta_{\mu}^{\rho})\,,
\eeq
plus a term that vanishes in radial gauge. So there are potentially new solutions to the equations of motion where the gauge-fixing auxiliary field picks up an imaginary nonzero profile. Indeed, this formalizes the intuitive observation that, when we fix a gauge, we are no longer free to vary with respect to the gauge-fixed components of the metric, and therefore are not obliged to solve the corresponding Einstein's equations. In this instance the $\mu\rho$ components of the metric cannot vary, and the forcing term is just the statement that the $\mu\rho$ components of the Einstein's equations may be violated. This violation in turn fixes the profile of $f_{\mu}$. 

We term these spacetimes $\lambda$-solutions. The logic behind the name is that these are solutions where there is external forcing from the gauge-fixing auxiliary field which acts as a Lagrange multiplier field in a delta function gauge, and we usually denote Lagrange mutlipliers as $\lambda$.

Indeed, the wormholes we described in the last Subsection all obey the radial gauge fixing condition and are precisely solutions of all but the $\rho\rho$ component of Einstein's equations. As such they are examples of $\lambda$-solutions. 

Of course it is more natural to study $\lambda$-solutions first in gauge theory than in gravity, where one might hope that these solutions correspond to some already understood phenomenon. However, for pure electromagnetism, the simplest $\lambda$-solutions are inconsistent with boundary conditions. For example, consider axial gauge in flat Minkowski space, whereby we set $A_0= 0$ and impose that $A_{\mu} $ is gauge-equivalent to zero at infinity. The Maxwell's equations, including the Lagrange mutliplier term, are $\partial_{\nu} F^{\mu\nu} = \lambda \delta^{\mu}_0$. Taking the divergence we see that the source $\lambda$ is time-independent, and indeed one can solve these equations for any $\lambda(\vec{x})$.  $\lambda$ appears as a static charge density, correspondingly the gauge field does not asymptote to the trivial vacuum in the infinite past or future, and so does not respect the boundary conditions. Another example is Coulomb gauge, where the Maxwell's equations now read $\partial_{\mu}F^{0\mu} = 0$ and $\partial_{\mu}F^{i\mu} = \partial^i \lambda$, taking the divergence implies that $\nabla^2\lambda = 0$, and so in flat space there are no solutions which die off at spatial infinity. There is a similar result in Lorenz gauge. 

In Yang-Mills theory we have some preliminary indications that the constrained instantons of Affleck~\cite{affleck1981constrained} for Yang-Mills theory in a Higgs phase can, in a radial gauge, be understood as $\lambda$-solutions. We hope to soon report on the study of such solutions in Yang-Mills theory. 

In field theory there is overwhelming evidence that we ought to sum over all field configurations consistent with the boundary conditions, and so we expect that $\lambda$-solutions have a role in field theory. In this work we are taking the effective field theory description of Einstein gravity (and supergravity) seriously, and so are of the opinion that these configurations may also be important in the weakly coupled, weakly curved regime of quantum gravity.

Note that these wormholes, realized as $\lambda$-solutions, have the same pseudomodulus $b$ that appeared when we stabilized the length of the wormhole. As a $\lambda$-solution however, $b$ is not stabilized and we must integrate over it.

This is a strange at first glance. These wormholes are solutions to the equations of motion of the gauge-fixed theory, and yet are labeled by a parameter $b$ on which the action depends. So these solutions are not exact saddle points of the gauge-fixed action. This seems self-contradictory, as solutions to the field equations \emph{are} saddle points. The fly in the ointment is that this equivalence between solutions and saddle points comes with a caveat: it is only true for fluctuations that fall off sufficiently fast near the boundary, and in this case a fluctuation of $b$ corresponds to a non-normalizable mode of the metric, as can be easily seen from~\eqref{E:torusWormhole} and~\eqref{E:S1xS3wormhole}. 

One way of thinking about $b$ is that there is an effective stress tensor on the right-hand-side of the field equations of the gauge-fixed theory, for radial gauge given in~\eqref{E:lambdaEinsteinEqs}, and that $\lambda$-solutions can be obtained for any profile of the auxiliary fields for which this stress tensor is conserved. For these wormholes, assuming translational symmetry along the boundary directions and isotropy, there is a one-parameter family of such stress tensors, with $T_{\rho \rho} \sim b^d/(G\sqrt{g})$.

The net result is that the one-loop approximation to the wormhole amplitude is an integral over the modulus $b$, as in our discussion of constrained instantons. However, while the wormholes~\eqref{E:torusWormhole} and~\eqref{E:S1xS3wormhole} can be understood as either constrained instantons or as $\lambda$-solutions, the spectrum of fluctuations around them is rather different in each case, as can be verified by studying the linearized approximation to the field equations around the wormhole, either with the constraint or with the gauge-fixing forcing term. We expect that both approaches compute the same integrated amplitude. Since the fluctuations are rather different in each case, it may be the case that the one-loop determinants at fixed $b$ also differ, and we conclude that the two approaches may produce rather different integral representations of the same amplitude.

Said another way, the individual wormhole geometries are unphysical.\footnote{Of course we could change the question being answered by e.g. adding a Gao-Jafferis-Wall~\cite{gao2017traversable} or Maldacena-Qi~\cite{maldacena2018eternal} like deformation for a large number of bulk matter fields, thereby obtaining a wormhole with some stabilized $b$. In this case the particular, stabilized wormhole so obtained would be physical.} The physical quantity is the wormhole amplitude, which receives contributions from all wormholes with allowed $b$. 

This fact is useful to bear in mind when working in a different gauge, thereby changing the form of the stress tensor that supports the geometry. For example, one might impose that the metric fluctuations around a given background $g_{\mu\nu}$ are transverse, with
\beq
	D_{\nu}h^{\mu\nu} - \frac{1}{2}D^{\mu} h = 0\,, \qquad h = g^{\mu\nu}h_{\mu\nu}\,.
\eeq
One can numerically find wormholes as $\lambda$-solutions supported by the corresponding gauge-fixing auxiliary fields, but they are not those studied above in~\eqref{E:torusWormhole} and~\eqref{E:S1xS3wormhole}. 

When we study $\lambda$-solutions below, we stick to wormholes in a radial gauge (or radial-like gauge in 10d supergravity), rather than another gauge, as a matter of computational convenience. The advantage of the radial gauge for us is that we can analytically solve the field equations of the gauge-fixed model in a number of cases, and more importantly, in those cases, the wormholes are the analytic continuation of Euclidean black holes.

\section{From wormholes to spectral statistics}
\label{S:fromWormholesToStatistics}

Let us take stock of what we have learned. We have two different methods for obtaining wormholes in Einstein gravity. Wormholes in Euclidean AdS are characterized by a length between the two asymptotic regions. With applications to black hole physics in mind, we considered ``vanilla'' boundary conditions where the boundaries are $\mathbb{S}^1\times\mathbb{T}^{d-1}$ or $\mathbb{S}^1\times\mathbb{S}^{d-1}$ with no other sources for bulk matter. In this setting there are no wormhole solutions to the Einstein's equations, so the saddle-point approximation to the two-boundary problem includes disconnected geometries where the length between the two boundaries is infinite. 

We can think of the length as a pseudomodulus with a runaway potential. We need to stabilize this modulus in order to find wormholes. One way is to fix the length ``by hand,'' and we showed how this can be done using the constrained instanton calculus. One then finds wormhole saddles of the constrained problem, and we discussed how this calculus leads to a strategy for approximating the full sum over wormholes. This strategy has two steps: first compute the amplitude at fixed length in a loop expansion, and then integrate over the length with the correct measure. Using a finite-dimensional example, we saw that this strategy breaks down if the loop expansion at fixed length breaks down, or if the underlying integral does not converge. Further, the constrained saddle must be perturbatively stable in the sense we explained in Subsection~\ref{S:formal}.

The other method to find solutions is to use the gauge-fixing auxiliary fields, which lead to an effective stress tensor on the right-hand-side of the Einstein's equations. This way one finds a family of wormhole solutions labeled by the length. Perturbations of the length are non-normalizable fluctuations of the metric, and so the natural way to compute the amplitude here is again to find the amplitude at fixed length, and then to integrate over the length at the end. If we fix a radial gauge, then these wormholes are precisely the same geometries obtained by directly fixing the length. A different constraint or gauge-fixing leads to different wormhole geometries. However this is not a problem: the full amplitude does not depend on the choice of gauge or how we slice the integration domain. One appealing feature of the wormholes attained by either fixing the length or using a radial gauge is that they are, in a sense we described, the analytic continuation of Euclidean black hole geometries with $\mathbb{S}^1\times\mathbb{T}^{d-1}$ or $\mathbb{S}^1\times\mathbb{S}^{d-1}$ boundary. Either way, we land on a new non-perturbative contribution to the sum over metrics.

We saw that the wormhole length, encoded through the parameter $b$, is equivalent to the energies perceived on the two boundaries, with $E_1 = E_2 = E \sim b^d/G$. We considered ``non-rotating'' wormholes, where the angular momenta perceived on the boundary vanish. For boundary geometries of the form $\mathbb{S}^1\times\mathbb{T}^{d-1}$ and $\mathbb{S}^1\times\mathbb{S}^3$ and assuming that the wormholes at fixed $b$ are perturbatively stable,  we learn that the amplitude has the schematic form
\beq
\label{E:wormholeAschematic}
	Z_{\rm wormhole}(\beta_1,\beta_2) = \int_{E_0}^{\infty} dE\,f(E;\beta_1,\beta_2)\, e^{-(\beta_1+\beta_2)E}\left( 1 + O(G)\right)\,,
\eeq
where $f(E)$ arises from the one-loop determinant at fixed $b$ and the corrections come from two- or higher loops at fixed $b$. The wormhole action $S_{\rm wormhole} = (\beta_1+\beta_2)E$ leads to the Boltzmann factor. This action is higher than that of disconnected saddles, so that $Z_{\rm wormhole}$ is a non-perturbative correction to the two-boundary amplitude. 

The integration domain is over $E\geq E_0$, where the lower endpoint $E_0$ is fixed by the requirement that the wormhole is everywhere smooth. In analogy with random matrix theory, we term this domain an \emph{emergent cut}. For the non-rotating wormholes with torus and $\mathbb{S}^1\times\mathbb{S}^3$ cross-section, we found that this cut agrees with the spectrum of non-rotating black holes with torus or $\mathbb{S}^3$ horizon. In particular, for the $\mathbb{S}^1\times\mathbb{S}^3$, the threshold energy $E_0$ is the energy of the lightest small black hole.

Wormhole amplitudes have been computed in Jackiw-Teitelboim gravity and for pure AdS$_3$ gravity. In both of those models the amplitude can be written in the form~\eqref{E:wormholeAschematic}.\footnote{The amplitude in 3d gravity with two torus boundaries and fixed boundary complex structures is most naturally written as a two-dimensional integral effectively over the energy and spin carried by the wormhole. That integration domain coincides with the domain of energies and spins of BTZ black holes. Fourier transforming to zero spin leads to a one-dimensional integral resembling~\eqref{E:wormholeAschematic}.} The one-loop factor $f(E)$ in both cases is such that the full integral does not admit a saddle-point approximation, and is dominated by wormholes with small bottleneck as $E\to E_0$. Crucially $f(E)$ includes the contribution from a zero mode, which corresponds to a translation of Euclidean time on one boundary relative to the other. This relative translation is called a twist. The zero mode volume goes as $\sqrt{\beta_1\beta_2}$ which leads to an important consequence. If one analytically continues $\beta_1$ and $\beta_2$ as in the computation of the spectral form factor, as $\beta_1 = \beta + i T$ and $\beta_2 = \beta - i T$, then for $T\gg \beta$ one has
\beq
\label{E:schematicRamp}
	Z_{\rm wormhole}(\beta+iT,\beta-iT) \sim \frac{T}{\beta} \,e^{-2\beta E_0} 
\eeq
which recalls the ``ramp'' we reviewed in Section~\ref{S:RMT}, a hallmark signature of level repulsion. 

In JT gravity the wormhole amplitude at hand (the ``double trumpet'' of~\cite{Saad:2019lba}) is related to the dual matrix model via
\beq
	 \left\langle \text{tr}\left( e^{-\beta_1H}\right)\text{tr}\left(e^{-\beta_2H}\right)\right\rangle_{\rm conn} \simeq Z_{\rm wormhole}(\beta_1,\beta_2)  + \hdots\,,
\eeq
where the dots indicate genus corrections and the average on the left-hand-side is over Hermitian matrices $H$ in the double-scaling limit with a particular leading density of states. So, the ramp obtained from JT gravity really is the ramp of the spectral form factor in the dual matrix model. As for AdS$_3$ gravity, it is not yet known if that model is a consistent theory of quantum gravity, but the evidence to date indicates that if it is, then it is dual to an ensemble, and the dictionary equates the $n$-boundary problem in gravity to an $n$-point average of CFT partition functions. In particular, the wormhole amplitude implies a ramp in the dual spectral form factor for the primary-counting partition function at fixed spin. 

Returning to higher-dimensional Einstein gravity, the two-boundary problem in gravity is most naturally interpreted as the two-point function of partition functions in the dual. (We are not claiming here that the dual description is an ensemble; it may well be that this two-point function factorizes.) The wormhole contributes to the connected part, so that
\beq
	\left\langle Z(\beta_1)Z(\beta_2)\right\rangle_{\rm conn} \simeq Z_{\rm wormhole}(\beta_1,\beta_2) + \hdots\,.
\eeq
Macroscopic wormholes have $E = O(1/G)$. Accordingly we expect that the amplitude is dominated by the ``lightest'' wormholes with $E\to E_0$, although it is possible that one-loop effects lead to a saddle-point approximation at some $E$ above $E_0$. Wormholes with small energy are strongly curved near the bottleneck, with curvatures blowing up as $R \propto \frac{1}{\left( \frac{E}{E_0}\right)^{\frac{1}{d}}-1}$\,. Under the assumption that the wormhole amplitude is dominated near there, we learn that the full amplitude is UV sensitive: it cannot be approximated in effective field theory, and in fact is only sensible in a UV completion like string theory. We observe that the strong coupling region corresponds to the edge of the black hole microstate spectrum, where the density of states is becoming small. As an aside we note that the $1/N$ expansion of random matrix theory breaks down near the spectral edge.

There is one feature of the amplitude that we expect to be robust even in the strongly coupled, small bottleneck region. As in JT and AdS$_3$ gravity there is a special zero mode of these wormholes, a relative temporal twist. If $\beta_1 = \beta_2=\beta$, then this zero mode has a volume $\propto \beta$. It seems reasonable that for $\beta_1 \neq \beta_2$ the properly computed volume (obtained from the Wheeler de Witt metric) is $\sim\sqrt{\beta_1\beta_2}$. (There are also spatial twists, whose volume is independent of $\beta_1$ and $\beta_2$.) Now suppose we analytically continue $\beta_1 = \beta+i T$ and $\beta_2 = \beta-iT$ and take the late time, low temperature limit, so that the wormholes become closer and closer to the double cone geometries of~\cite{Saad:2018bqo}. Then in the effective field theory approximation the relative twists are the only light degrees of freedom, in which case the integrand is $\sim T\, \tilde{f}(E)e^{-2\beta E}$ as long as $E-E_0 \gg \frac{1}{G}$. This suggests that the full amplitude takes the same form as~\eqref{E:schematicRamp} with a ``ramp,''\footnote{Indeed this is a refinement of the argument of~\cite{Saad:2018bqo} that the double cone encodes a semiclassical ramp in gravity.} although we stress that whether or not this is true depends on the strongly coupled regime of integration and on the precise one-loop factor.

Throughout this discussion we have been assuming that the wormholes under consideration are perturbatively stable at fixed $b$. In our earlier work~\cite{cotler2020gravitational} we showed that this was indeed true for wormholes with torus cross-section and $\beta_1 = \beta_2$. This assumption is crucial in writing the integral representation~\eqref{E:wormholeAschematic} for the wormhole amplitude, as the integrand arises from a saddle-point approximation at fixed energy. If the wormhole is perturbatively unstable with respect to fluctuations at fixed energy, then there should be another configuration at the same energy with lower action, and the integrand should entail an expansion around that background. In the next Section we continue this stability analysis, accumulating further evidence that these wormholes are indeed stable.

While the full wormhole amplitude is likely UV sensitive, and we expect only exists in UV complete theories like string theory, we now argue that we can reliably obtain some information within the effective field theory approximation. Roughly speaking, we would like to select energies which are far from the spectral edge. In JT gravity the double inverse Laplace transform of the wormhole amplitude gives the genus zero approximation to the connected two-point function of eigenvalue densities, $\rho_2(E_1, E_2)$. This quantity characterizes eigenvalue repulsion just as well as the ramp, with $\rho_2(E_1, E_2) - \rho(E_1) \rho(E_2) \sim -\frac{1}{(E_1-E_2)^2}$ to genus zero in random matrix theory. 

Now imagine taking a double inverse Laplace transform of the full wormhole amplitude (along with a Fourier transform to fixed, zero spin),\footnote{The wormhole is everywhere smooth and macroscopic as long as $E>E_0$ and $\beta_1, \beta_2>0$. Thus the one-loop factor $f$ ought to be nonsingular for $\text{Re}(\beta_1), \text{Re}(\beta_2)>0$, and so inverse Laplace transformation is essentially inverse Fourier transform with respect to $\beta_1$ and $\beta_2$.}
\beq
	\widetilde{Z}_{\rm wormhole}(E_1,E_2)=\frac{1}{(2\pi i)^2}\int_{\gamma_1-i\infty}^{\gamma_1 + i \infty} d\beta_1 \int_{\gamma_2-\infty}^{\gamma_2+i\infty} d\beta_2 \, e^{\beta_1 E_1+ \beta_2 E_2} \int_{E_0}^{\infty} dE\,f(E;\beta_1,\beta_2) e^{-(\beta_1+\beta_2)E}\,.
\eeq
Let us consider energies well away from the end of the cut, $E_1-E_0, E_2-E_0\gg \frac{1}{G}$, and relatively small energy differences, $E_1 - E_2=O(1)$. For a wide family of functions $f$ (including polynomials in $\beta_1$ and $\beta_2$) one can integrate in $\beta_1$ and $\beta_2$ first, with the result that the remaining integrand in $E$ is dominated near $E = \frac{E_1+E_2}{2}$, that is, by the wormholes we found with energies close to the average energy. The inverse Laplace transforms effectively stabilizes the wormhole length.

One way of thinking about this procedure is that the wormhole amplitude is naturally a canonical ensemble quantity, where we fix the inverse temperatures $\beta_1$ and $\beta_2$ on the two boundaries. Taking the inverse Laplace transform is akin to passing to microcanonical ensemble, enacting two constraints that fix the energies on the two boundaries.

One can envision fixing these energies as constraints from the get-go, leading to a problem in gravity with modified boundary conditions whereby one fixes the energies, spins (to vanish), and the spatial metrics, rather than boundary conformal structures. 

However it is not clear if the two-boundary problem at fixed energies admits a loop expansion. If the energies are the same, then there is a wormhole with that energy to expand around. However the physically interesting case is when the energies are almost, but not quite the same. Then the fixed energy boundary conditions would fix fluctuations around the wormhole, leading to a non-standard problem. Another potential pitfall is that the one-loop factor $f$ may be such that the integrals over $\beta_1$ and $\beta_2$ do not localize the energy to be near $\frac{E_1+E_2}{2}$. Yet another is that it is not clear if $\beta_1$ and $\beta_2$ are localized to nonzero values.

Fortunately there are other, closely related integral transforms of the full amplitude that are guaranteed to zoom in on macroscopic wormholes in such a way as to admit an effective field theory description. In statistical physics we often work with microcanonical ensembles where we fix the energy up to a resolution scale $\Delta$. Here, this can be accomplished by an integral transform
\beq
\label{E:fromWormholesToDoubleCone}
	\widetilde{Z}_{\Delta}(E_1,E_2) = \frac{1}{(2\pi i)^2} \int_{\gamma_1-i\infty}^{\gamma_1+i\infty} d\beta_1 \int_{\gamma_2-i\infty}^{\gamma_2+i\infty} d\beta_2 \,e^{\beta_1E_1 + \beta_2 E_2 +\Delta^2\frac{\beta_1^2+\beta_2^2}{2} }Z_{\rm wormhole}(\beta_1,\beta_2)\,.
\eeq
Now take $\Delta= O(G^{-n})$ and $E_1, E_2 = O(1/G)$ for some $1>n>0$. Then the integrals over $\beta_1$ and $\beta_2$ can be done by saddle-point approximation, leading to
\beq
	\widetilde{Z}_{\Delta}(E_1,E_2) \approx \frac{e^{-\frac{(E_1-E_2)^2}{4\Delta^2}}}{2\pi\Delta^2} \int_{E_0}^{\infty}dE\,f(E;\tilde{\beta}_1,\tilde{\beta}_2) \,e^{-\frac{\left( E - \frac{E_1+E_2}{2}\right)^2}{\Delta^2}}\,,
\eeq
with $\tilde{\beta}_1=\frac{E-E_1}{\Delta}$ and $\tilde{\beta}_2=\frac{E-E_2}{\Delta}$ the saddle-point values of $\beta_1$ and $\beta_2$. (We return to this shortly.) Because $\Delta =O( G^{-n})$, the effective Gaussian distribution for energy has a large variance and the integral over $E$ naively cannot be done by saddle-point approximation. However, this effective Gaussian distribution is centered at an enormous energy of $O(1/G)$, and since $\Delta \ll \frac{1}{G}$ we see that only configurations with large energies contribute. In gravity we have $E \sim \frac{b^d}{G}$ with $b$ the bottleneck size in AdS units, so that only a very narrow band of macroscopic wormholes contribute to $\widetilde{Z}_{\Delta}(E_1,E_2)$. Equating $dE \sim \frac{db}{b} \frac{b^d}{G}$ to the variance $\Delta$ we see that the variance of the bottleneck size, $db/b$, is of $O(G^{1-n})$ which is parametrically small. We can then trade the integration over energy for integration over $b$. The one-loop factor is, for macroscopic wormholes with $O(1)$ $ b$, homogeneous in the Newton's constant $G$, and so in fact the integral over $b$ can \emph{also} be done by saddle-point techniques, with the saddle point value $E = \frac{E_1+E_2}{2}$. Then we see that the saddle-point values of $\beta_1$ and $\beta_2$ are $\tilde{\beta}_1 = - \tilde{\beta}_2 = \frac{E_2-E_1}{2\Delta}$. This is undesirable, since we either have zero size or one of the inverse temperatures is negative at the saddle. A resolution to this quandary is to shift $\beta_1 \to \beta_1 + i T$ and $\beta_2 \to \beta_2 -i T$ before performing the integral transform. The double transform reads as before, but now the saddle-point values are $\tilde{\beta}_1 = \frac{E_2-E_1}{\Delta} +i T$ and $\tilde{\beta}_2 = \frac{E_1-E_2}{\Delta} -i T$, so that for general $E_1$ and $E_2$ there is a macroscopic configuration, albeit with either $\text{Re}(\tilde{\beta}_1)$ or $\text{Re}(\tilde{\beta}_2)$ negative, which is concerning. If we take $E_1 = E_2$ then we simply select the wormhole with $\tilde{\beta}_1 = -\tilde{\beta}_2 =iT$ and $E = E_1 = E_2$. But the wormhole with these parameters is nothing more than the double cone of~\cite{Saad:2018bqo}! 

In other words, this microcanonical version of the wormhole amplitude has a saddle point approximation in weakly curved, weakly coupled Einstein gravity, the double cone. The double cone comes with moduli with a flat potential at tree level -- the mass and angular momenta of the black hole before orbifolding -- which are stabilized precisely by this particular integral transform. The correct prescription for determining loop corrections also follows from this analysis. In addition to fluctuations of the quantum fields around the double cone, one integrates over fluctuations in $\beta_1$ and $\beta_2$ with the Gaussian measure in~\eqref{E:fromWormholesToDoubleCone}.

It is worth comparing the analysis here with the one performed in~\cite{Saad:2018bqo} to find a ramp in the SYK model. The two are very closely related, precisely because the wormhole amplitude in gravity takes a similar schematic form to the amplitude those authors found in the SYK model. Indeed, the authors of~\cite{Saad:2018bqo} define and use the same integral transform as that leading to $\widetilde{Z}_{\Delta}(E,E;T)$ to find a novel two-replica saddle in SYK describing a ramp in the spectral form factor. By studying complex metrics they argued that the same should hold true in gravity, a statement we have now demonstrated from an entirely Euclidean analysis.

A closely related option to that above is to inverse Laplace transform in the average energy alone, exchanging $\beta_1+\beta_2$ for an average energy $E_0$, again with a resolution scale $\Delta$, so that now one is a function of $E_0$, $\Delta$, and $\beta_1 - \beta_2=\Delta \beta$. This transform has a saddle-point approximation in all parameters where $\beta_1 + \beta_2 = 0$ and $E=E_0$. But then in order to have a macroscopic geometry we have $\beta_1 = -\beta_2 = i T$ for some $T$, which again recovers the double cone geometry. 

So we have seen how to realize the double cone as a genuine, stabilized saddle point of a particular two-boundary problem. The dual observable is a smeared version of the two-point function of the density of states, with some Lorentzian time evolution inserted,
\beq
	\widetilde{Z}_{\Delta}(E_1,E_2;T)\simeq \frac{1}{2\pi\Delta^2}\left\langle \text{tr}\left(e^{-\frac{(H-E_1)^2}{2\Delta^2}}e^{-iH T}\right) \text{tr}\left(e^{-\frac{(H-E_2)^2}{2\Delta^2}} e^{i HT}\right)\right\rangle_{\rm conn}\,.
\eeq
As noted in~\cite{Saad:2018bqo} the double cone has a zero mode, temporal twists, with a volume $\propto T$. Thus in the one-loop approximation one has $\widetilde{Z}_{\Delta}(E,E;T) \sim \frac{T}{\Delta^2} \tilde{f}(E;iT,-iT)$ where $\tilde{f}$ is the rest of the contribution to the one-loop determinant at fixed energy and spin coming from nonzero modes. If that factor is approximately independent of $T$ at large $T$, then this smeared, evolved two-point function has a ``ramp'' which we would interpret as a signature of eigenvalue repulsion.

One might wonder if there is an observable that admits a semiclassical expansion around a macroscopic \emph{Euclidean} wormhole, rather than the double cone with $\beta_1 = -\beta_2 = i T$. As we have discussed above, different choices of gauge-fixing or constraint likely lead to different integral representations of the gauge-invariant wormhole amplitude. So it is not sufficient to simply select a wormhole by hand. However, noting that boundary energy is a gauge-invariant observable in gravity, we can insert a gauge-invariant constraint fixing the average boundary energy to some fixed value. This constraint is different than passing to a microcanonical version of the amplitude as in our discussion above, as it leads to an observable which is a function of $\beta_1$, $\beta_2$, and the average boundary energy. This constraint selects a single wormhole from the family, and enforces a constraint on the spectrum of fluctuations in such a way that the result is gauge-invariant.

This is a success: we have found a way to gauge-invariant observable with a saddle-point approximation in terms of a wormhole at fixed energy. However this state of affairs is not entirely satisfactory, since we do not presently have a dual interpretation as we did with the double cone.

In sum, while the full wormhole amplitude is likely UV sensitive, we can consistently work away from the spectral edge within the effective field theory approximation. There are weakly curved saddle points with respect to all parameters, the double cone or a macroscopic wormhole, which one can analyze without ever making reference to the full amplitude.

In random matrix theory and in JT gravity the ``microcanonical'' quantity $\widetilde{Z}_{\Delta}(E_1,E_2;T)$ is a smeared, evolved version of the usual sine kernel and signifies eigenvalue repulsion just as well as the spectral form factor. It would be extremely interesting to obtain enough information about the one-loop factor $f$ for these wormholes in higher-dimensional AdS to approximate this quantity beyond the known zero mode corresponding to temporal twists, and diagnose whether there is indeed eigenvalue repulsion in the spectrum of AdS black hole microstates. It would also be very useful to understand the dual interpretation of the Euclidean amplitude at fixed bulk energy.

\section{Perturbative stability in Einstein gravity}
\label{S:stability}

We now undertake an analysis of the perturbative stability of the wormholes~\eqref{E:torusWormhole} and~\eqref{E:S1xS3wormhole} with $\mathbb{S}^1 \times \mathbb{T}^{d-1}$ and $\mathbb{S}^1\times\mathbb{S}^3$ cross-section, respectively.  For the first part of the Section we will perform our analysis where the wormholes have been stabilized as $\lambda$-solutions as in Subsection~\ref{S:complement}, and we have fixed a radial gauge. In that setting we can give a streamlined proof of stability for the $\mathbb{S}^1\times\mathbb{T}^{d-1}$ wormhole with $\beta_1=\beta_2$, and demonstrate that the most dangerous potential instabilities for the $\mathbb{S}^1\times\mathbb{S}^3$ wormhole are absent. We wrap up with the $\mathbb{S}^1\times\mathbb{S}^3$ realized as a constrained instanton at fixed length. There we can demonstrate complete stability at the quadratic level for $\beta_1=\beta_2$.

\subsection{$\mathbb{S}^1 \times \mathbb{T}^{d-1}$ wormholes in radial gauge}

Consider the wormhole metrics in~\eqref{E:torusWormhole} for $b > 0$.  Taking $\beta_1 = \beta_2 = \beta$, we further rescale $\tau$ by $\beta$ so that we have effectively set $\beta = 1$.  This gives us the metric
\begin{equation}
\label{E:torusequalbetas}
ds^2 = d\rho^2 + \frac{b^2}{4} \left(2\cosh\left(\frac{d \,\rho}{2}\right) \right)^{\frac{4}{d}} d\vec{y}^2
\end{equation} 
where $d\vec{y}^2 = d\tau^2 + d\vec{x}_\perp^2$.  In previous work~\cite{cotler2020gravitational}, we have established that these wormholes are perturbatively stable when realized as constrained instantons with a fixed length, in $R_{\xi}$ gauge.  Here, we give a simpler proof in radial gauge $g_{\mu \rho} = \delta_\mu^\rho$ that we expect to be useful in future work.

Stability in radial gauge amounts to considering all fluctuations of the $ij$ components of the metric.  Notice that the metrics in~\eqref{E:torusequalbetas} have $SO(d)$ symmetry, up to the identifications on the torus.  If we turn on momentum $\vec{k}$ on the torus, this breaks the symmetry down to $SO(d-1)$, again up to identifications (which only affect the allowed momenta).  The fluctuations thus decompose into irreducible representations of $SO(d-1)$.

First we consider $d = 2$, where it is convenient to parameterize the fluctuations in a different manner than in higher dimensions.  The fluctuations are all scalars, conveniently parameterized as
\begin{align}
h_{\mu \nu} dx^\mu dx^\nu &= 2 b \cosh(\rho) \sum_{\vec{k}} \left(s_{\vec{k}}(\rho) \frac{k_i k_j}{\vec{k}^2} +  t_{\vec{k}}(\rho) \,\delta_{ij} + \frac{1}{\sqrt{2}} \left( u_{i,\vec{k}}(\rho) \frac{k_j}{|\vec{k}|} + u_{j,\vec{k}}(\rho) \frac{k_i}{|\vec{k}|}\right)\right) e^{i \vec{k}\cdot\vec{x}}  dy^i dy^j\,.
\end{align}
The quadratic action then takes the simple form
\begin{align}
S_2 = \frac{1}{16 \pi G} \int d\rho \sum_{\vec{k}}\left(s_{\vec{k}}^*\left(t_{\vec{k}}''- t_{\vec{k}}\right)  - \left(\big|t_{\vec{k}}'\big|^2 + \big|t_{\vec{k}}'\big|^2 \right) + \big|u_{i,\vec{k}}'\big|^2 + \big|u_{i,\vec{k}}\big|^2 \right)\,,
\end{align}
where $'= \partial_{\rho}$\,.  We see that the $u_{i,\vec{k}}$ fluctuations have a manifestly positive-definite action, and the $s_{\vec{k}}$ fluctuations appear linearly; thus we should interpret $s_{\vec{k}}$ as a Lagrange multiplier.  Accordingly, we should deform the path integration contour of $s_{\vec{k}}$ or $t_{\vec{k}}$ or a combination of each so that the linear term is pure imaginary.  Having done this, integrating out $s_{\vec{k}}$ enforces $t_{\vec{k}}''(\rho) = t_{\vec{k}}(\rho)$.  Since there are no normalizable solutions to this constraint, we have $t_{\vec{k}}(\rho) = t_{\vec{k}}^*(\rho) = 0$ and so the remaining $t_{\vec{k}}$ terms drop out of the quadratic action.  As such, we have shown stability in the $d = 2$ case.

Now we turn to $d > 2$, and again turn on momentum $\vec{k}$ along the torus.  The fluctuations decompose into irreps of $SO(d-1)$, in particular: two scalars, a vector, and a symmetric traceless tensor.  We parameterize these as
\begin{align}
h_{\mu \nu} dx^\mu dx^\nu &= \left(\frac{b^{1/2}}{2^{1/4}} \, \cosh^{\frac{1}{d}}\!\left(\frac{d\,\rho}{2}\right)\right)^{4-d}\sum_{\vec{k}} \bigg(\sqrt{\frac{2(d-2)}{d-1}}\,s_{\vec{k}}(\rho) \frac{k_i k_j}{\vec{k}^2} + \sqrt{\frac{2}{(d-1)(d-2)}}\,t_{\vec{k}}(\rho)\,\delta_{ij} \nonumber \\
& \qquad \qquad \qquad \qquad +  \frac{1}{2}\sum_{i=2}^{d} \left(v_{i,\vec{k}}(\rho) \frac{k_j}{|\vec{k}|} + v_{j,\vec{k}}(\rho) \frac{k_i}{|\vec{k}|} \right) +  \sum_{i,j=2}^{d} T_{ij,\vec{k}}(\rho) \bigg)  e^{i \vec{k} \cdot \vec{x}} dx^i dx^j
\end{align}
The quadratic action corresponding to these fluctuations is
\begin{align}
\begin{split}
	S_2 =\,& \frac{1}{16 \pi G} \int d\rho \sum_{\vec{k}}\Bigg( s_{\vec{k}}^*\left(t_{\vec{k}}'' - \frac{d^2}{4}\,t_{\vec{k}}\right) 
-\Bigg(\big|t_{\vec{k}}'\big|^2 + \Bigg(\frac{d^2}{4} + \frac{2\vec{k}^2}{b^2} \frac{1}{\cosh^{\frac{4}{d}}\!\left(\frac{d\,\rho}{2}\right)}\Bigg)\big|t_{\vec{k}}\big|^2\Bigg) 
	\\
&+\sum_{i=2}^d \left(\big|v_{i,\vec{k}}'\big|^2 + \frac{d^2}{4}\,\big|v_{i,\vec{k}}\big|^2\right) + \sum_{i,j=2}^d \Bigg(\big|T_{ij,\vec{k}}'\big|^2 + \Bigg(\frac{d^2}{4} + \frac{2\vec{k}^2}{b^2} \frac{1}{\cosh^{\frac{4}{d}}\!\left(\frac{d\,\rho}{2}\right)}\Bigg)\big|T_{ij,\vec{k}}\big|^2\Bigg)\Bigg)
\end{split}
\end{align}
where we have integrated out the torus directions.  We immediately see that the quadratic actions for the vectors $v_{i,\vec{k}}$ and tensor $T_{ij,\vec{k}}$ fluctuations are manifestly positive semi-definite. As in $d = 2$, the $s_{\vec{k}}$ fluctuations appear linearly, and so we should deform the path integration contour of $s_{\vec{k}}$ or $t_{\vec{k}}$ or a suitable combination thereof.  Then the $s_{\vec{k}}$ fluctuations act as Lagrange multipliers imposing $t_{\vec{k}}''(\rho) = \frac{d^2}{4} \, t_{\vec{k}}(\rho)$ as well as the complex conjugate of this equation.  But there are no normalizable solutions, and so $t_{\vec{k}}(\rho) = t_{\vec{k}}^*(\rho) = 0$.  Since the remaining $t_{\vec{k}}(\rho)$ dependence drops out, the wormhole is stable against quadratic fluctuations.

While we have focused on Euclidean wormholes, there are similar simplifications in the quadratic action of metric fluctuations around empty AdS and global AdS. This leads to a strategy for evaluating the gravity path integral to one loop, as well as for identifying boundary graviton degrees of freedom. We expect that this sort of analysis allows one to make precise a notion of a theory of reparameterizations for the stress tensor both for AdS$_3$ gravity in the metric formalism, where there is already such a model obtained from the first-order formalism~\cite{Cotler:2018zff} and from CFT Ward identities~\cite{Haehl:2018izb}, and in higher-dimensional AdS, where we note the proposal~\cite{Haehl:2019eae}. It also allows for a computation of the $\mathbb{T}^2\times I$ wormhole amplitude in the metric formalism of AdS$_3$ gravity~\cite{upcoming}.

\subsection{$\mathbb{S}^1 \times \mathbb{S}^3$ wormholes in radial gauge}

Here we turn to the stability analysis for the $\mathbb{S}^1 \times \mathbb{S}^3$ wormholes~\eqref{E:S1xS3wormhole}, for general positive $\beta_1, \beta_2$.  Similar to the torus wormhole analysis above, we will consider the problem in the setting of $\lambda$-solutions, again in radial gauge. These wormholes have $U(1)\times SO(4)$ isometry. In order to assess stability, we will check the most dangerous sectors: the $s$-wave sector, and the sector with one unit of angular momentum on the $\mathbb{S}^3$, both with no momentum around the $\mathbb{S}^1$. We expect that turning modes with more angular momentum modes are even more stable than these.

\subsubsection{$s$-wave sector}

Let us begin by considering a more general metric, namely
\begin{equation}
	ds^2 = d\rho^2 + e^{2 \,f_1(\rho)} d\tau^2 + e^{2\,f_2(\rho)} d\Omega_3^2\,.
\end{equation}
Then the Einstein-Hilbert action evaluated on this metric can be written as
\begin{equation}
\label{E:swaveS}
	S[f_1, f_2] = \frac{3\pi}{4G}\int d\rho \, e^{f_1} \left(e^{3 f_2} \left(f_2'' + 2 f_2'^2 -2\right) - e^{f_2} \right)\,,
\end{equation}
where we have integrated over the $\mathbb{S}^1$ and $\mathbb{S}^3$.  Notice that $e^{f_1(\rho)}$ appears linearly in the action, and so $e^{f_1}$ acts as a Lagrange multiplier, enforcing the constraint 
\begin{equation}
	f_2'' + 2 f_2'^2 -2 - e^{-2 \, f_2} = 0\,.
\end{equation}
The solution to this constraint is
\beq
	e^{2f_2} = \frac{b^2\cosh(2(\rho-\rho_0)) - 1}{2}\,,
\eeq
labeled by the two parameters $b$ and $\rho_0$. Clearly the action evaluated on this vanishes.

Because the action takes this simple form it follows that, at the level of fluctuations in $f_1$ and $f_2$, the fluctuation of $f_1$ acts as a Lagrange multiplier (upon rotating its contour of integration) and the fluctuations of $f_2$ are accordingly constrained. The two allowed perturbations of $f_2$ are simply perturbations of the parameters $b$ and $\rho_0$ above. Both are non-normalizable. As such, upon integrating out fluctuations of $f_1$, there are no normalizable fluctuations of $ f_2$ that satisfy the constraint. The vanishing of the quadratic action after integrating out fluctuations of $f_1$ demonstrates stability in this sector, and the fact that there are no normalizable solutions to the constraint shows that there are no normalizable zero modes either.

\subsubsection{Turning on one unit of angular momentum on the $\mathbb{S}^3$}

Here we analyze the next most dangerous sector, those modes with the one unit of angular momentum on the $\mathbb{S}^3$ and none on the $\mathbb{S}^1$.  The fluctuations with these quantum numbers are, after a choice of normalization,
\begin{align}
\begin{split}
	h_{\tau\tau} &=\frac{2}{\sqrt{3}}\,\frac{1}{\sqrt{G}}\,g_{\tau\tau} \,h_1(\rho)\,Y^{\ell=1}(\Omega_3)\,,
	\\
	h_{ab} & =\frac{1}{\sqrt{3}} \,\frac{1}{\sqrt{G}}\,g_{ab}\,h_2(\rho)Y^{\ell=1}(\Omega_3)\,,
	\\
	h_{\tau a} & =\frac{2\,e^{f_1+f_2}}{\sqrt{G}}\sum_A v_A(\rho)\,Y^{\ell = 1,A}_a(\Omega_3)\,.
\end{split}
\end{align}
Here $x^a$ denotes the three angles on the $\mathbb{S}^3$, $e^{2f_1} = g_{\tau\tau}$ is the $\mathbb{S}^1$ warpfactor, $e^{2f_2} = \frac{b^2\cosh(2\rho)-1}{2}$ is the $\mathbb{S}^3$ warpfactor, $G = \frac{\sqrt{g}}{\sqrt{g_{\mathbb{S}^3}}}=e^{f_1+3f_2}$, $Y^{\ell}(\Omega_3)$ is a scalar harmonic on the $\mathbb{S}^3$, and $Y_a^{\ell,A}(\Omega_3)$ are the vector harmonics on $\mathbb{S}^3$ with $A=0,1,2$. The $A=0$ harmonics are the derivatives of the scalar harmonics, while the others are co-closed. With zero momentum around the $\mathbb{S}^1$ there is a time-reversal symmetry that ensures that the fluctuations $h_{\tau a}$ decouple from the others.

To proceed we recall that the scalar harmonics satisfy
\beq
	-\Box_{\mathbb{S^3}} Y^{\ell}(\Omega_3) = \ell(\ell+2)Y^{\ell}(\Omega_3)\,,
\eeq
and the vector harmonics obey (see e.g.~\cite{lindblom2017scalar}) 
\beq
	-\Box_{\mathbb{S}^3} Y^{\ell,A}_a(\Omega_3)=(\ell(\ell+2)-1-\delta_A^0)Y^{\ell,A}_a(\Omega_3)
\eeq
and exist for $\ell \geq 1$.  After simplification the quadratic action for these fluctuations reads
\begin{align}
	S_2 =\,& \frac{1}{16 \pi G} \int d\rho \left\{ h_1^* \bigg( h_2'' + \frac{2 \tanh (2 \rho )}{b^2 \cosh (2 \rho )-1}\,h_2' \right.
	\nonumber 
	\\
	& \qquad \qquad \qquad\qquad-\frac{\left(2 b^4 \cosh (4 \rho )-4 b^2 \cosh (2 \rho )+2 b^4+\text{sech}^2(2 \rho)-1\right)}{\left(b^2 \cosh (2 \rho )-1\right)^2}\,h_2\bigg) + \text{c.c.} 
	\nonumber 
	\\
	& \qquad \qquad \qquad- |h_2'|^2 - \frac{2 b^4 \cosh (4 \rho )-6 b^2 \cosh (2 \rho )+2 b^4+\text{sech}^2(2 \rho
   )+1}{\left(b^2 \cosh (2 \rho )-1\right)^2}\,|h_2|^2
	\\
	& \qquad \qquad \qquad +|v_0'|^2 + \frac{2 b^4 \cosh (4 \rho )-10 b^2 \cosh (2 \rho )+2 b^4-3 \,\text{sech}^2(2 \rho
   )+9}{\left(b^2 \cosh (2 \rho )-1\right)^2} \, |v_0|^2 
	\nonumber 
	\\
	&\qquad \qquad \qquad + \left.\sum_{A=1}^2\left(|v_A'|^2 + \frac{2 b^4 \cosh (4 \rho )-2 b^2 \cosh (2 \rho )+2 b^4-3 \,\text{sech}^2(2 \rho
   )+1}{\left(b^2 \cosh (2 \rho )-1\right)^2}\, |v_A|^2 \right)\right\}\,,
	\nonumber
\end{align}
where we have integrated over the $\mathbb{S}^1$ and $\mathbb{S}^3$.  Notice that the vector fluctuations decouple, and their action is positive semi-definite for $b > 1$. However, $h_1$ and $h_2$ couple but in a simple way.  Just like in the $s$-wave setting, $h_1$ appears linearly and multiplies a terms linear in $h_2$, and so we should rotate the path integration contour of $h_1$ or $h_2$ or an appropriate combination of each.  Then $h_1$ will act as a Lagrange multiplier and impose the constraint
\begin{align}
h_2'' + \frac{2 \tanh (2 \rho )}{b^2 \cosh (2 \rho )-1}\,h_2'-\frac{\left(2 b^4 \cosh (4 \rho )-4 b^2 \cosh (2 \rho )+2 b^4+\text{sech}^2(2 \rho)-1\right)}{\left(b^2 \cosh (2 \rho )-1\right)^2}\,h_2 = 0\,.
\end{align}
We find numerically that there are no non-trivial normalizable solutions to this equation for $b > 1$.  As such, the only normalizable solution is $h_2 = 0$.  Since the action for the $v_A$'s is positive semi-definite for $b > 1$, we have shown that all fluctuations with one unit of angular momentum on the $\mathbb{S}^3$ are quadratically stable.

\subsection{$\mathbb{S}^1 \times \mathbb{S}^3$ wormholes in $R_\xi$ gauge with length constraint}

Here we turn to an alternative stability analysis of the $\mathbb{S}^1 \times \mathbb{S}^3$ wormholes where the wormhole is stabilized as a constrained instanton with a fixed length.  In particular, we will consider Einstein gravity in $R_\xi$ gauge, and impose the length constraint in Eq.~\eqref{E:constraint} with $F(x) = \sqrt{g_{\mathbb{S}^3}}$\,.  Our stability analysis will follow along the lines of Appendix C of~\cite{cotler2020gravitational}, which treated the case of wormholes with $\mathbb{T}^d$ boundary and $\beta_1 = \beta_2$.

The 5d quadratic action for fluctuations $h_{\mu \nu}$ can be written as~\cite{bastianelli2013one, christensen1980quantizing}
\begin{align}
\label{E:qfluct1}
S_2 = \frac{1}{16 \pi G}&\int d^5 x \, \sqrt{g}\bigg(- \frac{1}{4} \,\overline{h}^{\mu \nu} \square \overline{h}_{\mu \nu} + \frac{3}{40} h \square h - \frac{1}{2}\left[D^\nu \overline{h}_{\nu \mu} - \frac{3}{10} D_\mu h\right]^2 - \frac{1}{2}\, \overline{h}^{\mu \lambda} \overline{h}^{\nu \sigma} R_{\mu \nu \lambda \sigma}  \nonumber \\
& \qquad \qquad \quad \, - \frac{1}{2}\left[\overline{h}^{\mu \lambda} \overline{h}_\lambda^\nu - \frac{1}{5} h \overline{h}^{\mu \nu}\right]R_{\mu \nu} + \frac{1}{4} \overline{h}^{\mu \nu} \overline{h}_{\mu \nu} (R - 2 \Lambda) - \frac{3}{200} h^2 R +\frac{3}{20} h^2 \Lambda \bigg) \nonumber \\
&\qquad \qquad \qquad \qquad \qquad \qquad \qquad \qquad \qquad \qquad \qquad \,\,\,\, - i \,\frac{\lambda}{8} \int d^{d+1} x \, \sqrt{g_{\mathbb{S}^3}}\, h_{\rho \rho}^2\,,
\end{align}
where we have used the notation $h_{\mu \nu} = \overline{h}_{\mu \nu} + \frac{1}{5} \, g_{\mu \nu} \, h$ with $\overline{h}_{\mu \nu}$ traceless, $\square = D_\mu D^\mu$, and $\lambda = - i \frac{3 b^4}{32\pi G}(\beta_1 + \beta_2)$.  Considering Eq.~\eqref{E:gaugeghost} with $\xi = 1$, we introduce the gauge fixing
\begin{equation}
\mathcal{G}_\mu = D^\nu \overline{h}_{\nu \mu} - \frac{3}{10} D_\mu h\,,
\end{equation}
and integrate out the gauge-fixing auxiliary field $f_\mu$.  Because there is a quadratic term in $f_{\mu}$ this adds to the quadratic action~\eqref{E:qfluct1} the term
\begin{equation}
S_{\text{gauge}} = \frac{1}{16\pi G} \int d^{5}x \, \frac{1}{2}\left[D^\nu \overline{h}_{\nu \mu} - \frac{3}{10} D_\mu h\right]^2
\end{equation}
and there is also a ghost action that is not relevant for addressing the question of perturbative stability.  This type of gauge-fixing term is referred to as $R_\xi$ gauge, where in our case $\xi = 1$.  Adding $S_{\text{gauge}}$ to $S_{2}$ cancels out an inconvenient term, giving us the following expression for $\widetilde{S}_2 = S_2 + S_{\text{gauge}}$\,:
\begin{align}
\label{E:qfluct2}
\widetilde{S}_2 = \frac{1}{16 \pi G}&\int d^5 x \, \sqrt{g}\bigg(- \frac{1}{4} \,\overline{h}^{\mu \nu} \square \overline{h}_{\mu \nu} + \frac{3}{40} h \square h - \frac{1}{2}\, \overline{h}^{\mu \lambda} \overline{h}^{\nu \sigma} R_{\mu \nu \lambda \sigma}  - \frac{1}{2}\left[\overline{h}^{\mu \lambda} \overline{h}_\lambda^\nu - \frac{1}{5} h \overline{h}^{\mu \nu}\right]R_{\mu \nu} \nonumber \\
& \qquad \qquad \,  + \frac{1}{4} \overline{h}^{\mu \nu} \overline{h}_{\mu \nu} (R - 2 \Lambda) - \frac{3}{200} h^2 R +\frac{3}{20} h^2 \Lambda \bigg) - i \,\frac{\lambda}{8} \int d^{d+1} x \,\sqrt{g_{\mathbb{S}^3}}\, h_{\rho \rho}^2\,.
\end{align}
Let us consider quantum fluctuations with $\omega$ units of angular momentum on the $\mathbb{S}^1$, and $\ell$ units of angular momentum on the $\mathbb{S}^3$.  Then the fluctuations $h_{\mu \nu}$ with these quantum numbers can be parameterized as
\begin{align}
\label{E:S1S3flucts}
\begin{split}
   h_{\rho \rho} &=\frac{2}{\sqrt{G}}\,h_1(\rho)\,e^{i \omega \tau}\,Y^{\ell}(\Omega_3)\,, \\
	h_{\tau\tau} &=\frac{2}{\sqrt{G}}\,g_{\tau\tau} \,h_2(\rho)\,e^{i \omega \tau}\,Y^{\ell}(\Omega_3)\,,
	\\
	h_{\mathbb{S}^3} &= \frac{1}{3}\,g^{ab} h_{ab}  = \frac{2}{\sqrt{3}} \frac{1}{\sqrt{G}}\,h_3(\rho)\,e^{i \omega \tau}\,Y^{\ell}(\Omega_3)\,,
	\\
	h_{\rho \tau} & =\frac{2\,e^{f_1}}{\sqrt{G}}\,u(\rho)\,e^{i \omega \tau}\,Y^{\ell}(\Omega_3)\,,
	\\
	h_{\rho a} & =\frac{2\,e^{f_2}}{\sqrt{G}}\sum_A v_A(\rho)\,e^{i \omega \tau}\,Y^{\ell,A}_a(\Omega_3)\,, \\
   h_{\tau a} & =\frac{2\,e^{f_1+f_2}}{\sqrt{G}}\sum_A w_A(\rho)\,e^{i \omega \tau}\,Y^{\ell,A}_a(\Omega_3)\,, \\
   	\overline{h}_{ab} & =\frac{\sqrt{2}\,e^{f_2}}{\sqrt{G}}\sum_A t_A(\rho)\,e^{i \omega \tau}\,Y^{\ell,A}_{ab}(\Omega_3)\,.
\end{split}
\end{align}
As before, $x^a$ denotes the angles of the $\mathbb{S}^3$, $e^{2 f_1} = g_{\tau \tau}$, $e^{2 f_2} = \frac{b^2 \cosh(2\rho) - 1}{2}$ is the $\mathbb{S}^3$ warpfactor, and $G = \frac{\sqrt{g}}{\sqrt{g_{\mathbb{S}^3}}}$.  We use the same notation as before for the $\mathbb{S}^3$ harmonics, and additionally introduce the tensor harmonics $Y^{\ell,A}_{ab}(\Omega_3)$.  The traceless tensor harmonics for $A = 1,2,3,4,5$ obey (see e.g.~\cite{lindblom2017scalar} for the traceful tensor harmonics)
\begin{align}
- \square_{\mathbb{S}^3} Y^{\ell,A}_{ab}(\Omega_3) &= \left(\ell(\ell + 2) - N_A \right)  Y^{\ell,A}_{ab}(\Omega_3)
\end{align}
where
\begin{equation}
N_A = \begin{cases} 
5 & \text{for }A = 1,2 \\
6 & \text{for }A = 3 \\
2 & \text{for }A = 4,5 \\
\end{cases}\,\,.
\end{equation}
We remark that in Eq.~\eqref{E:S1S3flucts}, the vector harmonics are only defined for $\ell \geq 1$, and similarly the tensor harmonics are only defined for $\ell \geq 2$.  That is, for small enough $\ell$ the vector and tensor fluctuations can be set to zero.

The advantage of this gauge choice is that modes in different representations of the $U(1)\times SO(4)$ isometry decouple at the quadratic level. Indeed, by inspecting Eq.~\eqref{E:qfluct2} and utilizing the orthogonality relations of the $\mathbb{S}^3$ harmonics, we see that the $h_{\rho \tau}$, vector, and tensor sectors all decouple.  The only coupled sector contains $h_{\rho \rho}$, $h_{\tau \tau}$, and $h$.  Let us begin by showing the stability of the decoupled sectors.  Letting $M_{S}(\ell) = \ell (\ell+2)$, $M_{V}(\ell,A) = \ell (\ell+2) - 1 - \delta_0^A$,  $M_{T}(\ell,A) = \ell (\ell+2) - N_A$, the quadratic actions are:
\begin{align}
	S_u = &\frac{1}{16 \pi G} \int d\rho \,\bigg(|u'|^2 \!+\! \bigg(\omega^2 \, e^{- 2 f_1} + M_S(\ell)\,e^{-2 f_2}  + e^{- 2 f_1 - 4 f_2} \frac{3 b^4}{64}  \big(-6 \beta_1^2 e^{4 \rho }+28 \beta_1 \beta_2
   -6 \beta_2^2 e^{-4 \rho } 
	\nonumber 
	\\
	& \qquad \qquad \qquad \qquad \,\,\, + b^2 (3 \beta_1^2 e^{6 \rho
   }+\beta_1 e^{2 \rho } (11 \beta_1-6 \beta_2)-\beta_2 e^{-2 \rho } (6 \beta_1-11 \beta_2)+3 \beta_2^2 e^{- 6 \rho})\big)   \bigg) |u|^2 \bigg) 
	\\
	S_v = &\frac{1}{16 \pi G} \int d\rho \,\bigg(|v_A'|^2 \!+\! \bigg(\omega^2 \, e^{- 2 f_1} + M_V(\ell,A)\,e^{-2 f_2}  + e^{- 2 f_1 - 6 f_2} \frac{b^4}{256} \big(4 \left(5 \beta_1^2 e^{4 \rho }-2 \beta_1 \beta_2+5
   \beta_2^2 e^{-4 \rho }\right)
	\nonumber 
	\\
	& \qquad \qquad \qquad \qquad \qquad \quad  -4 b^2 \left(3 \beta_1^2 e^{6 \rho }+\beta_1 e^{2 \rho } (11
   \beta_1+2 \beta_2)+\beta_2 e^{-2 \rho } (2 \beta_1+11 \beta_2)+3 \beta_2^2 e^{-6 \rho }\right)  	\nonumber 
   	\\
	& \qquad \qquad \qquad \qquad \qquad \qquad \qquad \quad \quad  + b^4 \big(5 \left(5 \beta_1^2-4 \beta_1 \beta_2+5
   \beta_2^2\right)+9 \beta_1^2 e^{8 \rho } +2 \beta_1 e^{4
   \rho } (\beta_1+11 \beta_2)
	\nonumber 
	\\
	& \qquad \qquad \qquad \qquad \qquad \qquad \qquad \qquad \qquad \qquad \qquad \quad +2 \beta_2 e^{-4 \rho } (11
   \beta_1+\beta_2)+9 \beta_2^2 e^{-8 \rho }\big) \bigg) |v_A|^2 \bigg) 
   	\\
	S_w = &\frac{1}{16 \pi G} \int d\rho \,\bigg(|w_A'|^2 \!+\! \bigg(\omega^2 \, e^{- 2 f_1} + M_V(\ell,A)\,e^{-2 f_2}  + e^{- 2 f_1 - 6 f_2} \frac{b^4}{64}  \big(5 \beta_1^2 e^{4 \rho }-2 \beta_1 \beta_2+5 \beta_2^2 e^{-4 \rho }
	\nonumber 
	\\
	& \qquad \qquad \qquad \qquad \qquad \quad  -3 b^2 \left(\beta_1^2 \beta_2^2 \left(e^{-6 \rho }+3 e^{-2 \rho
   }\right)+\beta_1^2 \left(3 e^{2 \rho }+e^{6 \rho }\right)\right) + b^4 \big(4 \beta_1^2-2 \beta_1 \beta_2+4 \beta_2^2 
   	\nonumber 
	\\
	& \qquad \qquad \qquad \qquad \qquad \qquad \quad \quad \quad  + \beta_1^2 e^{8 \rho }+2 \beta_1 e^{4 \rho } (\beta_1+\beta_2)+2 \beta_2 e^{-4 \rho } (\beta_1+\beta_2)+\beta_2^2\big)\bigg) |w_A|^2 \bigg) 
	\\
	S_t = &\frac{1}{16 \pi G} \int d\rho \,\bigg(|t_A'(\rho)|^2 \!+\! \bigg(\omega^2 \, e^{- 2 f_1} + M_T(\ell,A)\,e^{-2 f_2}  + e^{- 2 f_1 - 6 f_2} \frac{b^4}{64}  \big(-3 \beta_1^2 e^{4 \rho }-2 \beta_1 \beta_2-3 \beta_2^2 e^{-4 \rho }
	\nonumber 
	\\
	& \qquad \qquad \qquad \qquad \qquad \quad  -b^2 \left(\beta_1^2 e^{6 \rho }-\beta_1 e^{2 \rho } (\beta_1-4 \beta_2)+\beta_2 e^{-2 \rho } (4 \beta_1-\beta_2)+\beta_2^2 e^{-6 \rho }\right) 
	\nonumber 
	\\
	& \qquad \qquad \qquad \qquad \qquad \qquad \quad \,\,\,\,\,\,\,  + b^4 \left(\beta_1 \left(2 e^{4 \rho }+e^{8 \rho }\right)+\beta_2\right) \left(\beta_1+\beta_2 \left(e^{-8 \rho }+2 e^{-4 \rho
   }\right)\right)\big)\bigg) |t_A|^2 \bigg) \,.
\end{align}
Notice that the modes above are made more stable by turning on angular momentum on $\mathbb{S}^1$ and $\mathbb{S}^3$.  Thus if the sector with minimal angular momentum is stable, then so are the other sectors.  We have checked numerically for a large range of parameters $b > 1$, $\beta_1, \beta_2 > 0$ that for each action above the effective mass term is strictly positive for minimal $\omega$ and $L$.  An efficient way of doing so is to note that up to rescaling coordinates, the potential (for minimal $\omega$ and $L$) only depends on $b$, $\beta_1/\beta_2$ and $\rho$.  Then we can solve for $b$ as a function of $\beta_1/\beta_2$ and $\rho$, and check that the solutions are always less than or equal to unity. But, since $b>1$ in order for the wormhole to not pinch off, this implies that the potentials never cross zero. The potentials are clearly positive at large $\rho$, and thus putting the two together, are non-negative everywhere.  Thus the $h_{\rho \tau}$, $h_{\rho a}$, $h_{\tau a}$ and $\overline{h}_{ab}$ fluctuations are stable.

The fully coupled quadratic action for the $h_{\rho \rho}$, $h_{\tau \tau}$ and $h_{\mathbb{S}^3}$ is quite lengthy, and we do not write it down here.  To assess the stability of this coupled scalar sector, it is most convenient to search for normalizable zero modes of the equations of motion. This avoids the subtle question of deciding how to (or if to) deform the path integration contour of the fluctuations, as is common practice when treating fluctuations of the conformal mode~\cite{gibbons1978path}; while we had a sensible way of doing this in radial gauge, it is less clear how to proceed in $R_{\xi}$ gauge. As such, we performed a numerical search for normalizable zero modes for $\beta_1 = \beta_2$ and $b > 1$, and did not find such solutions.  Note that as $b$ goes to infinity the scalar sector becomes equivalent to the scalar sector of the wormhole with $\mathbb{T}^4$ cross-section; we have established the stability of the latter for $\beta_1 = \beta_2$ in $R_{\xi}$ gauge with a length constraint in~\cite{cotler2020gravitational}.  This means that we do not have to worry about the regime of very large $b$ in our $\mathbb{S}^1 \times \mathbb{S}^3$ wormhole analysis, and so our numerics for $b > 1$ up to a finite but large value of $b$ suffice to establish stability.  Taken together, our results imply the stability of the coupled scalar sector in the $\mathbb{S}^1 \times \mathbb{S}^3$ setting for $\beta_1 = \beta_2$.  Since we have already checked the stability of the other sectors, this indicates that the $\mathbb{S}^1 \times \mathbb{S}^3$ wormholes with $\beta_1 = \beta_2$ are completely stable.

We remark that the stability analysis in $R_{\xi}$ gauge with a length constraint is complementary to the stability analysis in radial gauge.  In particular, in $R_{\xi}$ gauge increasing angular momentum on the $\mathbb{S}^1$ and $\mathbb{S}^3$ only serves to increase the stability.  As such, it suffices to establish stability in sectors with minimal angular momentum.  Moreover the vector and tensor sectors decouple.  The most difficult sector to analyze in $R_{\xi}$ gauge is the coupled scalar sector; here we likewise only had to study the case of zero angular momentum, i.e.~the $s$-wave sector.  By contrast, in radial gauge it is quite easy to establish stability in the $s$-wave sector, whereas higher angular momentum sectors have coupling between various different fluctuations and so are more difficult to handle.

\section{Wormholes in Euclidean AdS$_5\times\mathbb{S}^5$
\label{S:stringwormholes}}

\subsection{Embedding into supergravity}
\label{S:embedding}

Now we turn our hand to the task of finding wormholes in supergravity and AdS/CFT. Rather than study many different compactifications, we elect to study wormholes in the original example of the AdS/CFT correspondence, type IIB supergravity on AdS$_5\times\mathbb{S}^5$ with $N$ units of Ramond-Ramond flux. IIB string theory on this background is dual to four-dimensional maximally supersymmetric Yang-Mills theory with gauge group $SU(N)$. 

We begin with a brief review of the basics of this compactification. The bosonic field content of 10d IIB supergravity is a metric $G_{MN}$, a dilaton $\Phi$, the Neveu-Schwarz B-field $B_2$, and the various Ramond-Ramond potentials $(C_0, C_2, C_4)$. The field strengths are
\begin{align}
\begin{split}
	H_3 & = dB_2 \,,
	\\
	F_1 & = dC_0\,,
	\\
	F_3 & = dC_2\,, \qquad \widetilde{F}_3  = dC_2 -C_0 \wedge H_3\,,
	\\
	F_5 & = dC_4\,, \qquad \widetilde{F}_5 = F_5 +\frac{1}{2}(B_2\wedge F_3 - C_2 \wedge H_3)\,,
\end{split}
\end{align}
and on-shell the five-form is self-dual, $\widetilde{F}_5 = \star \widetilde{F}_5$. The self-duality condition famously obstructs a simple action principle for IIB supergravity. The standard bosonic action used in the literature is, in Lorentzian signature
\begin{align}
\begin{split}
	S_{\rm IIB} &= S_{\rm NS} + S_{\rm RR} + S_{\rm CS}\,,
	\\
	S_{\rm NS} &= \frac{1}{2\kappa_{10}^2} \int d^{10}x\sqrt{-G}\,e^{-2\Phi}\left( R + 4 (\partial\Phi)^2-\frac{1}{2}|H_3|^2\right)\,,
	\\
	S_{\rm RR} &= -\frac{1}{4\kappa_{10}^2} \int d^{10}x \sqrt{-G} \left( |F_1|^2 + |\widetilde{F}_3|^2 + \frac{1}{2}|\widetilde{F}_5|^2\right)\,,
	\\
	S_{\rm CS} & = - \frac{1}{4\kappa_{10}^2} \int C_4 \wedge H_3 \wedge F_3\,,
\end{split}
\end{align}
with $2\kappa_{10}^2 = (2\pi)^7\ell_s^8$. To obtain the equations of motion one varies this action, and then by hand enforces the self-duality of $\widetilde{F}_5$. We have notated that for a $p$-form $V_p$,
\beq
	|V_p|^2 = \frac{1}{p!} V_{M_1\hdots M_p}V_{N_1\hdots N_p}G^{M_1N_1} \hdots G^{M_pN_p}\,.
\eeq
The action and fields above are written in the string frame. For our purposes it is better to work in the Einstein frame, with metric $G_{EMN} = e^{-\frac{\Phi}{2}}G_{MN}$, and to combine the axion and dilaton into the IIB axiodilaton $\tau = C_0 + i e^{-\Phi}$. Then the bosonic action becomes
\begin{align}
\begin{split}
	S_{\rm IIB} &= \frac{1}{2\kappa_{10}^2} \int d^{10}x \sqrt{-G_E} \left( R(G_E) - \frac{|\partial\tau|^2}{2(\text{Im}\,\tau)^2} - \frac{\mathcal{M}_{ij} }{2}F_3^i\cdot F_3^j - \frac{1}{4}|\widetilde{F}_5|^2\right) + S_{\rm CS}\,,
\end{split}
\end{align}
where $R(G_E)$ is the scalar curvature of $G_E$, $F_3^i = (H_3,F_3)$, the matrix $\mathcal{M}_{ij}$ is
\beq
	\mathcal{M}_{ij} = \frac{1}{\text{Im}\,\tau}\begin{pmatrix} |\tau|^2 & - \text{Re}\,\tau \\ -\text{Re} \,\tau & 1 \end{pmatrix}\,,
\eeq
and in writing $|\partial\tau|^2$, $F^i_3\cdot F_3^j$, and $|\widetilde{F}_5|^2$, the Einstein-frame metric $G_E$ is used to raise indices. Further, $\widetilde{F}_5$ is still self-dual with respect to the Hodge star obtained with $G_E$.

The AdS$_5\times\mathbb{S}^5$ vacuum of IIB supergravity is, in Einstein frame, the background
\begin{align}
\begin{split}
	ds^2_E &= \widetilde{L}^2 \left( ds^2_{\rm AdS} + d\Omega_5^2\right)\,, \qquad \tau= \frac{i}{g_s} \,,	\qquad F_5 = 4\widetilde{L}^4(\text{vol}_{\rm AdS} + \text{vol}_{\mathbb{S}^5})\,,
\end{split}
\end{align}
with vanishing 3-form flux. Here $ds^2_{\rm AdS}$ is the line element on a unit radius AdS$_5$, $\text{vol}_{\rm AdS}$ its volume form, $d\Omega_5^2$ is the line element on a unit radius $\mathbb{S}^5$, and $\text{vol}_{\mathbb{S}^5}$ its volume form. There are $N$ units of five-form flux through the $\mathbb{S}^5$, for which the flux quantization condition reads
\beq
	\int_{\mathbb{S}^5} \star F_5 = 2\kappa_{10}^2 \mu_3 N\,, 
\eeq
where $\mu_3$ is the RR charge of a D3-brane, characterized by the Wess-Zumino term in its action,
\beq
	S_{\rm WZ} = \mu_3 \int C_4\,, \qquad \mu_3 = \frac{1}{(2\pi)^3\ell_s^4}\,.
\eeq
This produces
\beq
	\frac{\widetilde{L}^4}{\ell_s^4} = 4\pi  N\,.
\eeq
The string frame radius is $L = e^{\frac{\Phi}{4}}\widetilde{L} = g_s^{\frac{1}{4}}\widetilde{L}$, so that one has the standard result $\frac{L^4}{\ell_s^4} = 4\pi g_s N$.

The AdS$_5\times\mathbb{S}^5$ vacuum can be understood as that of a simple $SO(6)$-invariant reduction on $\mathbb{S}^5$. Separate the ten dimensions as $x^M = (x^{\mu},y^{\alpha})$ with $y^{\alpha}$ the angles on $\mathbb{S}^5$. The nonzero bosonic fields in this simple reduction are
\begin{align}
\begin{split}
	ds^2_E &=  e^{-\frac{10}{3}\varphi(x)} g_{\mu\nu}(x)dx^{\mu}dx^{\nu} + \widetilde{L}^2e^{2\varphi(x)}d\Omega_5^2 \,,  \qquad \tau = \tau(x)\,,
	\\
	F_5 & = 4\widetilde{L}^4 \left( \widetilde{L}^{-5}e^{-\frac{40}{3}\varphi(x)} \text{vol}_5+\text{vol}_{\mathbb{S}^5} \right)\,.
\end{split}
\end{align}
where $g_{\mu\nu}$ is the 5d metric and $\text{vol}_5 $ its volume form. The Weyl shift in front of the 5d metric is there so that the 5d metric is in Einstein frame. The 5d action obtained upon reduction is
\beq
\label{E:5daction}
	S_5 = \frac{1}{16\pi G_5} \int d^5x \sqrt{-g} \left( R(g) -\frac{|\partial\tau|^2}{2(\text{Im}\,\tau)^2} + \frac{40}{3} (\partial\varphi)^2 - \mathcal{V}(\varphi)\right)\,, \qquad \mathcal{V}(\varphi) =\frac{8e^{-\frac{40}{3}\varphi}-20e^{-\frac{16}{3}\varphi} }{\widetilde{L}^2}\,, 
\eeq
and the potential has a minimum at $\varphi = 0$. For small fluctuations around $\varphi = 0$ one has $\mathcal{V}(\varphi) = -12 +\frac{40}{3}\times 32 \varphi^2 +O(\varphi^3)$. The 5d vacuum is AdS$_5$ with radius $L$, $ds^2_5 = \widetilde{L}^2 ds^2_{\rm AdS}$, with $\varphi = 0$ and constant $\tau$. The fluctuations of $\varphi$ around to the minimum comprise a massive scalar with $m^2L^2 = 32$, dual to a dimension 8 operator in the dual $\mathcal{N}=4$ super Yang-Mills, while those of the axiodilaton give two massless fluctuations dual to two marginal operators, the $\mathcal{N}=4$ Lagrangian and topological charge density.

We draw this out mainly to recall that the 5d Einstein frame metric comes from the 10d metric upon a suitable Weyl shift by the $\mathbb{S}^5$ warpfactor. Thus, we can simply embed our wormholes into IIB supergravity through geometries of the direct product form $(\text{5d wormhole})\times\mathbb{S}^5$. We first separate the non-compact space into a radial direction and the others, $x^{\mu} = (\rho,x^i)$, and then we have two options. The first is to use the constrained instanton calculus whereby we fix the 5d length of the wormhole with a constraint $\int d^{10}x \sqrt{g_{\rho \rho}} \,F(x^i,y^{\alpha})$ for a suitable $F$.  The other simple option is to make the wormhole a $\lambda$-solution as in the previous Subsection, by fixing a suitable radial gauge. The simplest option is to fix the radial components of the 5d metric as $g_{\mu\rho} = \delta_{\mu}^{\rho}$ and the mixed radial-angular components of the 10d metric to vanish, $g_{\rho\alpha} = 0$. 

In any case, we arrive at simple 10d versions of the 5d wormholes we studied above. In these backgrounds we set the dilaton equal to a constant, $\varphi$ to its minimum value at 0, the 5d metrics to those of the wormholes above, and the five-form flux is completely fixed by flux quantization and self-duality. In $\widetilde{L}=1$ units, the wormhole with torus cross-section is
\begin{align}
\begin{split}
\label{E:10dtorus}
	ds_{E}^2 &=\frac{b^2}{2}\cosh(2\rho) \left( \left( \frac{\beta_1 e^{2\rho}+\beta_2e^{-2\rho}}{2\cosh(2\rho)}\right)^2d\tau^2+ d\vec{x}_{\perp}^2\right) + d\rho^2 + d\Omega_5^2\,, \qquad	e^{\Phi}  =g_s\,,
	\\
	F_5 & = 4 (-i\text{vol}_5 + \text{vol}_{\mathbb{S}^5})\,,
\end{split}
\end{align}
where the $i$ arises from the continuation to Euclidean signature. If we stabilize this wormhole with a length constraint, then the function $F$ is $\sqrt{g_{\mathbb{S}^5}}$. Later when we study brane nucleation instabilities in this background we will require the four-form potential, which in a radial gauge reads
\beq
	C_4  = -i\frac{b^4}{16} \left( \beta_1e^{4\rho} - \beta_2 e^{-4\rho}+4(\beta_1+\beta_2)\rho\right) d\tau \wedge dx^1 \wedge dx^2 \wedge dx^3 + (\text{angular})\,.
\eeq
In the same way we can find a 10d wormhole with $\mathbb{S}^1\times\mathbb{S}^3$ cross-section, given by
\begin{align}
\label{E:10dS1xS3}
	ds_{E}^2 & = \frac{b^2\cosh(2\rho)-1}{2}\left( \left( \frac{\beta_1e^{2\rho}+\beta_2e^{-2\rho}}{2\left( \cosh(2\rho)-\frac{1}{b^2}\right)}\right)^2 d\tau^2 + d\Omega_3^2 \right) + d\rho^2 + d\Omega_5^2\,, \qquad	e^{\Phi} = g_s\,,
	\\
	\nonumber
	C_4 & =-i\left[\frac{b^4}{16} \left( \beta_1 e^{4\rho} -\beta_2 e^{-4\rho} + 4(\beta_1+\beta_2)\rho\right) - \frac{b^2}{4}(\beta_1 e^{2\rho}-\beta_2e^{-2\rho})\right] d\tau\wedge \text{vol}_{\mathbb{S}^3} + (\text{angular})\,,
\end{align}
where the five-form flux is given by the same expression as in~\eqref{E:10dtorus}, corresponding to the four-form potential indicated. If the wormhole is stabilized with a length constraint, then $F = \sqrt{g_{\mathbb{S}^3}}\,\sqrt{g_{\mathbb{S}^5}}$. 

One can readily verify that these are solutions to all of the IIB equations of motion except for the component of the Einstein's equations that corresponds to the variation with respect to $g_{\rho \rho}$, a combination of the $\rho \rho$ and angular components of the 10d Einstein's equations by virtue of $g_{\rho \rho} = G_{E\rho \rho} e^{\frac{10}{3}\varphi}$. 

As with the pure 5d analysis, if we had picked a different constraint or enforced a different gauge-fixing, the particular wormhole geometries we find would be different than those here. We stress again that the particular wormholes we study are not themselves of any particular significance: rather, the physical object is the wormhole amplitude, an appropriate integral over wormholes.

From this analysis it is clear how to uplift similar wormholes into other examples of the AdS/CFT correspondence. For instance, we can uplift Euclidean AdS$_3$ wormholes with torus cross-section to ones in the context of the $D1/D5$ system; Euclidean AdS$_4$ wormholes to ones in the context of coincident $M2$ branes; etc. 

\subsection{A partial stability analysis}

In this Subsection we initiate a study of the perturbative stability of the 10d wormhole with $\mathbb{S}^1\times\mathbb{S}^3$ cross-section in~\eqref{E:10dS1xS3}. For simplicity we stick to $\beta_1=\beta_2=\beta$, so that the wormhole is symmetric under a $\mathbb{Z}_2$ parity symmetry which sends $\rho\to - \rho$. However, we will only impose this when we numerically search for instabilities. By setting up our equations for general $\beta_1, \beta_2$, we can first check our analysis by setting $\beta_1=-\beta_2 = \beta$, so that the wormhole becomes a Euclidean black hole, and then reproduce known results for those.

The spectrum of fluctuations, and hence this stability analysis, depends on how we stabilize the wormholes. For the purposes of this Subsection we regard these wormholes as $\lambda$-solutions, whereby we fix a radial-like gauge. Denote the perturbation in the 5d metric be $h_{\mu\nu}$ and that in the 10d Einstein frame metric $H_{MN}$. At the linearized level this gauge choice reads
\beq
	h_{\rho\rho} = H_{\rho\rho} + \frac{1}{3} H_{\alpha}{}^{\alpha} = 0\,, \qquad h_{\rho i} = 0\,, \qquad H_{\rho\alpha} = 0\,,
\eeq
where we have separated $x^{\mu}=(\rho,x^i)$ and $y^{\alpha}$ denotes the angles on the $\mathbb{S}^5$. We also enforce as a gauge constraint on the 4-form potential that
\beq
	C_{\rho MNP} = 0\,.
\eeq
The axiodilaton is constant in the background and the 3-form fluxes vanish. Thus their perturbations decouple from those of the metric and four-potential at the quadratic level. The quadratic action for those fields is manifestly positive-definite on this background, and so fluctuations of those fields are perturbatively stable. 

The various perturbations can be organized according to the $U(1)\times SO(4)\times SO(6)$ symmetry of the background. On general grounds we expect that modes with lowest momenta are the most dangerous channels for a perturbative instability, and so we focus on low angular momentum. The modes with no angular momentum on the five-sphere are the fluctuations of the 5d metric and those of the $\mathbb{S}^5$ warpfactor. The latter is the fluctuation of $\varphi$ described by the effective action~\eqref{E:5daction}, and it decouples from those of the 5d metric. The quadratic action for $\varphi$ obtained from~\eqref{E:5daction} is manifestly positive-definite, and so those fluctuations are perturbatively stable. As for the 5d metric, the question of its perturbative stability is addressed by the 5d analysis of Section~\ref{S:stability}. There we accumulated evidence, though not complete, that those fluctuations are perturbatively stable for all angular momenta around $\mathbb{S}^1\times\mathbb{S}^3$ and for all $b$, when the wormhole is realized as a $\lambda$-solution. (We were able to completely demonstrate stability at $\beta_1=\beta_2$ when realized as a constrained instanton in $R_{\xi}$ gauge.)

The next lightest modes are those that carry a single unit of angular momentum on the $\mathbb{S}^5$. We focus on modes that carry no momentum around $\mathbb{S}^1\times\mathbb{S}^3$. Indeed, small black holes in global AdS$_5\times\mathbb{S}^5$ have an oblateness instability at small mass~\cite{Hubeny:2002xn} (see also~\cite{Prestidge:1999uq}) in which there is a perturbative instability of precisely this sort, and so this is a natural sector to study when assessing the wormholes.

The modes with these quantum numbers consistent with the gauge-fixing condition are\footnote{Note that the 5d metric perturbations are related to the 10d ones by $h_{\mu\nu} = H_{\mu\nu}+\frac{1}{3}g_{\mu\nu} H_{\alpha}{}^{\alpha}$.}
\begin{align}
\begin{split}
\label{E:10dflucs}
	h_{\tau\tau} &= g_{\tau\tau} h_1(\rho) Y^{k=1}(\Omega_5)\,, 
	\\
	 h_{ab} &= g_{ab} h_2(\rho) Y^{k=1}(\Omega_5)\,, 
	 \\
	H_{\alpha\beta} &= G_{\alpha\beta} h_3(\rho)Y^{k=1}(\Omega_5)\,,
	\\
	C_{\tau abc} &= -i\varepsilon_{abc} c_1(\rho) Y^{k=1}(\Omega_5)\,, 
	\\
	C_{\alpha\beta\gamma\delta} &= c_2(\rho)\varepsilon_{\alpha\beta\gamma\delta}{}^{\epsilon}Y_{\epsilon}^{k=1}(\Omega_5)\,,
	\\
	H_{\tau \alpha} &= \tilde{h}_1(\rho) Y_{\alpha}^{k=1}(\Omega_5)\,, 
	\\
	 C_{abc \alpha} &= \tilde{c}_1(\rho) \varepsilon_{abc} Y_{\alpha}^{k=1}(\rho)\,, 
	 \\
	 C_{\tau \alpha\beta\gamma} &= \tilde{c}_2(\rho)\varepsilon_{\alpha\beta\gamma}{}^{\delta\epsilon}\partial_{\delta}Y^{k=1}_{\epsilon}(\Omega_5) \,,
\end{split}
\end{align}
Here $X^M = (\rho,\tau,x^a,y^{\alpha})$, $x^a$ denotes angles on the $\mathbb{S}^3$ and $y^{\alpha}$ angles on the $\mathbb{S}^5$. Also $Y^k(\Omega_5)$ denotes a scalar harmonic on the $\mathbb{S}^5$ with $k$ units of angular momentum, $Y^{k}_{\alpha}(\Omega_5)$ the vector harmonics with $k$ units of angular momentum, $\varepsilon_{abc}$ and $\varepsilon_{\alpha\beta\gamma\delta\epsilon}$ are the epsilon tensors on $\mathbb{S}^3$ and $\mathbb{S}^5$. 

There are five different vector harmonics on $\mathbb{S}^5$, one of which is simply the derivative of the scalar harmonic, which also has the lowest eigenvalue under $-\Box_{\mathbb{S}^5}$. The perturbations in the first five lines mix when the vector harmonic in $C_{\alpha\beta\gamma\delta}$ is the derivative of the scalar. This sector of five mixed modes $(h_1,h_2,h_3,c_1,c_2)$ is even under the time reversal symmetry of the background and it comprises the most dangerous subsector of fluctuations. The other set couple to each other and are odd under time-reversal, and so we expect it to be less dangerous than the $T$-even sector. Thus, in the remainder of this Subsection we consider the $T$-even sector alone, when the vector harmonic in $C_{\alpha\beta\gamma\delta}$ is the derivative of a scalar harmonic.

Our strategy when assessing the stability of this sector is to scan for a normalizable zero mode of the linearized equations of motion as a function of the bottleneck size $b$. If there is such a mode, then it signals the onset of a perturbative instability, and if not, this sector is stable. 

The linearized equations of motion fall into two classes: those stemming from self-duality, and the linearized Einstein's equations. (The Maxwell's equations for the 5-form are automatically satisfied upon imposing self-duality.) Self-duality enforces two conditions on the RR fluctuations $c_1$ and $c_2$:
\begin{align}
\begin{split}
	c_1 & = - \frac{\sqrt{g}}{\sqrt{g_{\mathbb{S^3}}}} c_2'\,,
	\\
	c_1' & = -\frac{\sqrt{g}}{2\sqrt{g_{\mathbb{S}^3}}} \left( h_1+3h_2-\frac{40}{3}h_3 + 10 c_2 \right)\,,
\end{split}
\end{align}
where $' = \partial_{\rho}$. Solving the first for $c_1$ and plugging into the second gives a second order equation for $c_2$,
\beq
	\Box_5 c_2 = 5 c_2 + \frac{1}{2}\left(h_1+3h_2-\frac{40}{3}h_3\right)\,,
\eeq
and $\Box_5$ acts on $c_2$ as on a 5d scalar. Now we turn to the linearized Einstein's equations. Because of our gauge-fixing conditions the equations we have to solve are
\beq
	\mathcal{E}_{\tau \tau} =0\,, \qquad \mathcal{E}_a{}^a = 0\,, \qquad \mathcal{E}_{\rho \rho} - \frac{3}{5}\mathcal{E}_{\alpha}{}^{\alpha} = 0\,,
\eeq
where $\mathcal{E}_{MN} = R(G_E)_{MN} - \frac{R(G_E)}{2}G_{E,MN} = T_{MN}$ and $T_{MN}$ is the stress tensor of the 5-form flux. We do not have to satisfy the $\rho \rho$ and $\rho \alpha$ equations, since those components of the metric are fixed by the radial gauge. These linearized equations are rather complicated. They read
\begin{align}
	\nonumber
	\Box_5 h_1 & = \left( \ln \frac{b^2\cosh(2\rho)-1}{\beta_1e^{2\rho}+\beta_2e^{-2\rho}}\right)' (h_1'+2h_2') +5 h_1 + \frac{4h_2}{b^2\cosh(2\rho)-1} - \frac{40}{9}h_3 - \frac{80}{3}c_2\,,
	\\
	\Box_5 h_2 & = -\left( \ln \frac{b^2\cosh(2\rho)-1}{\beta_1e^{2\rho}+\beta_2e^{-2\rho}}\right)' h_2' -\left( \frac{4}{b^2\cosh(2\rho)-1}-5\right)h_3 - \frac{40}{9}h_3 - \frac{80}{3}c_2\,,
	\\
	\nonumber
	\Box_5 h_3 & = h_1+3h_2 -\left( \frac{95}{3}\frac{1}{b^2\cosh(2\rho)-1}  - \frac{b^2\left( 95(e^{4\rho}\beta_1 + e^{-4\rho} \beta_2 + (\beta_1+\beta_2)) - 24 (\beta_1+\beta_2)\right)}{6(b^2\cosh(2\rho)-1)(\beta_1e^{2\rho}+\beta_2e^{-2\rho})}\right)h_3 - 80 c_2\,.
\end{align}

Before embarking on solving these linearized equations numerically at fixed $b$, we note that this set of perturbations $(h_1,h_2,h_3,c_2)$ is slightly overcomplete. There is a residual gauge symmetry consistent with the gauge-fixing conditions with
\beq
	\xi^{\rho} = f'(\rho) Y^{k=1}(\Omega_5)\,, \quad \xi^{\alpha} = f(\rho) \partial^{\alpha}Y^{k=1}(\Omega_5)\,, \quad f(\rho) = c_+ e^{i w\rho} + c_- e^{-iw\rho}\,, \quad w = \sqrt{\frac{5}{3}}\,.
\eeq
We can use this residual symmetry to say set $c_2$ and its derivative to vanish at $\rho = 0$.

This residual symmetry is visible in the near-boundary solutions, which can be obtained by solving the linearized equations above at large $\rho$. That near-boundary solution reads
\begin{align}
\begin{split}
	  h_1 & =  c_1 e^{\rho} + c_{-5} e^{-5\rho} + c_+ e^{i w \rho} + c_- e^{-i w \rho} + \tilde{c}_+ e^{-(4-i w)\rho} + \tilde{c}_-e^{-(4+i w)\rho}\,,
	  \\
	  h_2 & = -\frac{c_1}{3}e^{\rho} - \frac{c_{-5}{3}}e^{-5\rho} + c_+ e^{i w\rho} + c_- e^{-i w \rho} + \tilde{c}_+ e^{-(4-i w)\rho} + \tilde{c}_- e^{-(4+i w)\rho}\,,
	  \\
	  h_3 & = c_5 e^{5\rho} + c_{-9} e^{-9\rho} + \frac{3-i w^{-1}}{8} c_+ e^{i w \rho} + \frac{3+i w^{-1}}{8}c_- e^{-i w \rho}
	  \\
	  & \qquad \qquad  \qquad \qquad+\frac{3-iw^{-1}}{8}\tilde{c}_+e^{-(4-i w)\rho} + \frac{3+i w^{-1}}{8} \tilde{c}_-e^{-(4+iw)\rho}\,,
	  \\
	  c_2 & = -\frac{c_5}{6}e^{5\rho} - \frac{c_{-9}}{6} e^{-9 \rho} +  \frac{3-i w^{-1}}{16} c_+ e^{i w \rho} + \frac{3+i w^{-1}}{16}c_- e^{-i w \rho}
	  \\
	  & \qquad \qquad \qquad \qquad+\frac{3-iw^{-1}}{16}\tilde{c}_+e^{-(4-i w)\rho} + \frac{3+i w^{-1}}{16} \tilde{c}_-e^{-(4+iw)\rho}\,,
\end{split}
\end{align}
labeled by the eight coefficients $(c_5,c_1,c_{-5},c_{-9},c_{\pm},\tilde{c}_{\pm})$. The modes $c_{\pm}$ are those generated by the residual gauge symmetry above and so are unphysical. The modes $\tilde{c}_{\pm}$ are their ``partners'' with exponents offset by $-4$, and these are also unphysical. The remaining four modes correspond to a dimension five operator with vev $\sim c_{-5}$ and source $\sim c_1$, the traceless fluctuation of the boundary stress tensor, and a dimension nine operator with vev $\sim c_{-9}$ and source $\sim c_5$, the fluctuation of the $\mathbb{S}^5$ warpfactor. Normalizable zero modes have $c_5 = c_1 = 0$.

To calibrate our analysis we first revisit this set of fluctuations in Euclidean black holes with $\beta_1 = - \beta_2 = \beta$. In that case, there is a far simpler fluctuation analysis in~\cite{Hubeny:2002xn}. Those authors employed a transverse gauge, for which a traceless fluctuation of the 5d metric decouples from the 4-form potential, leading to a single mode obeying a single second order equation. Translating from their conventions to ours, they found a zero mode at $b_{c,\rm them}=1.167$, in the small black hole regime. Small black holes with smaller mass are unstable. 

The radial gauge is a much less pleasant way of addressing the same problem for the black hole, but it calibrates us for our analysis of the wormhole, and so we proceed. We numerically solve the equations above subject to initial conditions that the fluctuation is regular at the bottom of the Euclidean cigar at $\rho = 0$. At general $b$ there are zero modes corresponding to the unphysical fluctuations $c_{\pm}$ and $\tilde{c}_{\pm}$. That is, there are no physical zero modes for general $b$. However at a critical value of $b_c = 1.178$ (which differs at the $1\%$ level from previous results; we have checked that in fact this is the correct critical value by also performing an analysis in transverse gauge, with the same result) we find a physical zero mode, signified by a degeneracy of initial conditions that lead to a normalizable mode. 

Now we turn our attention to the wormholes. We now set $\beta_1 = \beta_2 $, so that the wormhole enjoys a parity symmetry which flips $\rho \to - \rho$. We look for parity-even zero modes, obeying Neumann boundary conditions at $\rho = 0$. Using the residual gauge symmetry to set $c_2(0) = 0$ and fixing $h_1(0) = 1$, we then dial the initial conditions $h_2(0)$ and $h_3(0)$ and numerically shoot from the bottleneck at $\rho = 0 $ out to large $\rho$. For general $b$, as in the black hole, we find a normalizable zero mode corresponding to the unphysical fluctuations mentioned above. However we do not find a physical normalizable zero mode for \emph{any} value of $b$. We thereby conclude that this sector of perturbations is quadratically stable.

It is relatively straightforward to extend our analysis to one at higher angular momentum $k$, mode by mode, as well as to study the $T$-odd perturbation in~\eqref{E:10dflucs}. However we expect those sectors to be even more stable than this one.

\section{Brane nucleation}
\label{S:branes}

Having obtained wormholes in IIB supergravity, we study the prospect of non-perturbative instabilities, focusing on brane instantons. The basic idea is that these wormholes come with a five-form RR flux, and so if we were in Lorentzian signature, there would be the possibility of $D3-\overline{D3}$ pair production a la the Schwinger effect. We look for Euclidean versions of that process, studying probe branes in the wormhole backgrounds~\eqref{E:10dtorus} and~\eqref{E:10dS1xS3}. 

Similar nucleation instabilities have been studied in the context of wormholes with negatively curved boundaries~\cite{maldacena2004wormholes,Buchel:2004rr}. 

\subsection{The basic instability channel}
\label{S:basicBrane}

\subsubsection{$\mathbb{S}^1 \times \mathbb{T}^{d-1}$ wormholes}

Consider the wormhole with torus cross-section, which we recapitulate here in units where the AdS radius is unity, $\tilde{L}=1$,
\begin{align*}
	ds_E^2 & =g_{\mu\nu}dx^{\mu}dx^{\nu} + d\Omega_5^2=  \frac{b^2}{2}\cosh(2\rho) \left( \left( \frac{\beta_1 e^{2\rho} + \beta_2 e^{-2\rho}}{2\cosh(2\rho)}\right)^2 d\tau^2 + d\vec{x}_{\perp}^2 \right) + d\rho^2 + d\Omega_5^2\,, \qquad e^{\Phi} = g_s\,,
	\\
	C_4 &= - \frac{i b^4}{16}\left( \beta_1 e^{4\rho} - \beta_2 e^{-4\rho} + 4(\beta_1+\beta_2)\rho\right)d\tau\wedge dx^1 \wedge dx^2 \wedge dx^3 + (\text{angular})\,.
\end{align*}
The four-form potential satisfies $\partial_{\rho}C_{\tau123} = -4i\varepsilon_{\rho\tau 123}$ with $\varepsilon_{\mu_1\hdots\mu_5}$ the 5d epsilon-tensor. Now consider a $D3-\overline{D3}$ brane pair in this background, where we ansatz that the $D3$-brane sits at a point on the $\mathbb{S}^5$, is extended along the torus $(\tau,\vec{x}_{\perp})$, and is at a particular value of $\rho$ which we denote as $\rho_0$. 

If we were studying planar AdS$_5\times\mathbb{S}^5$, then this embedding solves the $D3$-brane embedding equations for any $\rho_0$: this solution describes a $D3$-brane parallel to the boundary, and parameterizes points on the Coulomb branch of the dual $\mathcal{N}=4$ SYM in which the $SU(N)$ gauge group is broken to $SU(N-1)\times U(1)$. This solution exists because there is an exact cancellation between the attractive force felt by the $D3$-brane, pulling it to the bottom of AdS, and the repulsive force from the 5-form flux background.

In global AdS$_5\times\mathbb{S}^5$ or for a black hole in AdS, the gravitational pull is stronger than the RR repulsion, and if a $D3$-brane was inserted at $\rho=\rho_0$, it would fall.

What we will see presently is that in the wormhole the gravitational pull is \emph{weaker} than the RR repulsion, at least near the bottleneck of the wormhole. In detail, consider the action of the $D3$ brane in the wormhole as a function of $\rho_0$. Using $\frac{\widetilde{L}^4}{\ell_s^4} = 4\pi N$ and $\mu_3 = \frac{1}{(2\pi)^3\ell_s^4}$ it is
\beq
\label{E:D3torusTake1}
	S_{D3} = \mu_3 \left( \int d^4 \sigma \sqrt{\text{P}[G_E]} -i \int \text{P}[C_4]\right) = \frac{b^4NV}{32\pi^2 }\left( 2\beta_2 e^{-4\rho_0} - (\beta_1+\beta_2) (4\rho_0-1)\right)\,,
\eeq
where $V$ is the volume of the spatial torus. This describes an effective potential for $\rho_0$, $V_{\rm eff} = 2\beta_2 e^{-4\rho_0} - (\beta_1 + \beta_2)(4\rho_0-1)$, and a quick check shows that this potential has no extremum and in fact tends to $-\infty$ as $\rho_0\to \infty$. 

If we analytically continued $\tau$ to real time, then this means that if we inserted a $D3$-brane at some fixed $\rho$, it would experience a force which pushes it out toward the AdS boundary as $\rho\to\infty$. A similar analysis for a $\overline{D3}$ brane shows that it would be pushed out to the other boundary as $\rho\to-\infty$. In Euclidean signature, we can imagine holding the $D3$ and $\overline{D3}$ branes fixed at large positive and negative $\rho$ respectively, leading to a configuration that satisfies the same boundary conditions as the original wormhole, but with lower action. This is a brane nucleation instability of the wormhole.

The fact that the effective potential has no extremum indicates a particularly fatal instability, albeit at tree level in the bulk, as we can lower the action by an arbitrarily large amount by inserting the $D3-\overline{D3}$ pair at arbitrarily large $|\rho_0|$. It may be that this runaway is stabilized by loop effects in $1/N$. For instance, the wormhole is a non-supersymmetric background, and one expects the tension of a $D3$ and $\overline{D3}$ brane in the wormhole to receive quantum corrections. (The brane charges are quantized and receive no corrections.) If quantum corrections increase the tension slightly, then this leads to a large positive correction to the brane action at large $|\rho_0|$, on account of the exponentially growing volume wrapped by the 3-branes, and this correction would compete against the tree-level runaway. However as of now we have no way to concretely address whether the effective potential for $\rho_0$ is stabilized or not. 

In order to understand the physics of this instability, let us briefly consider wormholes with torus boundary in other ``lamppost'' examples of AdS/CFT. These examples (that do not simply involve orbifolds of the $R$-symmetry of $\mathcal{N}=4$ in such a way as to preserve at least $\mathcal{N}=1$ SUSY) are ABJM theory~\cite{Aharony:2008ug}, dual to M-theory on AdS$_4\times\mathbb{S}^7/\mathbb{Z}_k$; the mysterious $\mathcal{N}=(2,0)$ theory on coincident $M5$ branes, dual to M-theory on AdS$_7\times\mathbb{S}^4$; and the $D1/D5$ theory, dual to IIB string theory on AdS$_3\times\mathbb{S}^3\times\mathcal{M}_4$ with $\mathcal{M}_4$ equal to $\mathbb{T}^4$ or $K3$.

For M-theory on AdS$_4\times\mathbb{S}^7$ or AdS$_7\times\mathbb{S}^4$, one can imitate the uplift of AdS$_5$ wormholes we used in the last Section to obtain 11d geometries of the direct product form (wormhole)$\times$(sphere). The line element and flux background of the AdS$_4\times\mathbb{S}^7$ wormhole are
\begin{align}
\begin{split}
	ds_{11}^2 & =\frac{1}{4}\,ds_{\rm wormhole}^2 + d\Omega_7^2 
	\\
	&=  \frac{1}{4}\left( d\rho^2 + \frac{b^2}{4}\left( 2\cosh\left(\frac{3\rho}{2}\right)\right)^{\frac{4}{3}}\left( \left( \frac{\beta_1 e^{\frac{3\rho}{2}} + \beta_2 e^{-\frac{3\rho}{2}}}{2\cosh\left(\frac{3\rho}{2}\right)}\right)^2d\tau^2 + d\vec{x}_{\perp}^2\right)\right) + d\Omega_7^2\,,
	\\
	C_3 & =- \frac{i}{64}b^3\left( \beta_1 e^{3\rho} - \beta_2 e^{-3\rho} + 3(\beta_1+\beta_2)\rho\right)d\tau\wedge dx^1 \wedge dx^2\,,
\end{split}
\end{align}
and the action of a probe $M2$ brane inserted at $\rho = \rho_0$, at a point in $\mathbb{S}^7$, and extended along the torus, is given by
\beq
\label{E:M2action}
	S_{\rm M2}  = \mu_2 \left( \int d^3\sigma \sqrt{\text{P}[G_{11}]}- i \int \text{P}[C_3]\right) = \frac{\mu_2 b^3 V}{64}\left( 2 \beta_2 e^{-3\rho_0} - (\beta_1+\beta_2)(3\rho_0+1)\right)\,.
\eeq
Similarly, for a probe $M5$ brane in the AdS$_7\times\mathbb{S}^4$ wormhole
\begin{align}
\begin{split}
	ds_{11}^2 & = 4\,ds_{\rm wormhole}^2 + d\Omega_4^2\,,
	\\
	& = 4 \left( d\rho^2 + \frac{b^2}{4}\left( 2 \cosh\left( 3\rho\right)\right)^{\frac{2}{3}}\left( \left( \frac{\beta_1 e^{3\rho}+\beta_2 e^{-3\rho}}{2\cosh(3\rho)}\right)^2d\tau^2 + d\vec{x}_{\perp}^2\right)\right) + d\Omega_4^2\,,
	\\
	C_6 & = -i b^6 \left( \beta_1 e^{6\rho}-\beta_2 e^{-6\rho} + 6 (\beta_1+\beta_2)\rho\right)\,,
\end{split}
\end{align}
reads
\beq
\label{E:M5action}
	S_{M5} = \mu_5 \left( \int d^6 \sigma \sqrt{\text{P}[G_{11}]} - i \int \text{P}[C_6]\right) = \mu_5 b^6 V \left( 2\beta_2 e^{-6\rho_0} - (\beta_1+\beta_2)(6\rho_0+1)\right)\,.
\eeq

Comparing the action~\eqref{E:D3torusTake1} of a probe $D3$ brane in the AdS$_5\times\mathbb{S}^5$ wormhole, the action~\eqref{E:M2action} of a probe $M2$ brane in the AdS$_4\times\mathbb{S}^7$ wormhole, and the action~\eqref{E:M5action} of a probe $M5$ brane in the AdS$_7\times\mathbb{S}^4$ wormhole, all with torus boundary, we see that they all take the parametric form
\beq
	S_{\rm probe} \propto b^d V \left( 2 \beta_2 e^{-d\rho_0} - (\beta_1+\beta_2)(d\rho_0+1)\right)\,,
\eeq	
with $d=3,4,6$. In each of these cases the potential has a runaway to large $\rho_0$. Moreover, in each case there is a cancellation between large $O(e^{d\rho})$ terms in the tension and Wess-Zumino terms of the probe brane action. This cancellation occurs for the following reason. These geometries are asymptotically supersymmetric as $\rho\to\infty$. A probe brane inserted at fixed $\rho$ in this asymptotic region is a BPS configuration; the tension and Wess-Zumino contributions to its action are both proportional to $e^{d\rho}$, and these cancel by the BPS condition. 

It is easy to check that if the probe branes were slightly non-BPS with a tension slightly bigger than the charge, then the effective potential for $\rho$ is stabilized, with a minimum at large but finite $\rho$. (If the tension is slightly below the charge, then this only serves to worsen the runaway.)

A simple arena with nearly BPS branes is the $D1/D5$ system. Consider $N_1$ $D1$ branes and $N_5$ $D5$ branes wrapping $\mathbb{T}^4$, intersecting along a string. The near-horizon geometry is AdS$_3\times\mathbb{S}^3\times\mathbb{T}^4$, and for generic moduli of the compactification there are no BPS branes. There are various branes which give effective strings in the AdS$_3$, with an effective action of the qualitative form
\beq
\label{E:braneSx}
	S = \mu \left( x \int d^2\sigma \sqrt{\text{P}[G_E]} - i \int \text{P}[C]\right)\,,
\eeq
where $x$ is the effective mass-to-charge ratio of the brane. Thanks to the BPS of the supersymmetric background, this ratio is bounded below as $x\geq 1$ with $x=1$ corresponding to an exactly BPS brane.

We can find wormholes in this setting of the form (wormhole)$\times\mathbb{S}^3\times\mathbb{T}^4$. The action of a string sitting at fixed $\rho$ becomes
\beq
	S =  \frac{\mu b^2 V}{4}\left( \beta_1(x-1)e^{2\rho_0} + \beta_2(x+1)e^{-2\rho_0} -(\beta_1+\beta_2)(2\rho_0-x)\right)\,.
\eeq
Note that the runaway to large $\rho$ is stabilized when the brane is non-BPS. This potential has a minimum at 
\beq
	e^{2\rho_0} = \frac{\beta_1+\beta_2+\sqrt{(\beta_1-\beta_2)^2+4x^2\beta_1\beta_2}}{2\beta_1(x-1)}\,.
\eeq
There are similar results for the anti-string. The total action of the string pair is
\begin{align}
\begin{split}
\label{E:torusGeneralS}
	S_{\rm pair} &= \frac{\mu b^2 V}{4}\left( (\beta_1+\beta_2)\left( \ln \left( \frac{x-1}{x+1} \frac{-(\beta_1+\beta_2)+\sqrt{(\beta_1-\beta_2)^2+4x^2\beta_1\beta_2}}{\beta_1+\beta_2+\sqrt{(\beta_1-\beta_2)^2+4x^2\beta_1\beta_2}}\right)+2x\right)\right.
	\\
	&\hspace{3in} \left.\phantom{\left(\frac{\beta_1}{\beta_2}\right)}+2 \sqrt{(\beta_1-\beta_2)^2+4x^2\beta_1\beta_2}\right)\,.
\end{split}
\end{align}

There are two questions at hand. The first is whether or not the action of the string pair is negative. The second question is whether or not the additional degrees of freedom stemming from the string pair are perturbatively stable. If the action is lower and the setup is perturbatively stable, then the wormhole with a string-pair inserted is a more dominant saddle than the original wormhole, and there is a nucleation instability. 

For a simplifying case take $\beta_1 = \beta_2 = \beta$. The action of the pair becomes
\beq
\label{E:torusPairaction}
	S_{\rm pair} = 2\mu b^2V\beta \left( x -\text{arccoth}\,x\right)\,,
\eeq
which is negative when $x<x_0$ with $x_0 = \coth\,x_0= 1.1997$. In fact, this form of the pair action, proportional to $x-\text{arccoth}\,x$, is what one finds for a non-BPS brane in any AdS$_{d+1}$ wormhole with torus boundary. Another simple case is to use that the renormalized mass-to-charge ratio $x$ is close to $1$. Then at fixed $\beta_1,\beta_2$, 
\beq
\label{E:torusPairsmallx}
	S_{\rm pair} = \frac{\mu b^2V}{2}\left( \ln (x-1) + \ln \frac{\sqrt{\beta_1\beta_2}}{\beta_1+\beta_2} +2 \right) (\beta_1+\beta_2)+ O(x-1)\,,
\eeq
which is large and negative. As for the question of perturbative stability, the potential instabilities come from the sigma model fields on the string worldvolume. However it is easy to check that these are all stable: there are zero modes stemming from where the strings are inserted on the transverse space, and all other bosonic modes are massive. (There is a massless $U(1)$ gauge theory on each brane, but the modes of the gauge field are massive on account of finite worldvolume.)

The physics of this instability is that of screening. It is probabilistically favorable for string pairs to nucleate, with the positively charged string at positive $\rho$ and the negatively charged string at negative $\rho$. For a single pair, the region in between has $Q-1$ units of flux and that outside the pair has $Q$ units. (We are being agnostic as to precisely what string is under consideration.) The pair is therefore likely to nucleate and partially screen the RR flux supporting the wormhole. It is not clear what the endpoint of this instability is, although one expects that it involves enough nucleated pairs so that there is a geometric transition and one is again dealing with fluxes and geometry.

To recap, for the wormholes with torus boundary, we find rather different behavior depending on whether the AdS vacuum at hand includes BPS branes or not. If there are BPS branes in the AdS vacuum, then there is a runaway nucleation instability of the corresponding Euclidean wormhole with torus boundary. If there are no BPS branes in the AdS vacuum, then the wormhole with torus boundary is unstable to brane nucleation, but there is no runaway. In the dual descriptions, it is worth noting that the existence of BPS branes in the vacuum corresponds to whether the dual CFT has a moduli space of vacua dual to separated branes in AdS. If the CFT has such a moduli space and is placed on a spatial torus with supersymmetry-preserving boundary conditions, then this moduli space remains even in the finite volume theory, where it leads to a continuous density of states labeled by the moduli. However, even when this moduli space is absent, it is worth noting that even when the $D1/D5$ CFT does not have BPS branes, it has a continuous density of states above a certain threshold dual to long strings in AdS$_3$~\cite{Seiberg:1999xz}. 

This suggests a mechanism to stabilize the runaway in AdS$_5\times\mathbb{S}^5$. Suppose we deform by a SUSY-breaking mass that lifts the moduli space, dual to metric fluctuations of the $\mathbb{S}^5$. This would lead to a new wormhole with torus boundary, which we expect to still be unstable to 3-brane nucleation. However, this deformation should stabilize the brane action at large positive $\rho$. This possibility is certainly testable in the limit where the deformation is very weak, so that the wormhole geometry is approximately the one studied above plus small perturbations.

\subsubsection{$\mathbb{S}^1 \times \mathbb{S}^3$ wormholes}

Now consider the $\mathbb{S}^1\times\mathbb{S}^3$ wormhole~\eqref{E:10dS1xS3} in AdS$_5\times\mathbb{S}^5$, which we recapitulate as:
\begin{align*}
	ds_E^2 & = \frac{b^2\cosh(2\rho)-1}{2}\left( \left(\frac{\beta_1e^{2\rho}+\beta_2e^{-2\rho}}{2\left( \cosh(2\rho)-\frac{1}{b^2}\right)}\right)^2d\tau^2 + d\Omega_3^2\right) + d\rho^2 + d\Omega_5^2\,, \qquad e^{\Phi} = g_s\,,
	\\
	C_4 &= - i\left[ \frac{ b^4}{16}(\beta_1 e^{4\rho} - \beta_2 e^{-4\rho} + 4(\beta_1+\beta_2)\rho) - \frac{b^2}{4}(\beta_1e^{2\rho} - \beta_2 e^{-2\rho})\right] d\tau\wedge \text{vol}_{\mathbb{S}^3} + (\text{angular})\,,
\end{align*}
where $\partial_{\rho}C_{\tau abc} = -4i\varepsilon_{\rho\tau abc}$ with $a,b,c$ angles on the $\mathbb{S}^3$. Consider a $D3-\overline{D3}$ brane pair as in our analysis above, where the 3-branes are extended along $\mathbb{S}^1\times\mathbb{S}^3$, sit at points in the $\mathbb{S}^5$, and at points in $\rho$. While in our analysis of the torus wormhole we needed to use that there ought to be radiative corrections to the 3-brane tension, in this wormhole we can well-approximate $x=1$ from the beginning as we now see.

A probe $D3$-brane sitting at $\rho=\rho_0$ has an action
\beq
	S_{D3} = \frac{b^4N}{16}\left( \Big(2\beta_2 e^{-4\rho_0} - (\beta_1+\beta_2)(4\rho_0-1)\Big) + \frac{2}{b^2}(\beta_1 e^{2\rho_0} - 3 \beta_2 e^{-2\rho_0})\right)\,.
\eeq
Comparing with the expression~\eqref{E:D3torusTake1}, we see that the $O(b^4)$ terms agree upon substituting $V = 2\pi^2$, the volume of the $\mathbb{S}^3$, but now there is an $O(b^2)$ term. This additional term stabilizes $\rho_0$. The solution for $\rho_0$ is somewhat complicated and so we do not present it.\footnote{Solving $\frac{\partial S_{D3}}{\partial\rho_0}$ leads to three different solutions, only one of which is real. It is of course this real solution that we have in mind.} After performing the corresponding analysis for the $\overline{D3}$ brane, one finds a rather complicated action for the brane pair. As for perturbative stability, there are zero modes of the 3-brane sigma models stemming from the location where the branes are inserted on the $\mathbb{S}^5$, while all other bosonic modes are massive.

These expressions simplify enormously when $b\gg 1$, whereby
\begin{align}
\begin{split}
\label{E:largebSpair}
	e^{2\rho_0} &= \frac{\beta_1+\beta_2}{\beta_1} b^2 + O(b^0)\,,
	\\
	S_{D3-\overline{D3}} & = \frac{b^4N}{4} \left( -\ln \,b^2 + \ln \frac{\sqrt{\beta_1\beta_2}}{\beta_1+\beta_2} - \frac{3}{2}\right) (\beta_1+\beta_2) + O(b^2)\,.
\end{split}
\end{align}
The action is of the qualitative form as~\eqref{E:torusPairsmallx}, in particular it is large and negative. So we find a brane nucleation instability at large $b$. 

\begin{figure}[t]
\begin{center}
\includegraphics[width=3in]{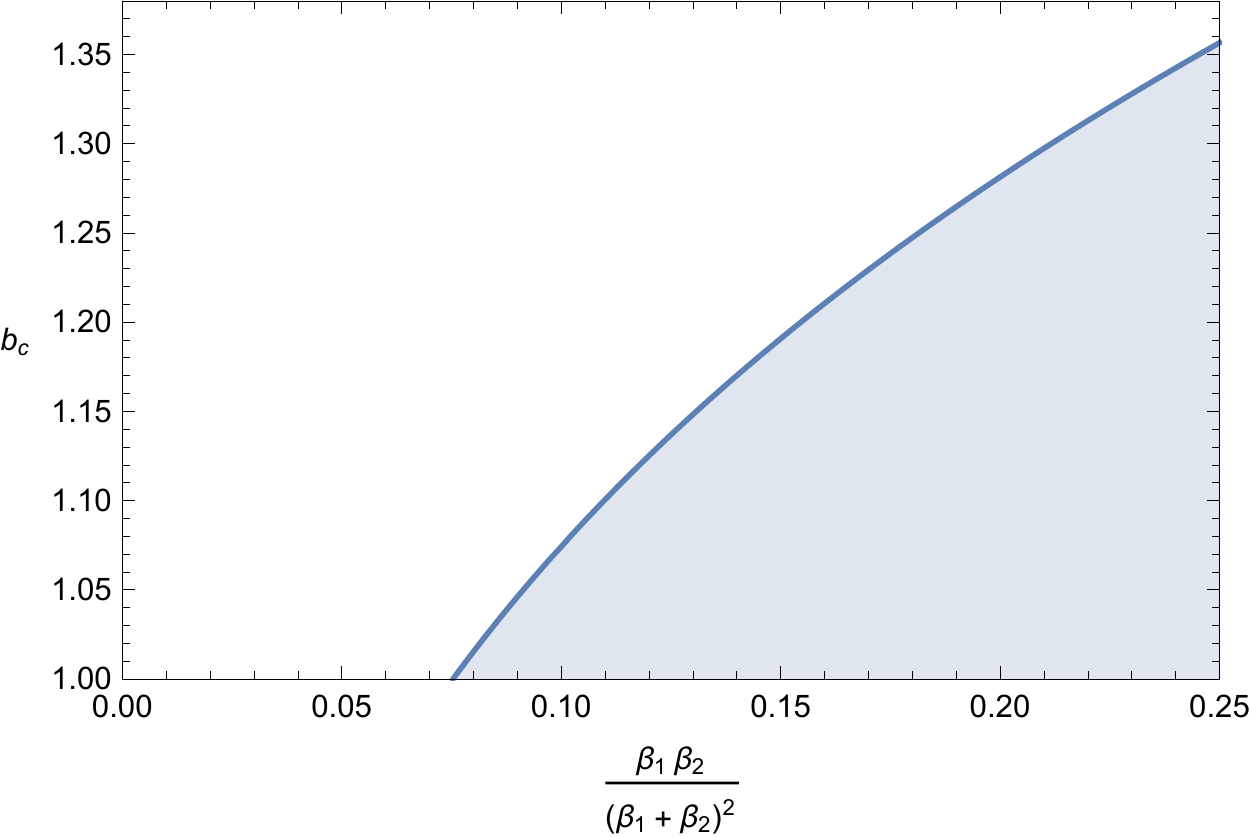}
\caption{\label{F:branebc} The critical value of $b$ as a function of $\frac{\beta_1\beta_2}{(\beta_1+\beta_2)^2}$. The shaded region indicates wormholes with $b<b_c$, which are stable against the nucleation of 3-brane pairs.}
\end{center}
\end{figure}

For general $b$ we must resort to numerics. We find that for fixed $\beta_1$, $\beta_2$, the action of the brane pair is negative for $b>b_c$ where $b_c$ is plotted in Fig.~\ref{F:branebc}. For $\beta_1=\beta_2$ we find $b_c = 1.3565$ and for other ratios $b_c$ is diminished. So wormholes with $b<b_c$ are stable against brane nucleation! We find that $b_c$ passes below the minimum possible value of $b$, namely 1, when the dimensionless ratio $\frac{\beta_1\beta_2}{(\beta_1+\beta_2)^2}$ is below $0.075$ and so all wormholes in that region are unstable to brane nucleation.

To summarize, the wormholes with torus cross-section are always unstable to the nucleation of 3-brane pairs, while there are wormholes with relatively small, but still macroscopic $\mathbb{S}^1\times\mathbb{S}^3$ cross-section are stable against brane nucleation.

We conclude this Subsection with a curious observation. The existence of positions in the wormhole where 3-branes can rest in equilibrium implies that we can form wormholes with $N$ units of 5-form flux in one asymptotic region and $M$ in the other with $|N-M| = O(1)$, by simply inserting $|N-M|$ $D3$ or $\overline{D3}$ branes in the wormhole. It is presently unclear what this would correspond to in the dual description.

\subsection{Another instanton}
\label{S:otherBrane}

We have also found a brane instanton in these wormhole backgrounds which is analogous to the worldline instanton in the Schwinger effect~\cite{affleck1982pair}. If we analytically continued to Lorentzian signature, this instanton encodes a dynamical, non-perturbative instability of the wormhole to the production of brane pairs, analogous to the instability of a constant electric field in QED to the production of $e^+-e^-$ pairs, and compute the decay rate. It is the instanton that mediates the transition between the wormhole, and the wormhole with a brane pair that we found in the last Subsection.

\begin{figure}[t]
\begin{center}
\includegraphics[width=4in]{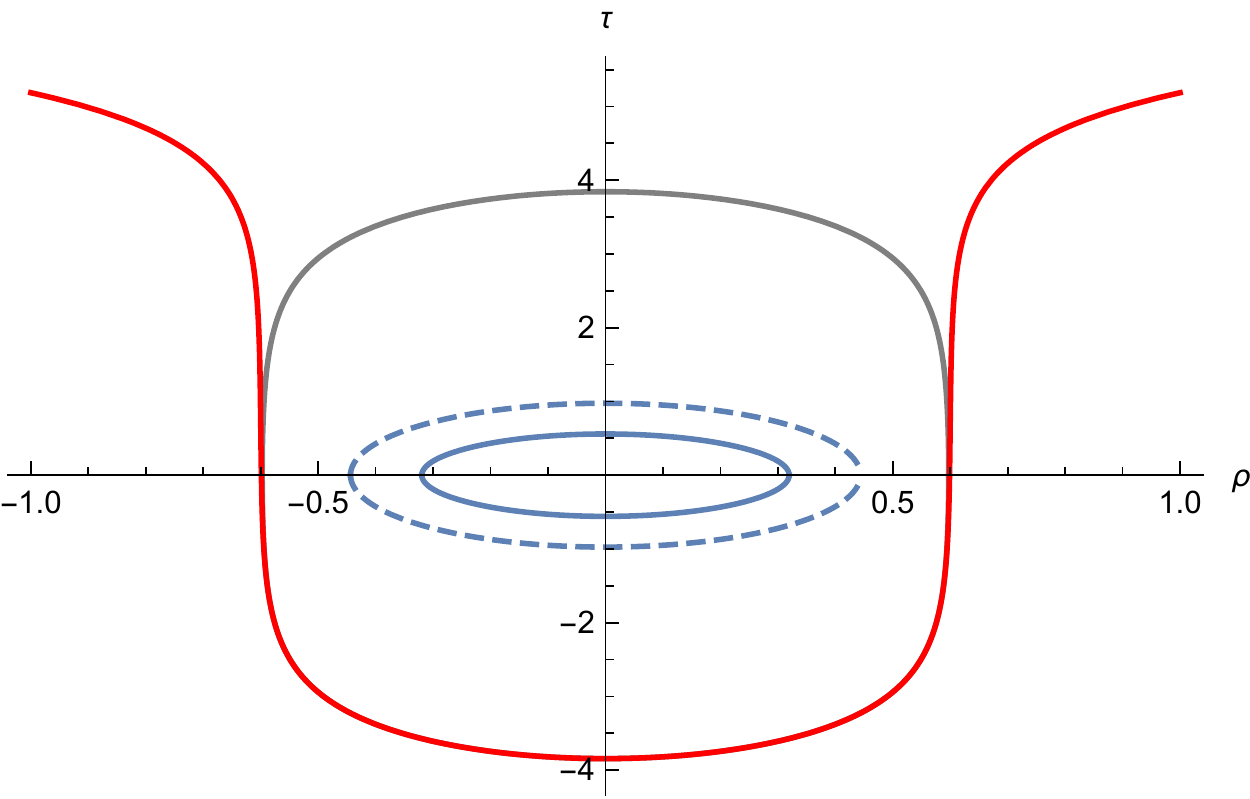}
\caption{\label{F:otherInstanton} 3-brane embeddings in the wormhole with torus cross-section and $\beta_1 = \beta_2$ (upon rescaling $\tau \to \frac{\tau}{\beta b}$). The innermost solid loop is the trajectory of the $D3$-brane instanton, with $x= 1$. The dashed loop indicates the trajectory for a toy model of a 3-brane with a mass-to-charge ratio $1.15$, the outermost loop is the trajectory for the toy model with mass-to-charge ratio $x=1.19967$, just below the critical value of $x_0 = 1.19967864..$. The red line is the trajectory when $x=1.19968$, just above the critical value.}
\end{center}
\end{figure}

Let us begin with the simplest example where we can study this instanton, namely a wormhole with torus cross-section. In this setting we saw that there is a runaway instability for BPS branes, but a ``finite'' nucleation instability for nearly BPS branes in the $D1/D5$ system. As a phenomenological exercise, let us suppose that there was an AdS$_5$ compactification with nearly BPS 3-branes characterized by a mass-to-charge ratio $x$ as in~\eqref{E:braneSx}. (One could perform the following exercise for effective strings in AdS$_3$, with essentially the same results.) For simplicity let us take $\beta_1 = \beta_2 = \beta$, so that the wormhole is symmetric around $\rho = 0$. Now consider a $3$-brane embedded in such a way that it wraps the spatial torus $\mathbb{T}^3$ and is extended along a curve in Euclidean time $\tau$ and the radial coordinate $\rho$. We parameterize the embedding through $\tau(\rho)$. With a general mass-to-charge ratio $x$ (which for a $D3$ brane equals one, but which we take to be general in order to explore the brane physics), the effective action for $\tau(\rho)$ stems from the Dirac-Born-Infeld and Wess-Zumino terms on the brane, reading
\beq
	S = \frac{N V}{2\pi^2} \int d\rho\left\{ x \left( \frac{b^2}{2}\cosh(2\rho)\right)^{\frac{3}{2}}\sqrt{1+ \frac{b^2}{2}\cosh(2\rho)\tau'(\rho)^2} -  \frac{b^4}{8}(\sinh(4\rho) + 4\rho)\tau'(\rho)\right\}\,.
\eeq
Rescaling $\tau \to \tau/b$ we see that $b$ scales out the action. We now look for solutions for $\tau(\rho)$ which are closed loops that are symmetric across the wormhole bottleneck. Because $\tau$ does not appear directly, only its derivative, its conjugate momentum is constant,
\beq
	\frac{\partial P}{\partial\rho} = 0\,, \qquad P =x \sqrt{\frac{\cosh^5(2\rho)\tau'(\rho)^2}{1+\cosh(2\rho)\tau'(\rho)^2}} - \frac{1}{2}(\sinh(4\rho)+4\rho)\,.
\eeq
Paths which are symmetric around $\rho = 0$ have $\tau'(0) = 0$ and thus $P=0$. Solving for $\tau'(\rho)$ we have
\beq
	\tau'(\rho) = \frac{\sinh(4\rho)+4}{\sqrt{ 4x^2 \cosh^5(2\rho) - \cosh(2\rho)(\sinh(4\rho)+4\rho)^2}}\,.
\eeq
Numerically integrating from $\rho = 0$, we find dramatically different solutions for $x$ depending on whether $x$ is below the value $x_0=1.1997$ uncovered in~\eqref{E:torusPairaction}. There we noted that for $1<x<x_0$ there is a nucleation instability to a brane pair which lowers the action. For $x=1$, i.e. for BPS branes, this instability has a runaway. When it comes to these solutions, we find that in the same regime $x<x_0$ (to seven digits of numerical accuracy), that this trajectory for $\tau(\rho)$ is a closed loop, while for $x>x_0$ this trajectory goes to the AdS boundary. See Fig.~\ref{F:otherInstanton} for a few brane embeddings.

These trajectories are genuine instantons only when they close into a loop. For this symmetric wormhole, we find such loops to always be symmetric and only exist for $x<x_0$. We have numerically obtained the action of these instantons and found it to always be positive. However, when inspecting the spectrum of fluctuations at the physically relevant case of $x=1$, there is a negative mode, roughly the radius of the ellipse. (We have not investigated the spectrum for $x\in (1,x_0)$, but expect there to be a single negative mode in general.) As such this instanton gives an instability of the wormhole: its contribution to the amplitude is exponentially suppressed, but its one-loop determinant is imaginary, coming from an analytic continuation of the unstable mode. This imaginary correction to the wormhole amplitude indicates, if we were to analytically continue to real time, an instability of the wormhole to brane-antibrane nucleation. This instability would proceed with a decay time given by the the exponential of minus the instanton action, of $O(\exp(-b^3 N V))$. The initial condition of the nucleated $D3-\overline{D3}$ brane pair is the one prepared by the portion of the Euclidean instanton with $\tau <0$, glued to real time along the $\tau = 0$ surface. The $D3$ brane is then prepared at rest at $\rho \approx 0.3$ and the $\overline{D3}$ brane at $\rho \approx -0.3$. The 5-form flux then expels the 3-branes from the bottleneck at $\rho = 0$, sending them into the near-boundary region. If the branes were exactly BPS, then they would be sent all the way to the AdS boundaries. If we are discussing nearly BPS branes in the $D1/D5$ system, so that $x$ is just slightly above unity, then they are sent to large but finite $\rho$, near the value $\rho_0$ we found in the last Subsection.

Note that the periodicity of Euclidean time does not matter for the purpose of finding these instantons. When these trajectories close, they do so irrespective of $\beta$.

\begin{figure}[t]
\begin{center}
\includegraphics[width=4in]{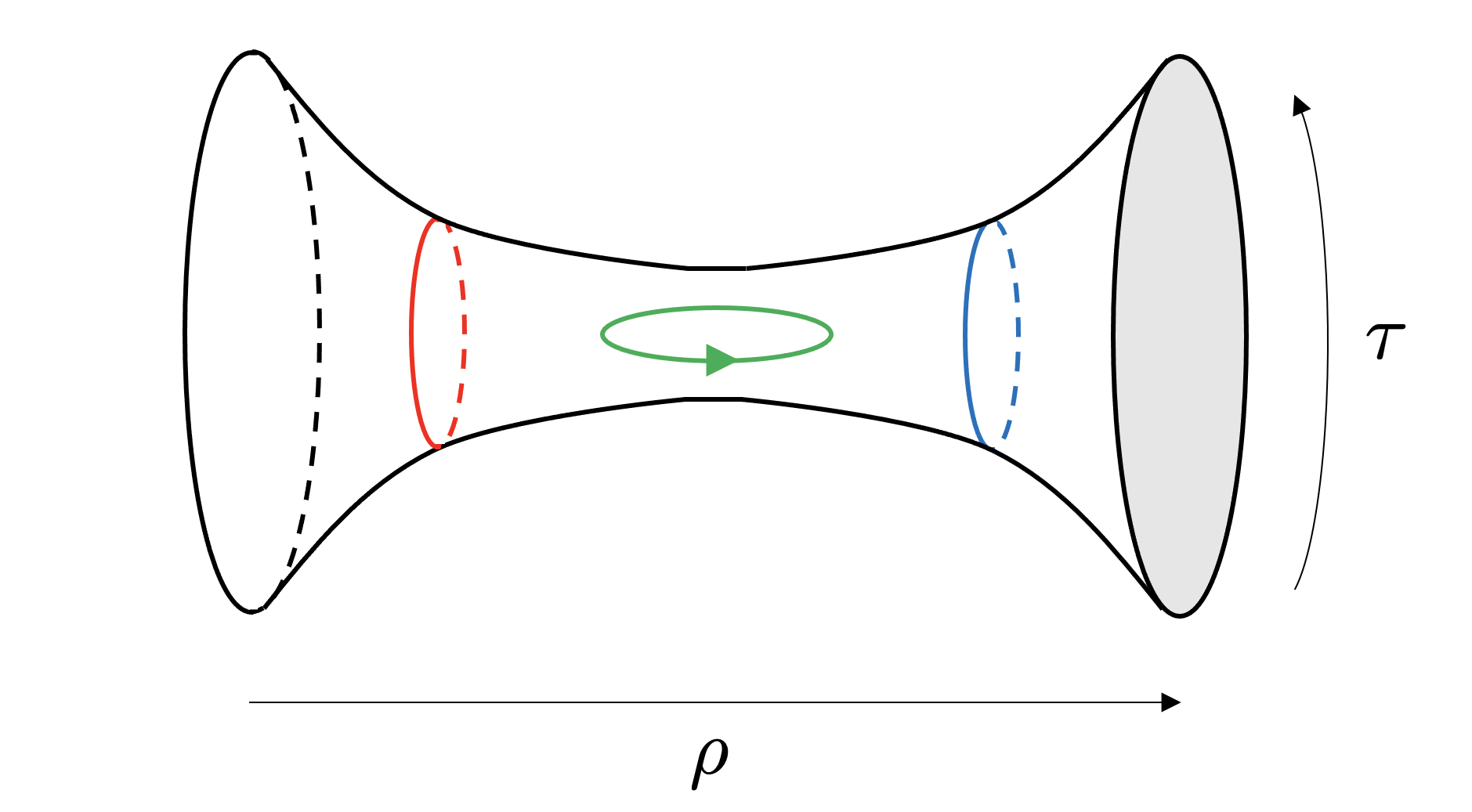}
\caption{\label{F:braneSummary} A visual summary of the two brane instantons found in this Section. The circle is the Euclidean time direction, and we have suppressed the other directions. The blue/red pair wrapping the time circle are the $D3-\overline{D3}$ brane pair at constant $\rho$ from Subsection~\ref{S:basicBrane}. For the $\mathbb{S}^1\times\mathbb{S}^3$ wormhole, this arrangement is a saddle of probe 3-branes and, if $b>b_c$, placing the branes there lowers the wormhole action. To the left of the $\overline{D3}$-brane, and to the right of the $D3$-brane, there are $N$ units of 5-form flux, while in between there are $N-1$. The green circle is the instanton from Subsection~\ref{S:otherBrane}. This configuration leads to an instanton correction to the wormhole amplitude, again for $b>b_c$. If we analytically continued $\tau$ to real time then this instanton leads to a dynamical instability of the wormhole to the production of a 3-brane pair.}
\end{center}
\end{figure}

We find very similar results for $D3$-brane trajectories in the $\mathbb{S}^1\times\mathbb{S}^3$ wormhole again with $\beta_1 = \beta_2 = \beta$, in which the $D3$-brane wraps $\mathbb{S}^3$, sits at a point in $\mathbb{S}^5$, and sweeps out a curve in Euclidean time $\tau$ and the radial coordinate $\rho$. Earlier we found that there were configurations with nucleated $D3-\overline{D3}$ brane pairs that lowered the wormhole action, as long as the bottleneck size $b$ was above a critical value $b_c$. We again find closed loop trajectories for $D3$-branes near the bottleneck, only when $b>b_c$, which qualitatively resemble those in Fig.~\ref{F:otherInstanton}. However for $b<b_c$ these trajectories are no longer closed loops, instead going out to the boundary. 

These instanton corrections when $b>b_c$ again have positive action, and so give exponentially suppressed contributions to the wormhole amplitude. However, for those configurations we have studied, we have seen that fluctuations of the ``radius'' of the trajectory are wrong-sign, leading to an imaginary one-loop determinant. The physics of these instantons is then the same as those in the torus wormhole.

We summarize both types of brane instantons in Fig~\ref{F:braneSummary}.

\subsection{Lorentzian evolution and the spectral form factor}

As we reviewed in Section~\ref{S:RMT}, the spectral form factor is a useful analytic continuation of the two-point function of partition functions. Under the provisional assumption that the two-boundary problem in AdS gravity computes the two-point function of partition functions in the dual description, we see that we can get the spectral form factor of the dual to AdS gravity by taking the two-boundary problem and performing an analytic continuation $\beta_1 \to \beta+i T$ and $\beta_2 \to \beta - i T$. The disconnected contribution from Euclidean black holes quickly decreases. What about connected geometries, i.e. wormholes, and wormholes with nucleated brane pairs? 

For real $\beta_1, \beta_2>0$, we saw in Subsection~\ref{S:basicBrane} that the AdS$_5\times\mathbb{S}^5$ torus wormhole has a runaway instability to 3-brane nucleation. This instability was absent for the $D1/D5$ system at generic points in the moduli space. In that case there are still nucleation instabilities, as there are wormholes with nucleated string pairs that obey the same boundary conditions as the wormhole with no string pairs and have lower action. These string pairs partially screen the flux supporting the geometry, and dominate over the wormhole in the connected two-boundary amplitude.

However, this is no longer the case when we analytically continue $\beta_1$ and $\beta_2$ and work at late times. Take the full action~\eqref{E:torusGeneralS} of the string pair in an AdS$_3$ wormhole at general $\beta_1, \beta_2$, obtained from the Euclidean computation, and continue $\beta_1 \to \beta + i T, \beta_2 \to \beta-iT$. Let us hold $\beta$ fixed while taking $T$ large. Then
\beq
\label{E:torusBranelateTime}
	S_{\rm pair} = \mu b^2V\sqrt{x^2-1}\,T + O(T^0)\,,
\eeq
which is large and positive at late times. So the nucleation instability is lifted by finite Lorentzian time evolution. 

This is consistent with the early time result when the mass-to-charge ratio $x$ approaches 1. This can be obtained from~\eqref{E:torusPairsmallx} upon analytic continuation, giving
\beq
	S_{\rm pair} = \frac{\mu b^2 V \beta}{2}\left( \ln(x-1) + \ln \left(\frac{T}{\beta}\right) + \hdots\right)\,,
\eeq
for $x-1\ll \beta/T\ll  1$. In this limit evolution increases the action of the brane pair.

\begin{figure}[t]
\begin{center}
\includegraphics[width=3in]{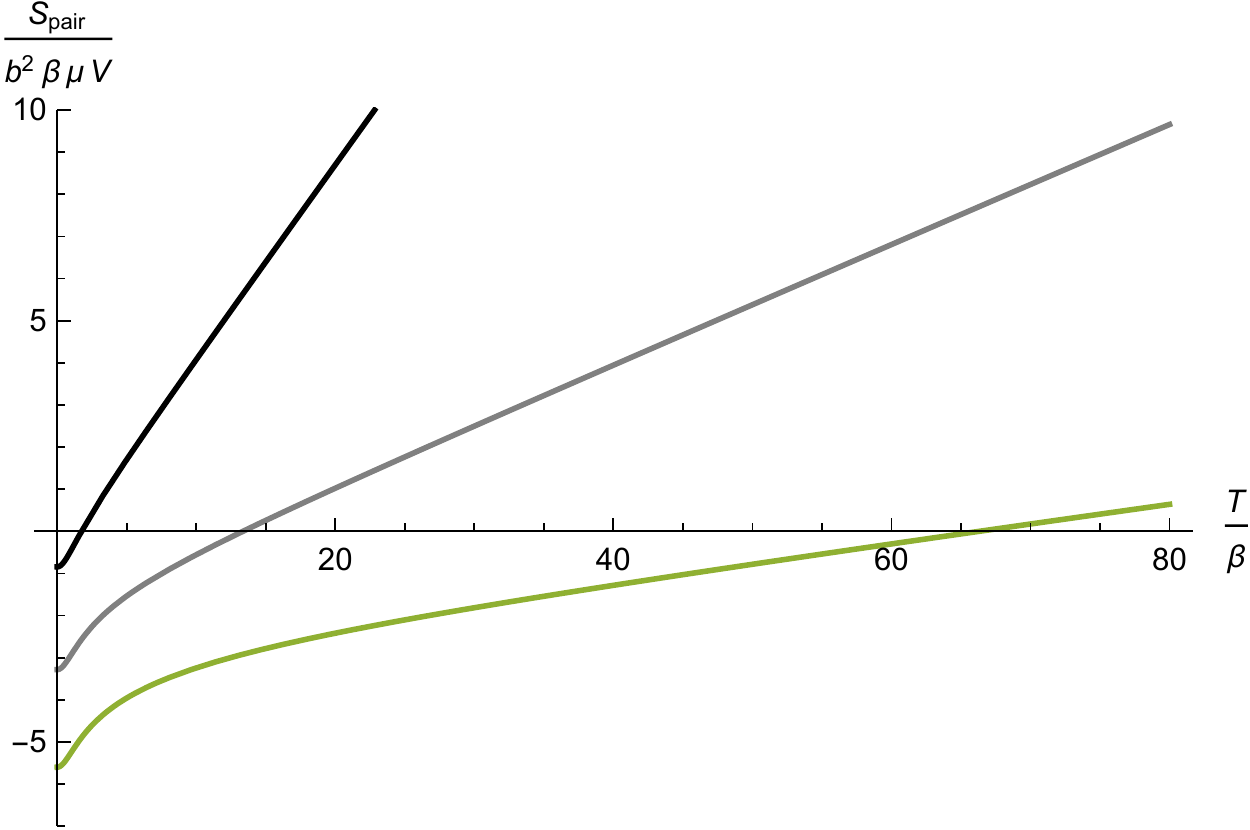} \quad \includegraphics[width=3in]{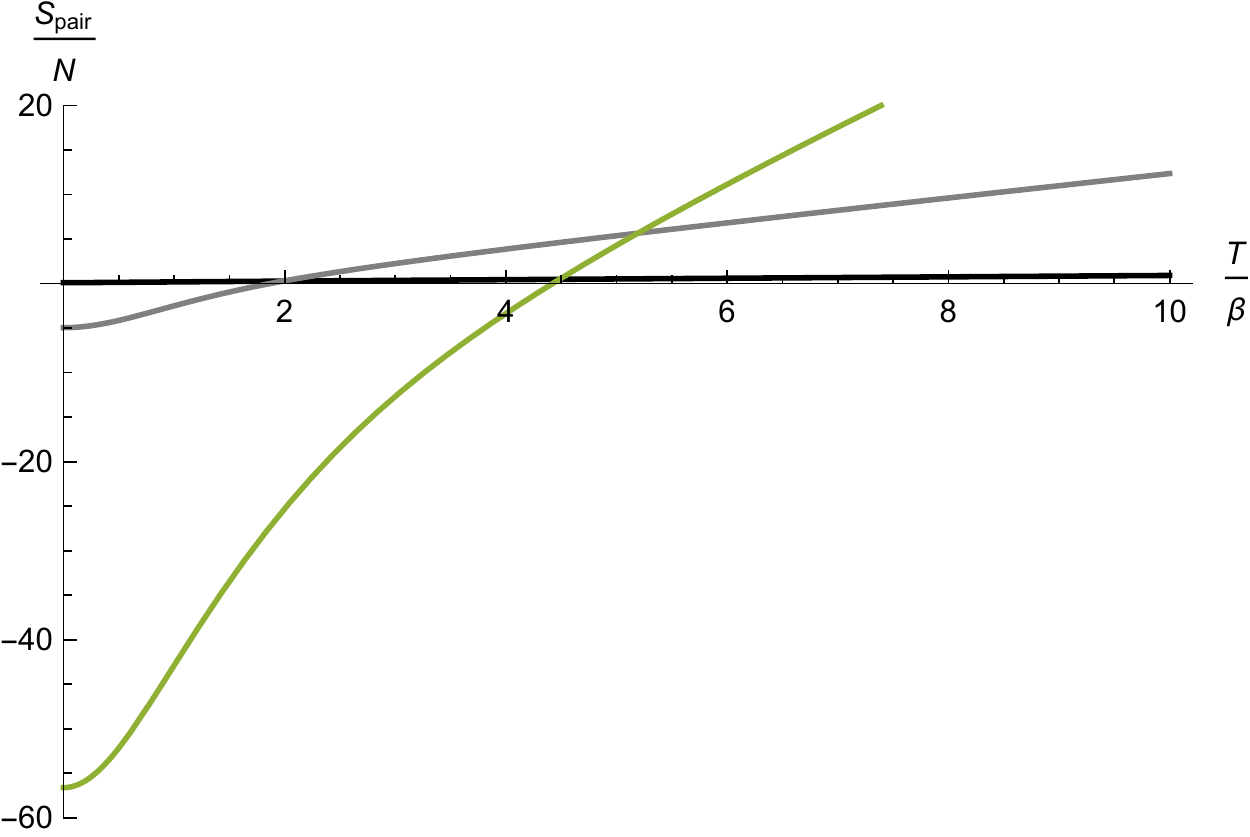}
\caption{\label{F:torusLift} (Left) The action of the 3-brane pair in the torus wormhole, analytically continued to real time. The upper black curve corresponds to a mass-to-charge ratio $x=1.1$, the middle gray curve to $x=1.01$, and the lower curve to $x=1.001$. The time $T_c$ at which the action passes through zero denotes the time after which the brane nucleation instability is lifted. 
(Right) The action of the non-BPS string pair in an AdS$_3\times\mathbb{S}^3\times\mathbb{T}^4$ wormhole with torus boundary, analytically continued to real time. The black curve corresponds to $b=1.2$, for which the action is always positive; the gray curve to $b=2$, and the dark green curve to $b=3$. The wormhole is then stable against brane nucleation after sufficient real time evolution.}
\end{center}
\end{figure}

At intermediate times it is instructive to simply plot the action of the brane pair~\eqref{E:torusGeneralS} as a function of time. We do so in Fig.~\ref{F:torusLift} for a few small values of $x$. The time $T_c$ where the action of the brane pair crosses zero is the time after which the wormholes dominate over wormholes with nucleated brane pairs. Numerically we find that $T_c$ is of order $\frac{-\ln(x-1)}{\sqrt{x^2-1}}$ for small $x-1$. 

We may also consider the effect of Lorentzian time evolution on the 3-brane pair in the $\mathbb{S}^1\times\mathbb{S}^3$ wormhole. Our results are qualitatively similar to those of the torus wormhole, with small $x$ replaced by large $b$. For large $b$ and fixed $\beta_1$, $\beta_2$, our result for the action of the pair~\eqref{E:largebSpair} becomes for $b\gg \frac{T}{\beta} \gg 1$
\beq
	S_{D3-\overline{D3}} = \frac{b^4N}{4}\left( - \ln b^2 + \ln \left(\frac{T}{\beta}\right)+ \hdots\right)\,,
\eeq 
so that early time evolution increases the action of the brane pair. We can also perform an asymptotic analysis at late time with $b$ held fixed. One finds an action of the form
\beq
	S_{D3-\overline{D3}} = f(b)NT+ O(T^0)\,, \qquad f(b) = \frac{3^{\frac{3}{2}}}{2^{\frac{11}{3}}}b^{\frac{8}{3}} + O(b^{\frac{4}{3}})\,,
\eeq
where we have indicated the form of $f(b)$ at large $b$, but more generally one has $f(b)\geq 0$.  
So the nucleation instability is again lifted by finite time evolution. We find that the time it takes to lift the instability scales with $b$, becoming larger as $b$ increases. 

The punchline from this analysis is that, while our wormholes are subject to brane nucleation instabilities, these instabilities disappear when computing the spectral form factor at late enough Lorentzian time. (For the torus wormholes this only occurs when there are no exactly BPS branes in the bulk.) This timescale is much shorter than that at which we might hope to see a ramp in the spectral form factor, with $T = O(e^{N^2})$, and so the 3-brane pairs contribute to the very early time physics.

\section{Discussion}
\label{S:discussion}

In this manuscript we have argued that Euclidean wormholes encode the coarse-grained energy level statistics of black hole microstates in AdS. Using either the method of constrained instantons or via the $\lambda$-solutions described in Subsection~\ref{S:complement}, these wormholes give connected contributions to the two-boundary problem in AdS quantum gravity. We studied simple and rather generic wormholes where the boundary geometry has a thermal circle, like Euclidean black holes. We found an integral representation for the wormhole amplitude where the integration variable is the energy carried by the wormhole. When this energy is large enough to correspond to a macroscopic geometry, the integrand has a saddle-point expansion around a particular macroscopic wormhole. However, there is a Boltzmann-like suppression of macroscopic wormholes, suggesting that the full wormhole amplitude is UV dominated.

This full amplitude depends on the sizes $\beta_1$ and $\beta_2$ of the thermal circles on the two asymptotic boundaries of the geometry, and as such is naturally in canonical ensemble. Performing a transformation to microcanonical ensemble (with a little smearing and some additional Lorentzian time evolution as described after~\eqref{E:fromWormholesToDoubleCone}), this transformed version of the amplitude now admits a saddle-point expansion where loops can in principle be computed. The saddle point is a particular analytic continuation of these Euclidean wormholes, and is none other than the double cone geometry of Saad, Shenker, and Stanford~\cite{Saad:2018bqo}, with moduli fixed by the integral transform. Further, we can identify the dual observable to which this saddle contributes, namely a modified version of the connected two-point function of the density of states.  Our results give new evidence for level repulsion in the microstate spectrum, although a more complete one-loop analysis is required to support this claim. Physically this quantity is a smeared version of the two-point function of the density of states, far away from the edge of the black hole microstate spectrum. We can also find a bulk observable with a saddle point approximation, dominated by a macroscopic Euclidean wormhole, by simply inserting a delta function constraint on the boundary energy into the integral over wormholes, although it is not yet clear what this observable means.

While we initially worked in Einstein gravity with cosmological constant or effective field theories of Einstein gravity coupled to matter, we adapted the logic to find wormholes and wormhole amplitudes in type IIB supergravity on Euclidean AdS$_5\times\mathbb{S}^5$, that is, in the AdS/CFT correspondence. Our methods can be easily adapted to any of the standard ``lampposts'' of holographic duality, like the $D1/D5$ system, 11d supergravity on Euclidean AdS$_4\times\mathbb{S}^7$, etc. By embedding these wormholes into full AdS/CFT we were able to study non-perturbative decay channels of these Euclidean wormholes, like the nucleation of $D3-\overline{D3}$ brane pairs a la the Schwinger effect. While the $\mathbb{S}^1\times\mathbb{S}^3$ Euclidean wormholes were generically unstable to brane nucleation, we found that the instability goes away under the analytic continuation required to obtain the Lorentzian spectral form factor, at least after enough real time evolution. This tells us that (i) the nucleation instability is unimportant when computing the late time spectral form factor, and (ii) the nucleated 3-brane pairs give corrections to the early time behavior of the spectral form factor.

So we see that wormholes generically and robustly give a connected contribution to the two-boundary problem in AdS quantum gravity, in the ``vanilla'' instance where the boundaries are flat or positively curved and no sources are turned on for operators dual to bulk matter. Crucially, from the full amplitude we obtain observables which have a saddle point approximation accessible in gravitational effective field theory.

While in the present work we have focused on Euclidean wormholes in the AdS$_5\times\mathbb{S}^5$ near-horizon geometry supported by $N$ units of 5-form flux, we have also found analogous wormholes in the full asymptotically flat geometry. We also expect that the method of constrained instantons may be used to locate wormholes with more than two boundaries, or with more complicated topologies.

The AdS wormholes we studied, however, raise a problem by giving a controlled contribution to the connected part of the two-boundary problem. The problem pointed out in~\cite{maldacena2004wormholes} is that we expect the dual description of stringy examples of AdS/CFT to be single CFTs, for which the partition function on a space $\mathcal{M}_1\cup \mathcal{M}_2$ factorizes by locality, 
\beq
	Z_{\rm CFT}[\mathcal{M}_1\cup \mathcal{M}_2] = Z_{\rm CFT}[\mathcal{M}_1] Z_{\rm CFT}[\mathcal{M}_2]\,.
\eeq
By the usual AdS/CFT dictionary we expect that the CFT on the disconnected space $\mathcal{M}_1\cup \mathcal{M}_2$ is dual to string theory on a space with two asymptotic boundaries $\mathcal{M}_1$ and $\mathcal{M}_2$. But then we run into a factorization paradox: if the dual description is a single CFT, then the connected part of the two-boundary problem must identically vanish. There is an obvious tension, since we can obtain quantities from the connected part that, while being non-perturbatively small compared to $Z_{\rm CFT}$, admit a semiclassical approximation.

We would like to emphasize the generic nature of the wormholes studied in this work. These wormholes arise when the boundary is positively curved are not supported by special matter profiles. Further, because of the twist zero mode volume, which depends on the data held fixed on the boundaries and which can be dialed to be quite large, the wormhole amplitudes can be continuously scanned in such a way as to be relatively large, while yet being suppressed relative to disconnected contributions.

One way out is that perhaps there is a ``stringy exclusion principle'' that, for reasons unknown, forbids these wormhole geometries. Of course we cannot rule out such a possibility, but it seems difficult to imagine a rule for quantum gravity which allows different topologies to contribute sometimes, as in the Hawking-Page transition, the genus expansion of worldsheet string theory, or the more recent replica wormhole story in JT gravity~\cite{penington2019replica,almheiri2020replica}, but not these. 

Another possibility is that there are more complicated topologies and perhaps non-geometric, ``stringy'' contributions to the two-boundary problem which exactly cancel the wormhole contributions. We do not know how to address this possibility concretely. However, we would like to stress two points here. The first is that the wormholes studied here lead to a very particular (and likely quite intricate) form of the factorization-violating amplitude as a function of boundary data. It seems to be a tall order to expect this precise form to be canceled by a set of unknown contributions. The second is that the factorization-violating amplitude appears to encode something physically sensible, namely level repulsion in the black hole microstate spectrum. While we expect a theory of quantum gravity with a single CFT dual to have a disconnected spectral form factor with a ramp upon performing a smoothing or running time average of the wildly fluctuating curve~\cite{Cotler:2016fpe}, the wormhole does not contribute to this disconnected part at all.  As such, if the wormhole was canceled out by another contribution, then paradoxically it would not account for level repulsion in the final answer.

If factorization is restored in some yet-unknown way, then it would be nice if there is some modified question to which the wormholes give the answer. In particular, imagine taking an ensemble average over the Yang-Mills coupling of $\mathcal{N}=4$ super Yang-Mills, with a narrow variance around a mean value.  Or alternatively we could turn on other couplings and disorder average them (see e.g.~\cite{gao2017traversable, maldacena2018eternal}).  One might hope that the factorization-healing contributions disappear under such averages, while the wormhole survives.

However, we would also like to consider the prospect that the effective field theory description of gravity is telling us something worth listening to, take the wormhole answer seriously, and see where it leads.  In particular, the wormholes suggest that the spectrum of black hole microstates is ensemble-averaged even in AdS$_5\times\mathbb{S}^5$. How can this possibly be?

An immediate objection to the possibility of disorder in the standard AdS/CFT duality is that $\mathcal{N}=4$ SYM is essentially unique. It is completely specified by the field content, symmetries, gauge group, and the complexified Yang-Mills coupling. So to have a disorder average where each member of the ensemble obeys the superconformal symmetry, it seems one can only have a distribution for the Yang-Mills coupling. This distribution would have to be extremely narrow in order to be consistent with the large number of precision tests of the duality. 

Yet there is good reason to expect that a distribution for the Yang-Mills coupling cannot be alone responsible for the wormholes. Our methods in AdS$_5\times\mathbb{S}^5$ can be adapted to the AdS$_7\times\mathbb{S}^4$ near-horizon geometry of $N$ coincident $M5$-branes, leading to a factorization paradox. However the $M5$-brane theory has no marginal couplings to average over.

So suppose now that there are distributions for the couplings of other operators. These couplings necessarily break conformal invariance, and so this possibility is tightly constrained by the holographic dictionary. A particular stringent test is the following. Consider type IIB string theory on global AdS$_5\times\mathbb{S}^5$. Conformal invariance on $\mathbb{R}\times\mathbb{S}^3$ forbids primary operators from acquiring a vev. Boundary conformal invariance is dual to the isometries of AdS$_5$, and so we expect that the bulk computation of the vev vanishes to all orders in bulk perturbation theory. Now suppose that a source $\lambda$ (in units of the $\mathbb{S}^3$ radius) for a non-marginal operator $W$ is turned on to some small value. Then the vev of $V$ is approximately given by a second-order result in conformal perturbation theory of the form $\langle V\rangle \sim C_{VWW}\lambda^2$ where $C_{VWW}$ is the OPE coefficient of $V$ with two $W$'s. If in gravity somehow $\lambda$ was drawn from a narrow distribution with a width $\delta\lambda$, then gravity would give us a vev $ \langle V\rangle \sim C_{VWW}\delta \lambda^2$. In order to be consistent with bulk effective field theory we conclude that the variance must be smaller than any power of $1/N$, a non-perturbatively suppressed effect. 

Nonetheless, if there is some ensemble averaging taking place, then we would want the variances to be large enough to generate the wormhole amplitudes studied in this work. These are suppressed by factors of $O(e^{-N^2})$ relative to disconnected contributions. This appears to be too large to be generated by a small number of sources drawn from distributions with non-perturbatively small width. In $\mathcal{N}=4$ SYM, if we pick operator dimension cutoff $\Lambda = O(N^2)$, there are $O(e^{N^2})$ different operators with dimensions below $\Lambda$, nearly all of which correspond to black hole microstates. A possibility that we have not yet been able to rule out is that this large number of operators have sources with exponentially small variances generated by spacetime wormholes. We wonder if this is enough to parametrically generate wormhole amplitudes of the size found in this work, while remaining consistent with precision tests of the holographic correspondence.

While our discussion here is speculative, we would like to emphasize that the question at hand -- is there a mechanism for ensemble-averaging $\mathcal{N}=4$ SYM which is consistent with precision tests of holography and can generate connected correlations of the right size to correspond to wormhole amplitudes? -- is relatively concrete, and can be parametrically addressed before making a concerted search for factorization-restoring amplitudes.

Finally, we note that if indeed there is intrinsic disorder in standard AdS/CFT along the lines discussed above, then a version of the factorization paradox would still remain.  Namely one would expect a single realization of the disorder to have a gravity dual, but now with additional non-perturbatively suppressed corrections that restore factorization.

Whether factorization is miraculously restored by other contributions, or is broken by intrinsic disorder even in AdS$_5\times\mathbb{S}^5$, we conclude by stressing the genericity and robustness of the paradox, made all the more strange by the fact that the wormholes appear to encode the reasonable physics of level repulsion in the spectrum of black hole microstates.

\subsection*{Acknowledgments}
We would like to thank O.~Aharony, A.~Maloney, M.~Mueller, S.~Rezchikov, S.~Shenker, D.~Stanford, and A.~Strominger for enlightening discussions. We would also like to thank R.~Mahajan, D.~Marolf, and J.~Santos for coordinating joint submission to arXiv of their work~\cite{Mahajan2021wormholes}. JC is supported by a Junior Fellowship from the Harvard Society of Fellows, as well as in part by the Department of Energy under grant DE-SC0007870. KJ is supported in part by start-up funds from the University of Victoria.

\bibliography{refs}
\bibliographystyle{JHEP}

\end{document}